\documentclass[8.5pt,twoside,twocolumn]{article}
\usepackage[super,sort&compress,comma]{natbib} 
\usepackage{times,mathptmx}
\usepackage{sectsty}
\usepackage{balance} 

\usepackage{graphicx} %eps figures can be used instead
\usepackage{lastpage}
\usepackage[format=plain,justification=raggedright,singlelinecheck=false,font=small,labelfont=bf,labelsep=space]{caption} 
\usepackage{fancyhdr}
\usepackage{amsmath}
\usepackage{nccmath}
\usepackage{graphicx}
\usepackage{color}
\usepackage{upgreek}
\usepackage[linkcolor = blue, citecolor = blue, urlcolor = blue, colorlinks = true]{hyperref}
\usepackage[version=3]{mhchem}
\usepackage{commath,amssymb}
\usepackage{bm}
\usepackage[capitalise]{cleveref}
\usepackage{textcomp}
\usepackage{ushort}
\usepackage[dvipsnames]{xcolor}
\usepackage{tikz}
\usepackage{amssymb}
\usepackage{wasysym}
\usepackage{widetext}

\newcommand{\nm}{\, \mathrm{n m}}

\newcommand{\mM}{\, \mathrm{m mol/L}}

\newcommand{\umpers}{\, \mathrm{\upmu m s^{-1}}}
\newcommand{\pers}{\, \mathrm{s^{-1}}}

\newcommand{\um}{\, \mathrm{\upmu m}}

\newcommand{\M}{\, \mathrm{mol/L}}

\newcommand{\matrify}[1]{\ushort{\ushort{ #1}}}
\newcommand{\vectify}[1]{\ushort{ #1}}

\begin{document}

\thispagestyle{plain}
\fancypagestyle{plain}{
\fancyhead[L]{\includegraphics[height=8pt]{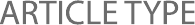}}
\fancyhead[C]{\hspace{-1cm}\includegraphics[height=20pt]{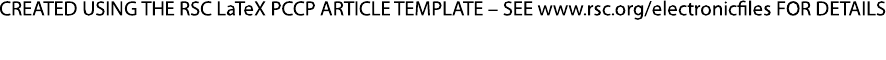}}
\fancyhead[R]{\includegraphics[height=10pt]{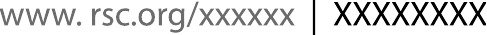}\vspace{-0.2cm}}
\renewcommand{\headrulewidth}{1pt}}
\renewcommand{\thefootnote}{\fnsymbol{footnote}}
\renewcommand\footnoterule{\vspace*{1pt}% 
\hrule width 3.4in height 0.4pt \vspace*{5pt}} 
\setcounter{secnumdepth}{5}

\makeatletter 
\def\subsubsection{\@startsection{subsubsection}{3}{10pt}{-1.25ex plus -1ex minus -.1ex}{0ex plus 0ex}{\normalsize\bf}} 
\def\paragraph{\@startsection{paragraph}{4}{10pt}{-1.25ex plus -1ex minus -.1ex}{0ex plus 0ex}{\normalsize\textit}} 
\renewcommand\@biblabel[1]{#1}            
\renewcommand\@makefntext[1]% 
{\noindent\makebox[0pt][r]{\@thefnmark\,}#1}
\makeatother 
\renewcommand{\figurename}{\small{Fig.}~}
\sectionfont{\large}
\subsectionfont{\normalsize} 

\newtoks\rowvectoks
\newcommand{\rowvec}[2]{%
 \rowvectoks={#2}\count255=#1\relax
 \advance\count255 by -1
 \rowvecnexta}
\newcommand{\rowvecnexta}{%
 \ifnum\count255>0
 \expandafter\rowvecnextb
 \else
 \begin{pmatrix}\the\rowvectoks\end{pmatrix}
 \fi}
\newcommand\rowvecnextb[1]{%
 \rowvectoks=\expandafter{\the\rowvectoks&#1}%
 \advance\count255 by -1
 \rowvecnexta}

\newcommand{\matr}[1]{\boldsymbol{#1}} % undergraduate algebra version
\newcommand*{\balancecolsandclearpage}{
  \cleardoublepage
  \twocolumngrid
}

\definecolor{ABpurple}{RGB}{128, 0, 128}
\definecolor{ABred}{RGB}{255, 0, 0}
\definecolor{ABgreen}{RGB}{0, 255, 0}
\definecolor{ABbrown}{RGB}{128, 64, 0}

\fancyfoot{}
\fancyfoot[LO,RE]{\vspace{-7pt}\includegraphics[height=9pt]{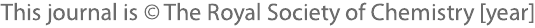}}
\fancyfoot[CO]{\vspace{-7.2pt}\hspace{12.2cm}\includegraphics{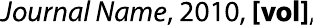}}
\fancyfoot[CE]{\vspace{-7.5pt}\hspace{-13.5cm}\includegraphics{RF}}
\fancyfoot[RO]{\footnotesize{\sffamily{1--\pageref{LastPage} ~\textbar  \hspace{2pt}\thepage}}}
\fancyfoot[LE]{\footnotesize{\sffamily{\thepage~\textbar\hspace{3.45cm} 1--\pageref{LastPage}}}}
\fancyhead{}
\renewcommand{\headrulewidth}{1pt} 
\renewcommand{\footrulewidth}{1pt}
\setlength{\arrayrulewidth}{1pt}
\setlength{\columnsep}{6.5mm}
\setlength\bibsep{1pt}

\twocolumn[
  \begin{@twocolumnfalse}
\noindent\LARGE{\textbf{Ionic Screening and Dissociation are Crucial for Understanding Chemical Self-Propulsion in Water}}
\vspace{0.6cm}

\noindent\large{\textbf{Aidan T. Brown,$^{\ast}$\textit{$^{a}$} Wilson C. K. Poon,\textit{$^{a}$}, Christian Holm,\textit{$^{b}$} and
Joost de Graaf\textit{$^{a, b}$}}}\vspace{0.5cm}
%Please note that \ast indicates the corresponding author(s) but no footnote text is required. 

\noindent\textit{\small{\textbf{Received Xth XXXXXXXXXX 20XX, Accepted Xth XXXXXXXXX 20XX\newline
First published on the web Xth XXXXXXXXXX 200X}}}

\noindent \textbf{\small{DOI: 10.1039/b000000x}}
\vspace{0.6cm}
%Please do not change this text.

\noindent \normalsize{Water is a polar solvent and hence supports the bulk dissociation of itself and its solutes into ions, and the re-association of these ions into neutral molecules in a dynamic equilibrium, e.g., \ce{H2O2 <=> H+ + HO2-}. Using continuum theory, we study the influence of these reactions on the self-propulsion of colloids driven by surface chemical reactions (chemical swimmers) in aqueous solution. The association-dissociation reactions are here shown to have a strong influence on the swimmers' behaviour, and must therefore be included in future modelling. In particular, such bulk reactions permit charged swimmers to propel electrophoretically even if all species involved in the surface reactions are neutral. The bulk reactions also significantly modify the predicted speed of chemical swimmers propelled by ionic currents, by up to an order of magnitude.  For swimmers whose surface reactions produce both anions and cations (ionic self-diffusiophoresis), the bulk reactions produce an additional reactive screening length, analogous to the Debye length in electrostatics. This in turn leads to an inverse relationship between swimmer radius and swimming speed, which could provide an alternative explanation for recent experimental observations on Pt-polystyrene Janus swimmers [S. Ebbens~\textit{et al.}, Phys. Rev. E \textbf{85}, 020401 (2012)]. We also use our continuum theory to investigate the effect of the Debye screening length itself, going beyond the infinitely thin limit approximation used by previous analytical theories. We identify significant departures from this limiting behavior for micron-sized swimmers under typical experimental conditions, and find that the limiting behavior fails entirely for nanoscale swimmers.}
\vspace{0.5cm}
 \end{@twocolumnfalse}
  ]

%Footnotes
%\footnotetext{\dag~Electronic Supplementary Information (ESI) available online. See Appendix~\ref{videos} %for captions to Supplementary Videos 1-4. See DOI: 10.1039/b000000x/.}

%Please use \dag to cite the ESI in the main text of the article.
%If you article does not have ESI please remove the the \dag symbol from the title and the above footnotetext.

\footnotetext{\textit{$^{a}$~SUPA, School of Physics and Astronomy, The University of Edinburgh, King's Buildings, Peter Guthrie Tait Road, Edinburgh, EH9 3FD, United Kingdom, E-mail: abrown20@staffmail.ed.ac.uk}}
\footnotetext{\textit{$^{b}$~Institute for Computational Physics, Stuttgart University, Pfaffenwaldring 27, D-70569 Stuttgart}}

%%%%%%%%%
\section{\label{sec:intro}Introduction}

The 20$^\mathrm{th}$ century witnessed a revolution in condensed matter physics, due to the ready availability of well-characterised colloidal particles ($1$~nm to $10$~$\upmu$m in size). These particles are often viewed as `large atoms': they are small enough to be subject to Brownian motion, and thus to all the machinery of equilibrium statistical physics, but large enough that their microscopic dynamics and interactions can be observed and tuned. Studying colloidal particles has led to fundamental breakthroughs. Most notably, the observation and subsequent understanding of Brownian motion in colloidal systems~\cite{einstein05} led to acceptance of the molecular picture of matter.

\begin{figure}
  \centering
  \includegraphics[width=8.5 cm]{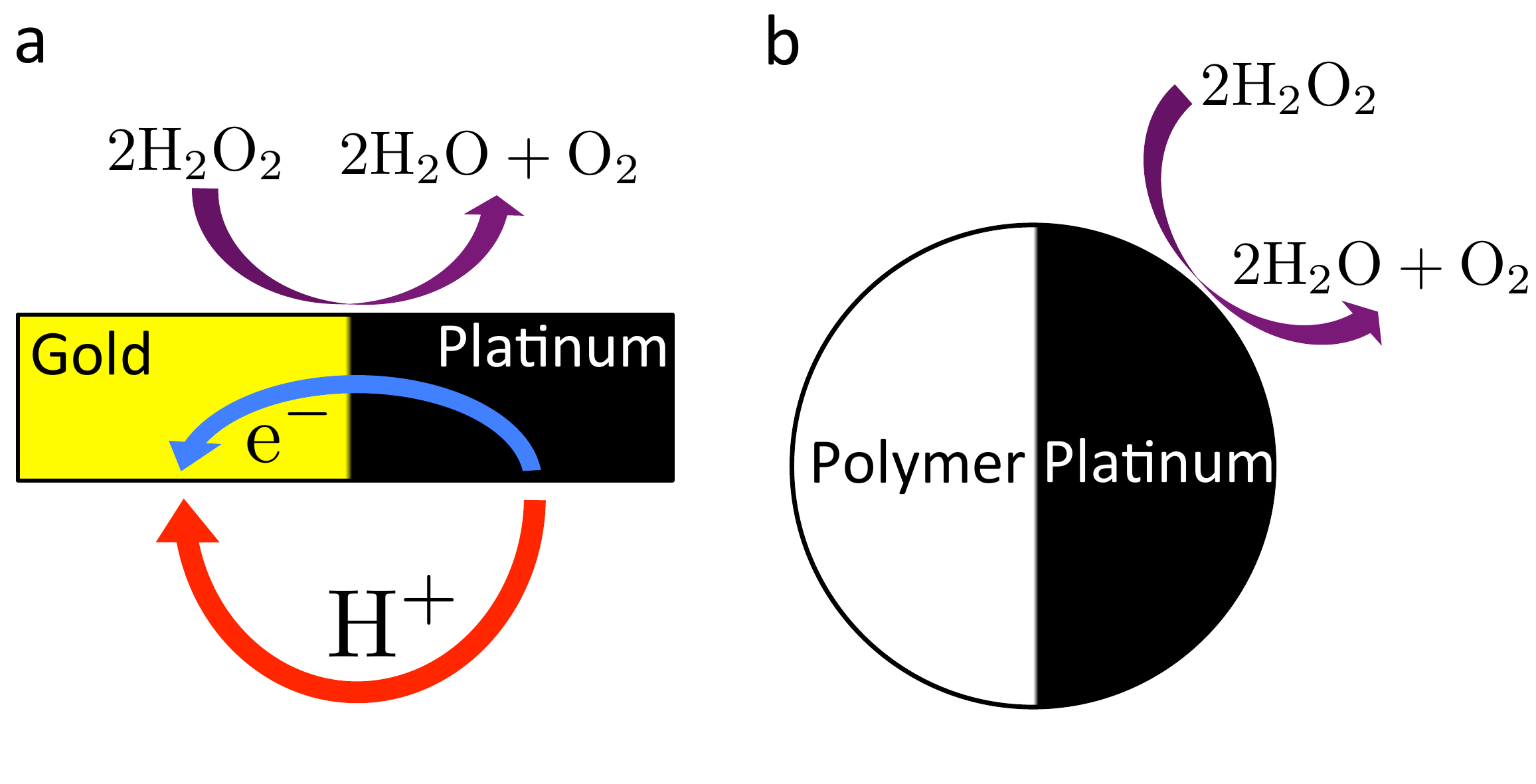}
  \caption{\label{fig:example}Cartoon of the two paradigmatic chemical swimmers discussed in the text. Both swimmers move at a few $\umpers$ in 10\% \ce{H2O2} solution, powered by the decomposition of \ce{H2O2} on their surfaces.  a)  Bimetallic (typically gold-platinum) rod~\cite{paxton04}, of typical length 2$\um$ and width 300~nm. The accepted propulsion mechanism for these swimmers is via a \ce{H+} current, as shown. b) Platinum-polymer (usually polystyrene) Janus sphere~\cite{howse07}, of typical radius 1$~\um$. }
\end{figure}

Moving into the 21$^\mathrm{st}$ century, physicists have started to recruit colloids to tackle systems that are intrinsically out-of-equilibrium, specifically where the components are themselves self-propelled. This is the field of `active matter'. A wide range of novel, self-propelled colloids~\cite{paxton04,wang06,howse07,solovev09,yoshinaga10,buttinoni12,ahmed13, ebbens10} have been synthesised --- see example sketches in Fig.~\ref{fig:example} for two designs relevant to this work. Such self-propelled colloids are intrinsically out of equilibrium --- they continuously transform chemical, thermal or electromagnetic energy into directed motion --- and recent work has focussed on using these systems to experimentally explore exciting non-equilibrium phenomena such as phase separation and collective motion~\cite{theurkauff12,palacci13,buttinoni13, ginot15}. 

In parallel with this research, much work has gone into understanding the experimental propulsion mechanisms at the level of surface chemical reactions. Working out how these tiny motors function is a fundamental problem in its own right. However, understanding the propulsion mechanism is also an essential first step in understanding the experimental collective behaviour. This is because unlike biological swimmers such as \textit{E. coli}, where the biochemical reactions responsible for propulsion take place internally, synthetic swimmers are usually propelled by external chemical reactions that directly modify the chemical, electrostatic, or temperature fields of their surroundings. These fields modify the propulsion speed of other swimmers, generating so-called `phoretic' swimmer-swimmer interactions in addition to the hydrodynamic and contact interactions experienced by all swimmers~\cite{soto14, uspal15, banigan16, brown16}. As these phoretic interactions are directly coupled to the chemical reactions responsible for propulsion, knowing how artificial swimmers self-propel is essential for understanding their collective behaviour.

This bottom-up approach contrasts with the tactic employed by most theoretical modellers of active matter, which is to explore the phenomenology arising from minimal or effective models of swimmer-swimmer interactions. For example, much theoretical work is based on the simple Vicsek model~\cite{vicsek95}, the active Brownian model~\cite{Stenhammar13,zheng13} which considers only contact forces, or hydrodynamic~\cite{matas14,zottl14,yang15} and generic phoretic interactions~\cite{thakur12,pohl14,bickel14,soto14} in isolation. While this research is valuable in unlocking generic non-equilibrium physics principles, it is dangerous to rely too much on this approach when making comparisons with experiment. To be specific, these minimal models can often qualitatively reproduce experimental behavior such as phase separation, but it is rarely clear whether this is by chance, i.e., a result of judicious tuning of free parameters, or whether the success of the effective model is because the complicated microscopic behaviour really can be reduced down to generic, coarse-grained interactions. If experimental non-equilibrium physics is to have a future, then modellers must not shy away from accounting for real, microscopic physics. Where a microscopically justified model fails is where one can discover \textit{new} physics. 

That said, current understanding of self-propulsion mechanisms is often very incomplete, and hence the necessary foundations for models which can reproduce multiparticle behaviour from a bottom-up perspective are lacking. In particular, most research so far has focussed on unravelling the surface chemistry of the swimmer itself~\cite{paxton04,paxton06,golestanian07,moran10,moran11,ebbens12,brown14,ebbens14}. This is understandable and necessary, but it has meant that other aspects, such as the chemistry of the bulk solvent, have been neglected. In this article, we show that taking into account the chemistry of the most common solvent --- water --- significantly, and often qualitatively, modifies the predicted propulsion behaviour of almost all self-propelled synthetic swimmers. We focus on propulsion here, because understanding propulsion is a necessary first step in understanding more complex behaviour. However, the mathematical model that we present is general enough for investigating interparticle interactions and collective behaviour, which we intend to explore in future work.

%%%%%%%%%
\section{Chemical Propulsion}

%%%%%%%%%%%%
\subsection{Self-electrophoresis}

We discuss here the most experimentally typical self-propelled colloids, which we term `chemical swimmers'. They are most easily defined by example. Fig.~\ref{fig:example} shows two chemical swimmers, both powered by the catalytic decomposition of hydrogen peroxide on their surfaces. Because the colloid surface is anisotropic, this reaction produces chemical gradients which, via interaction with the particle surface, eventually lead to self-propulsion.

We say `eventually' because the propulsion mechanism of these swimmers is somewhat involved. For the example given in Fig.~\ref{fig:example}a, \ce{H2O2} decomposition does not occur just by the simple chemical reaction
\begin{fleqn}
\begin{align}
  \nonumber \mathrm{(R1)} &\quad \ce{2H2O2 -> 2H2O + O2},
\end{align}
\end{fleqn}
but also occurs partially electrochemically, with two half reactions taking place preferentially on the Au or Pt surfaces
\begin{fleqn}
\begin{align}
\nonumber \mathrm{(R2)} &\quad
  \begin{array}{rcl}
    \ce{H2O2}              & \overset{\ce{Pt}}{\longrightarrow} & \ce{O2 + 2H+ + 2e-} , \\
    \ce{2e- + 2H+ + H2O2}  & \overset{\ce{Au}}{\longrightarrow} & \ce{2H2O} ,
  \end{array}  
\end{align}
\end{fleqn}
producing a proton gradient outside the colloid, which generates a local electric field. The colloid surface, like most surfaces in water, is charged, so this electric field causes electroosmotic flow over the colloid surface, leading to self-propulsion. This propulsion mechanism is called `self-electrophoresis'~\cite{wang06}. The electric field also generates a proton current outside, which is balanced by an electric current inside the conductive swimmer.

A large body of experimental evidence confirms that self-electrophoresis is the appropriate propulsion mechanism for these bimetallic swimmers. For example, their propulsion speed scales inversely with salt concentration~\cite{palacci13,brown14}, which is expected from a simple application of Ohm's law. Recent results strongly indicate that self-electrophoresis is also the appropriate propulsion mechanism for the type of colloid shown in Fig.~\ref{fig:example}b, which has a single metallic coating. This is at first surprising because there is no obvious mechanism for producing the ionic gradient needed for self-electrophoresis. However, geometrical differences between the equator and pole of the catalytic coating, such as thickness variation, may couple to the half-reaction rates in (R2) and so provide the necessary asymmetry~\cite{ebbens14}. In this paper, we go further and show that these effects are not limited to swimmers that can support ionic currents. All swimmers in aqueous solution are likely to be self-electrophoretic to a major degree, whatever their surface reaction mechanism.

%%%%%%%%%%%%
\subsection{Surface Chemistry}

Before we discuss the effects of the bulk solution, we point out one general difficulty with self-electrophoresis that will also apply to other complex propulsion mechanisms. This is that the relevant surface reaction rates are extremely hard to measure. The overall reaction rate (R1) can be easily obtained by measuring reactant or product concentrations~\cite{brown14,ebbens14}, but for self-electrophoresis the important rate is the proton production rate (R2), and this might make up only a tiny proportion of the overall reaction, most of which proceeds simply via (R1). Measuring the rate of an individual reaction pathway like (R2) is challenging, and has not yet been done, to our knowledge, for any self-propelled particle. This would not be a problem if we could predict these rates, or even assume them to be constant, but surface catalysis is a notoriously sensitive phenomenon, and these surface reaction rates are likely to vary unpredictably with almost every parameter, e.g., pH, ionic strength, and surface roughness~\cite{gileadi11}. Reversing the argument, the only currently available method of estimating these reaction rates is from the particle propulsion velocity itself. That is, with a sufficiently accurate microscopic model, the surface reaction rates can be inferred from the propulsion speed~\cite{paxton06}. The catalytic chemistry of micro-and-nano particles is of huge industrial importance, so this provides another major motivation for obtaining a detailed theoretical understanding of self-electrophoresis.

%%%%%%%%%%%%
\subsection{\label{bulk chemistry}Bulk Chemistry}

At first glance, the chemistry of the bulk solution is much simpler than that of the surface. However, a polar solvent such as water presents two complications which have not so far been taken into account. The first of these is electrostatic screening. In so-called phoretic mechanisms, such as self-electrophoresis, fluid flow is generated in a layer around the particle surface. In self-electrophoresis, the thickness of this interaction layer is given by the electrostatic screening or Debye length $\kappa^{-1}$: outside this screening layer, the free charge density, which is responsible for fluid flow, decays rapidly to zero. Theoretical studies typically make use of the thin screening assumption $\kappa a\gg 1$, where $a$ is the swimmer radius, because this dramatically simplifies the calculation of propulsion speed, flow fields, etc.~\cite{golestanian07,sabass12b,ebbens14, brown14}. However, this assumption is not generally valid: $\kappa^{-1}$ is of order 100 nm for an experimentally typical 3~M \ce{H2O2} solution~\cite{brown14}, and active colloids typically range in size from 10~nm~[\!\!\citenum{lee14}] to 10~$\um$~[\!\!\citenum{ebbens12}]. In this paper, we do not use the thin screening assumption, and we show that this dramatically reduces the predicted propulsion speeds, by up to several orders of magnitude for nanoscale swimmers.

The second complication of the aqueous environment is that water is not chemically inert. It is an `active fluid' that can be driven out of chemical equilibrium by the reactions on the particle surface. This consideration has been appreciated for biological fluids such as the cytoplasm~\cite{banigan16}, where biomolecules are continuously synthesized and broken down, but it is also true for a simple fluid such as water. This is because water permits the ionic dissociation of both itself and any polar solutes, e.g., \ce{H2O <=> H+ + OH-} and \ce{H2O2 <=> H+ + HO2-}. The implications of these reactions for self-electrophoresis are the main focus of this paper. The most striking implication is that a gradient of a neutral molecule like \ce{H2O2} will result in ionic gradients, here of \ce{H+} and \ce{HO2-} ions, which will themselves produce electric fields. This means that a surface reaction with only uncharged species like \ce{H2O2} can itself generate self-electrophoretic propulsion. Taking the bulk chemistry into account thus requires a careful reanalysis of the surface chemistry too. 

\begin{figure}
  \centering
  \includegraphics[width=8.5 cm]{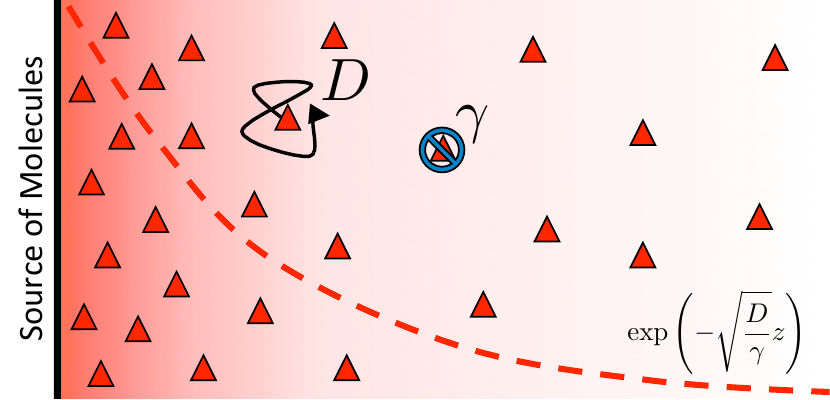}
  \caption{\label{fig:screen}Schematic representation of the simple 1D molecular screening model introduced in the text. A wall (green) releases molecules (red triangles) which diffuse with $D$ (black arrow) are consumed in the bulk with a rate $\gamma$ (blue symbol). This leads to an exponential decay of the concentration, as indicated using the dashed red line and continuum-level red gradient.}
\end{figure}

It might be argued that such effects, though large, are `merely quantitative', and introduce no new physics. It is therefore worth highlighting that these bulk reactions also introduce qualitatively new phenomena. To demonstrate this, we describe an effect called `reactive screening'~\cite{banigan16}, which will underlie our later, more detailed discussion. We illustrate this effect with a simple 1D model, see Fig.~\ref{fig:screen}. Let an uncharged molecule of diffusivity $D$ be produced uniformly at a plane surface $z=0$ and consumed in the bulk ($z>0$) with rate $\gamma$. The steady-state concentration profile $c(z)$ then obeys
\begin{align}
	D\frac{\partial^2 c}{\partial z^2} &= \gamma c , \label{simple model}
\end{align}
where diffusion, on the left, is balanced against consumption, on the right. The solution to Eq.~\eqref{simple model} is ${c=c_0\exp(-qz)}$, with $c_0$ the concentration at the surface, and the reactive screening length  $q^{-1}=(D/\gamma)^{1/2}$. This uncharged model has been applied to the diffusiophoresis of small particles inside a biological cell, where the relevant bulk reactions are the breakdown of biomolecules in the cytoplasm~\cite{banigan16}. With self-electrophoresis, as we shall see, reactive screening can also screen the electrostatic potential, effectively turning off the long-range electrostatic interactions, which would otherwise be inevitable. This reactive screening is a qualitatively new effect of bulk reactions, which cannot be ignored \textit{a priori} even at the level of phenomenological theories.

%%%%%%%%%
\section{Overview of Main Results \label{Summary}}

The theory of self-electrophoretic propulsion is mathematically involved, even without the introduction of additional bulk reactions, so we will use this section to sketch out our main results in advance. This will necessarily skim over or simplify many relevant details, but these will be addressed in later sections. This section will also aid readers who are not interested in engaging with the mathematical details to pick out the most relevant expressions in the more mathematical sections.

%%%%%%%%%%%%
\subsection{Overall Framework}

Our main mathematical result is that the self-electrophoretic propulsion speed of an arbitrary, uniformly charged, sperical swimmer can be written, if a suitable linearization is applied, in the form
\begin{align}
  U &= U^\mathrm{SM}(j^\mathrm{s},c^\mathrm{salt}, \sigma, \dots) F(\kappa a) B(qa, \dots) , \label{general result}
\end{align} 
where $U^\mathrm{SM}$ is the `standard model' propulsion speed assuming the thin screening limit without bulk reactions~\cite{golestanian07}. $U^\mathrm{SM}$ depends on, among other parameters, the surface reaction rates $j^\mathrm{s}$, the salt concentration $c^\mathrm{salt}$, and the surface charge density $\sigma$. We introduce the factors $F$ and $B$ to account for realistic electrostatic screening and bulk reactions, respectively. These factors both depend on dimensionless parameters, respectively $\kappa a$ and $qa$, with $\kappa$ and $q$ the inverse electrostatic and inverse reactive screening lengths respectively, and $a$ the swimmer radius.

\begin{figure}
  \centering
  \includegraphics[width=8.5 cm]{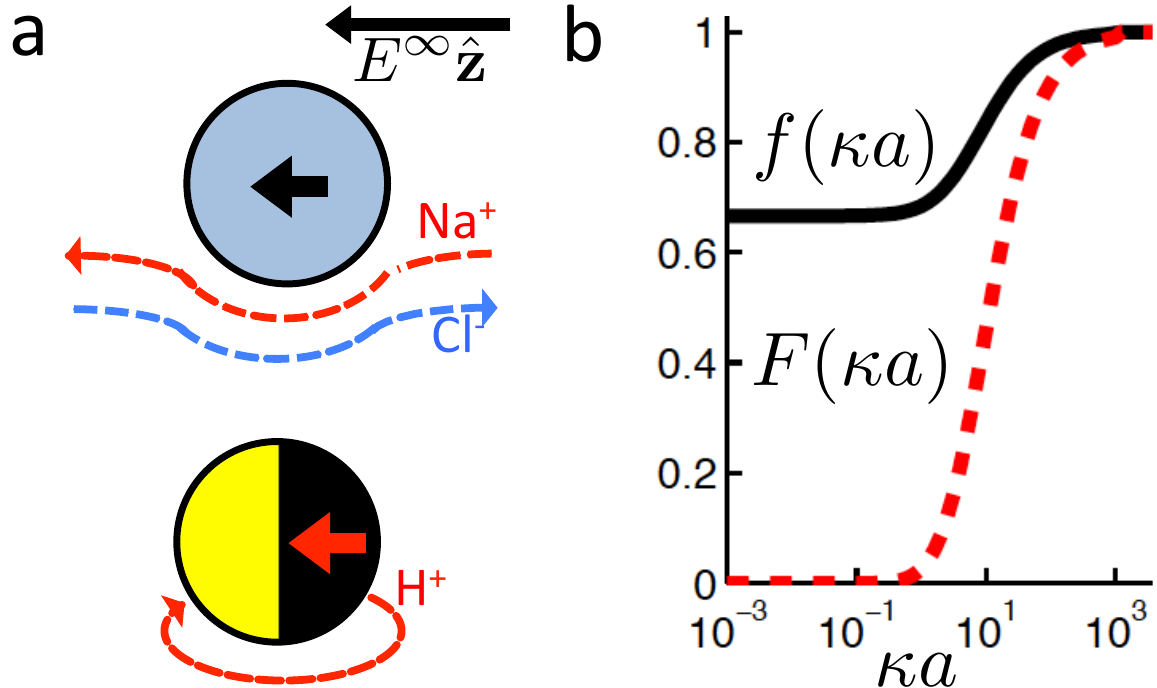}
  \caption{\label{henry} (a) Schematic showing the difference in boundary conditions between external electrophoresis (upper) and self electrophoresis (lower). Thick, colored arrows show the direction of motion of positively charged particles. (b) Henry's function $f(\kappa a)$ which determines the mobility of a particle in an external electric field (\textbf{---}), and $F(\kappa a)$, the equivalent function for self-electrophoresis (\textcolor{ABred}{\textbf{\--\--\--}}).}
\end{figure}

%%%%%%%%%%%%
\subsection{Electrostatic Screening}

The new dimensionless factor $F(\kappa a)$ is exactly analogous to the well-known function $f(\kappa a)$ (Henry's function~\cite{henry31}) that controls the speed of a particle undergoing electrophoresis in an external field via $U_\mathrm{ext} = \mu_\mathrm{E} E^\infty$, with $\mu_\mathrm{E} = \zeta \epsilon f(\kappa a)/\eta$ the electrophoretic mobility, $\epsilon$ the dielectric constant, $\zeta$ the particle's surface potential, $\eta$ the solution viscosity, and $E^\infty$ the external electric field~\cite{henry31, kim13}. Both $f$ and $F$ are plotted in Fig.~\ref{henry}. For $\kappa a\ll 1$, i.e., for small particles or low salt concentration, $F$ decreases rapidly, scaling as $(\kappa a)^3$. This is different from external electrophoresis because of the different geometries of the driving fields --- a uniform field for external electrophoresis compared with a dipole for self-electrophoresis. 

The implication of this $a^3$ scaling is that, other things being equal, nanoswimmers should swim much slower than microswimmers. Experimentally, however, nanoswimmers are found to swim faster than equivalent microswimmers~\cite{lee14}. From this we conclude that other things are not equal: either the surface reaction rates are much larger for nanoswimmers, or the standard self-electrophoresis theory does not apply for these small swimmers. If this issue can be resolved, which we do not attempt here, it will likely also give insight into the related phenomenon of directed motion in nanoscale biological enzymes~\cite{muddana10, sengupta13}.

\begin{figure}[bh!]
  \centering
  \includegraphics[width=8.5 cm]{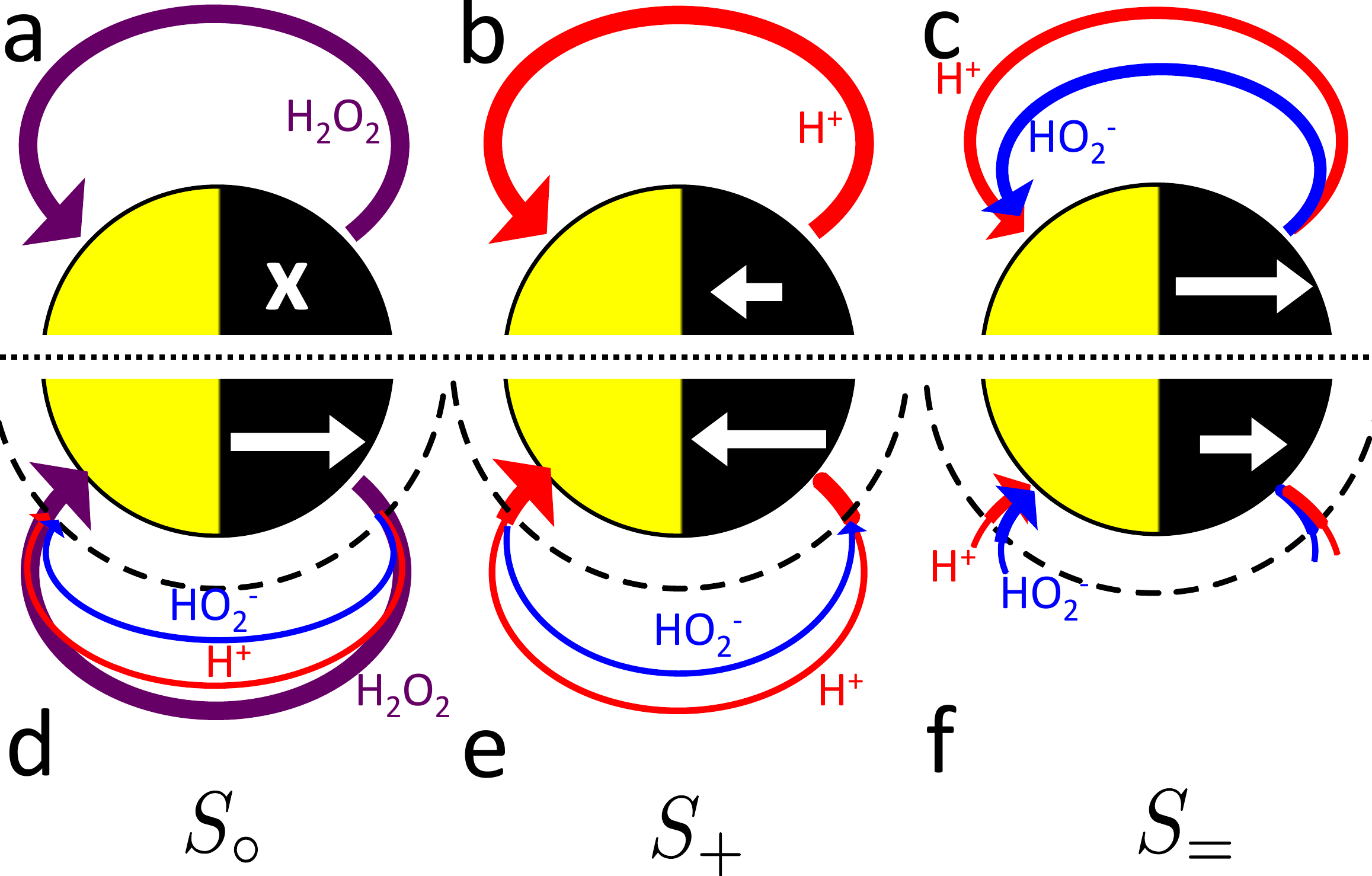}
  \caption{\label{fig:reactionEffect}Schematic of the effect of bulk ionic reactions on the propulsion of three model swimmers. The upper panel shows the system without, and the lower panel with, bulk reactions. Coloured arrows indicate fluxes of three chemical species \ce{H2O2} (purple), \ce{H+} (red) and \ce{HO2-} (blue). The thickness of the arrows corresponds very roughly to the relative intensity of the fluxes. White arrows denote the direction of particle propulsion (\textbf{x} = no propulsion). Arrow length indicates relative speed. Dashed semicircles show the approximate extent of the reactive screening length $q^{-1}$.}
\end{figure}

%%%%%%%%%%%%
\subsection{Bulk Reactions}

The effects of ionic dissociation depend upon the nature of the surface reaction responsible for propulsion. A common feature is the importance of the reactive screening length $q^{-1}$ which controls the propulsion behaviour through the parameter $qa$. We can understand why $qa$ is the relevant parameter as follows: for $qa\ll 1$, the swimmer is smaller than the reactive screening length, so any molecules produced at the swimmer surface will diffuse away or return to the swimmer surface before they have time to react. In this `reactionless limit', the swimmer will behave as though there are no bulk reactions, which is the usual, tacit assumption. For swimmers larger than the reactive screening length, $qa\gg 1$, we are in a `reactive limit' where the bulk ionic reactions dominate the behaviour. Crucially, for typical experimental conditions, e.g., 3 $\M$ [\ce{H2O2}], the reactive screening length $q^{-1}\approx 70$~nm, which is in the centre of the experimental range of swimmer radii~\cite{lee14, ebbens12}. Both the reactionless and reactive limits, and the intermediate regime ($qa\approx 1$), are therefore experimentally relevant.

We now explain the effect of bulk reactions on specific types of swimmer. The overall surface reaction we focus on is the \ce{H2O2} decomposition reaction (R1). As we have seen, this overall reaction can occur through several different pathways. We therefore define three model swimmers, shown in Fig.~\ref{fig:reactionEffect} with surface reactions that are representative of these different pathways. A real swimmer might exhibit any or all of these. 

The upper panels (a-c) of Fig.~\ref{fig:reactionEffect} show these model swimmers without bulk reactions. In (a), there is a single surface flux of neutral \ce{H2O2} molecules. This models the purely neutral decomposition of \ce{H2O2} in reaction (R1). Here we first make three general points: First, \ce{O2} and \ce{H2O} are not included here because the model is chiefly representative of the more complex physical reality: in this case we consider a reaction that produces and consumes only a single neutral species. Second, the chemical fluxes are dipolar rather than monopolar, i.e., they both leave and enter the particle surface. For a uniformally charged particle, only the dipolar component of the flux contributes to the propulsion speed~\cite{golestanian07}, so our choice of a dipolar surface-flux profile does not affect our results, and simplifies the argument. Third, we consider only self-electrophoresis here, and ignore `neutral-self-diffusiophoresis', which is propulsion generated by a direct non-electrostatic interaction between a neutral species, such as \ce{O2} and the swimmer surface~\cite{golestanian05, howse07}, and which is typically much weaker than self-electrophoresis~\cite{brown14, degraaf15}. Hence the model swimmer in (a) does not move because, without bulk reactions, a surface reaction involving only uncharged species cannot generate electric fields, and therefore cannot produce self-electrophoresis. 

In (b) a surface proton flux generates an electric field via reaction (R2). We assume the particle is positively charged, and it then swims in the direction indicated by the white arrow. This corresponds to the standard self-electrophoresis model, e.g., for Au-Pt swimmers~\cite{paxton04}. The electric field has a dipolar form, like the proton flux.

In (c), we have a new mechanism, with equal fluxes of \ce{H^+} and \ce{HO2-} ions. There is no net electrical current for this swimmer since there are equal positive and negative fluxes. However, self-propulsion still occurs. This is because the two ions diffuse at different rates (\ce{H+} faster than \ce{HO2-}), and this creates a so-called diffusion potential, which acts to prevent net charge separation. The diffusion potential leads to an associated (self-generated) electric field, which then generates motion via electrophoresis in the usual way (white arrow). This propulsion mechanism is called `ionic diffusiophoresis'~\cite{anderson89}, and is typically used to model swimmers composed of solid salts, which generate propulsion through dissolution of the swimmer itself, e.g., \ce{AgCl (s) -> Ag+ (aq) + Cl- (aq)}~\cite{duan12}. We include this model here because ionic diffusiophoresis may contribute to the propulsion of Pt-Janus swimmers, for example via
\begin{fleqn}
\begin{align}
\nonumber \mathrm{(R3)} &\quad
  \begin{array}{lcl}
    \ce{2H2O2}             & \overset{\ce{Pt}}{\longrightarrow} & \ce{2H+ + 2OH- + O2}
  \end{array}
\end{align}
\end{fleqn}
with subsequent recombination of \ce{H+} and \ce{OH-} in the bulk. However, note that reaction (R3) is \textit{not} that shown in Fig.~\ref{fig:reactionEffect}c, where \ce{HO2-} is used instead of \ce{OH-} for modelling simplicity.

The lower half of Fig.~\ref{fig:reactionEffect}(d-f) shows the effect on each of these swimmers of a single ionic reaction occuring in the bulk, aqueous phase
\begin{fleqn}
\begin{align}
  \nonumber \mathrm{(R4)} &\quad \ce{H2O2 <=> H+ + HO2-} .
\end{align}
\end{fleqn}
As we mentioned before, this reaction will only begin to have a significant effect when we are in the reactive, $qa>1$ regime. In Fig.~\ref{fig:reactionEffect}, the reactive screening length $q^{-1}$ is indicated by the dashed line: for these swimmers, $qa\approx 3$. The white arrows show the effect of this reaction on the propulsion speed, which is different for each of the swimmers. For (a$\rightarrow$d), the reaction generates propulsion, for (b$\rightarrow$e) the reaction increases the propulsion speed, and for (c$\rightarrow$f) the propulsion speed falls. We now briefly explain the reason for these effects.

In the absence of a swimmer, reaction (R4) is in a state of dynamic equilibrium. If a swimmer consumes or produces molecules on either side of this equilibrium, then this will push the reaction out of equilibrium, and the system will respond so as to reduce the effect of that perturbation: this is Le Chatelier's principle. Thus, in Fig.~\ref{fig:reactionEffect}d, the \ce{H2O2} flux injected from the particle surface is to the left of the equilibrium, so the \ce{H2O2} partially disocciates into ions producing ionic fluxes. In Fig.~\ref{fig:reactionEffect}e the proton flux is to the right of equilibrium, so there is net ionic recombination in the bulk to give an \ce{H2O2} flux (this small \ce{H2O2} flux is not shown because it does not significantly contribute to self-electrophoresis) and an \ce{HO2-} flux in the \textit{opposite direction} to the original proton flux. In Fig.~\ref{fig:reactionEffect}f, the proton and \ce{HO2-} fluxes are both to the right of equilibrium, so these both recombine with counterions in the bulk to give an \ce{H2O2} flux instead of the ionic fluxes.

\begin{figure}[t]
  \centering
  \includegraphics[width=8.5 cm]{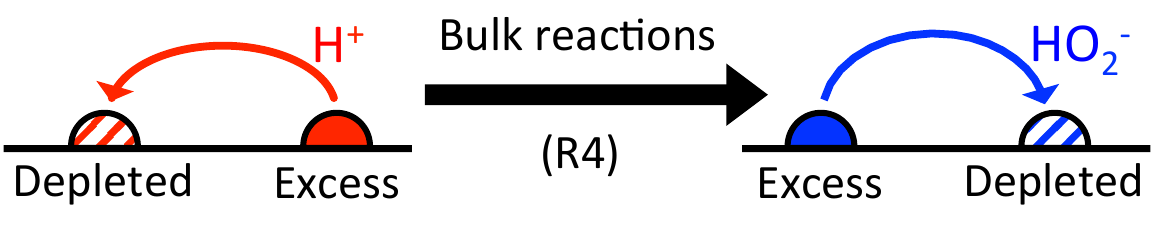}
  \caption{\label{fig:kappaq} The effect of bulk reactions on ionic currents. Regions with excess \ce{H+} ions become depleted in \ce{HO2-} ions and vice versa. This means that an initial \ce{H+} current flowing from excess to depleted regions (left) is partially replaced by a \ce{HO2-} current flowing in the opposite direction (right).}
\end{figure}

For (a$\rightarrow$d), the new ionic fluxes produce a diffusion potential, which generates motion. Hence a swimmer without any electrochemical reactions on its surface can still exhibit self-electrophoretic propulsion. Crucially, it will also display the experimental behaviour that would be expected of a self-electrophoretic swimmer, e.g., propulsion speed scaling inversely with salt concentration (via the $U^\mathrm{SM}$ factor in Eq.~\eqref{general result}). This means that the kind of ionic behaviour observed in Refs.~[\!\!\citenum{brown14, ebbens14, paxton06}] is not \textit{a priori} a signature of an electrochemical surface reaction. In practice, however, we find that, because of the weak dissociation of \ce{H2O2}, the simple non-electrochemical surface reaction mechanism in Fig.~\ref{fig:reactionEffect}a, d cannot account for the magnitude of the experimentally observed propulsion in, e.g., Ref.~[\!\!\citenum{brown14}]: genuine self-electrophoretic propulsion is still required.

In (b$\rightarrow$e), the important point is that chemical reactions conserve charge, and this also implies the conservation of electrical current. There is a net electrical current in (b), and because this current is conserved it will have the same magnitude with or without bulk reactions. It is only the \textit{identity} of the current-carrying ions which changes: in this case, the current becomes partially carried by \ce{HO2-} ions travelling in the opposite direction, see Fig.~\ref{fig:kappaq}. As we discuss later, the propulsion speed scales inversely with the diffusivity of the current-carrying ion. \ce{HO2-} diffuses approximately 10 times slower than \ce{H+}, and this is why the speed increases. In fact, in the appropriate environment of high pH ($=$ high \ce{HO2-} concentration), the predicted speed increases ten-fold because the current becomes entirely carried by \ce{HO2-} ions.

In (c$\rightarrow$f), on the other hand, there is no net electric current to be conserved and both anions and cations react freely with their counterions in the bulk. Hence, far from the swimmer, the ionic gradients become vanishingly small, with a resultant drop in propulsion speed compared to the case without bulk reactions. In detail, the presence of ions in the bulk, due to the surface reactions, generates a diffusion potential (similar to the situation in Fig.~\ref{fig:reactionEffect}d). However, since the ions in Fig.~\ref{fig:reactionEffect}f can recombine through bulk reactions, the further one is from the swimmers surface, the fewer ions generated by the surface reaction remain to induce the diffusion potential. This shows up as an exponentially screened potential, with screening length $q^{-1}$. This also affects the swimming speed, because we find that the magnitude of the diffusion potential scales with the thickness of the screening layer, leading to a scaling of $U\propto 1/(qa)$. This can lower the predicted propulsion speed by up to a factor of approximately 100 for realistic parameters. A similar $U \propto 1/a$ scaling has been observed with Pt-polymer Janus particles~\cite{ebbens12}, and we suggest that this could be a direct observation of these bulk reactions in action.

%%%%%%%%%%%%
\subsection{Summary}

In brief, we find that taking account of ionic reactions and realistic electrostatic screening has very significant and system-dependent effects on the propulsion of a wide range of chemical swimmers. These effects include increasing or decreasing the predicted speed by several orders of magnitude, as well as the qualitatively new behaviour of reactive screening. In addition, we find that even swimmers with no ionic surface reaction can behave as though they are self-electrophoretic. Taking these effects into account will be crucial if we are to correctly interpret experiments on self-propelled particles and develop new, non-equilibrium physics. The remainder of this paper therefore gives a detailed account of the theory that gives rise to the results summarized thus far, and compares it to experiments as far as current data allow.

%%%%%%%%%
\section{\label{sec:math}Theoretical Model}

\begin{figure}
  \centering
  \includegraphics[width=8.5 cm]{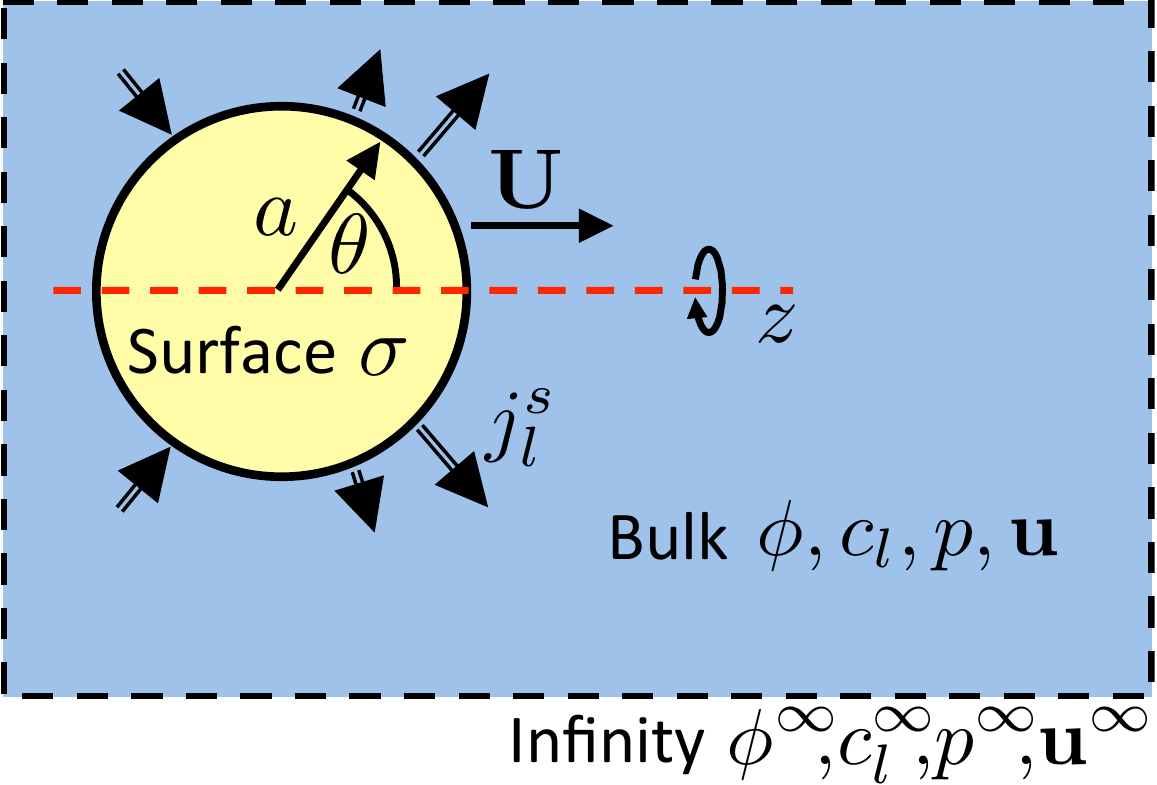}
  \caption{\label{fig:diagram}Diagram of a model swimmer, highlighting the distinction between bulk and surface parameters. In the bulk, we have an electrostatic potential field $\phi$, chemical concentration fields $c_l$, a pressure field $p$, and fluid velocity $\boldsymbol{u}$. Far from the particle, these fields approach uniform values (superscript $\infty$). On the particle surface, the uniform surface charge density $\sigma$ and the nonuniform molecular fluxes out of the surface $j^\mathrm{s}_l$ set boundary conditions for the bulk potential and concentration fields, respectively. The particle, of radius $a$, is axisymmetric around the $z$ axis, so the surface fluxes are parametrized by the polar angle $\theta$. The swimming velocity, which we calculate, is $\boldsymbol{U}=U\hat{\boldsymbol{z}}$.}
\end{figure}

In this and the following two sections, we present a detailed quantitative model of self-electrophoresis. Here in Section~\ref{sec:math}, we will lay out the general theoretical model and detail how this will be applied to the specific \ce{H2O2} reaction system described above. In order to obtain analytical results we also linearize our theory. We will then apply this model to obtain explicit results, first for a system with only surface reactions (Section~\ref{sec:ephorNBR}), and then with bulk reactions too (Section~\ref{sec:elphoWB}).

%%%%%%%%%%%%
\subsection{\label{Gen Mod}General Model} 

The standard theoretical approach to self-electrophoresis involves coupling the chemical fluxes arising from reactions on the particle surface to bulk differential equations (Nernst-Planck, Poisson, and Navier-Stokes)~\cite{moran11}. This treatment generally ignores bulk chemical reactions by assuming that each chemical species is conserved. We adopt the standard approach, but include bulk reactions by coupling chemical fluxes to local reaction rates. We solve this model numerically using COMSOL. Separately, and in common with previous work~\cite{golestanian07}, we also linearize the model to obtain an analytical approximation. Unlike in previous work, the analytical solution does not require the assumption of a thin electrostatic screening layer.

We consider a spherical swimmer of radius $a$ and uniform surface charge density $\sigma$, see Fig.~\ref{fig:diagram}. The electrostatic boundary condition of such a particle is
\begin{align}
	\hat{\boldsymbol{n}}\cdot\nabla\phi(\boldsymbol{s}) &= -\frac{\sigma}{\epsilon} \label{G} ,
\end{align}
with $\phi$ the electrostatic potential field and $\epsilon$ the dielectric constant of the fluid (the dielectric constant of the particle is assumed to be zero). Here $(\boldsymbol{s})$ and $\hat{\boldsymbol{n}}$ indicate evaluation at, and the normal out of, the particle surface. We have chosen a uniform, dielectric boundary condition for simplicity. However, in Appendix~\ref{sec:electro} we show formally that, with an appropriate choice of surface potential, an equipotential (conducting) surface gives the same swimming speed as a dielectric. We do not deal with mixed dielectric/conducting particles here, but this should not qualitatively affect the basic physics.

Propulsion is generated by reactions on the swimmer surface. These reactions produce and consume $N$ different chemical species, labelled $l=1, \dots, N$. The surface production (or, if negative, consumption) rate per unit area of each species is $j^\mathrm{s}_l(\theta)$, and is a function only of $\theta$, the polar angle with respect to the symmetry axis $\hat{\boldsymbol{z}}$, see Fig.~\ref{fig:diagram}. The surface reaction rates can be equated to the bulk flux $\boldsymbol{j}_l$ of each species out of the particle surface, giving the boundary condition
\begin{align}
  \hat{\boldsymbol{n}}\cdot\boldsymbol{j}_{l}(\boldsymbol{s}) &=  j^\mathrm{s}_{l}(\boldsymbol{s}) . \label{j boundary}
\end{align} 
These bulk fluxes obey the classical Nernst-Planck equation
\begin{align}
  \boldsymbol{j}_{l} = c_{l}\boldsymbol{u} -D_{l}\nabla c_{l} - \frac{D_{l} z_{l} e}{k_BT} c_{l} \nabla\phi , \label{flux}
\end{align}
with $z_l$, $D_l$, and $c_l$ respectively the valency, diffusivity, and concentration field of each chemical species, $\boldsymbol{u}$ the fluid flow field, $e$ the fundamental charge, $k_B$ Boltzmann's constant, and $T$ temperature. Physically, Eq.~\eqref{flux} expresses the bulk fluxes as linear sums of advective, diffusive, and conductive terms respectively. Eq.~\ref{flux} is the standard flux expression used in studies of self-electrophoresis. Its main simplification is the neglect of cross-coupling terms between the molecular fluxes, and this is valid as long as we are in the dilute limit with relatively small ionic gradients~\cite{degraaf15}, which is true here.

Without bulk reactions, conservation of chemical species would require that the bulk fluxes are incompressible vector fields, i.e., $\nabla\cdot \boldsymbol{j}_l=0$. Bulk chemical reactions can be incorporated by writing instead
\begin{align}
	\nabla\cdot\boldsymbol{j}_l=R_l(c_1, \dots, c_N) , \label{production}
\end{align}
where $R_l$ is the local rate at which each chemical is produced (if negative, consumed), in chemical reactions. In general, $R_l$ depends on the local concentration of all chemical species involved in reactions with species $l$. 

Chemical reactions are charge-conserving, i.e., 
\begin{align}
  \sum_l z_lR_l=0 , \label{charge conservation}
\end{align}
 everywhere, and combining this condition with Eq.~\eqref{production} implies the conservation of electrical current
\begin{align}
  \nabla\cdot\boldsymbol{i} &= 0 . \label{div current}
\end{align} 
where the electrical current $\boldsymbol{i}=e\sum_lz_l\boldsymbol{j}_l$. 

Infinitely far from the particle, the chemical concentrations are labelled $c_l^\infty$, and are determined by equilibrium equations and charge neutrality. The other boundary conditions at infinity are $\boldsymbol{j}_l^\infty=0$, $\phi^\infty=0$, $\boldsymbol{u}^\infty=0$ (in the lab frame), and $p^\infty=p_\mathrm{atm}$, where $p$ is the hydrostatic pressure field and $p_\mathrm{atm}$ is the atmospheric pressure, whose absolute value does not affect the calculations.

The electrostatic potential $\phi$ is determined by the Poisson equation
\begin{align}
  \epsilon \nabla^2 \phi = - \rho_\mathrm{e} , \label{rho}
\end{align}
with charge density $\rho_\mathrm{e} = e \sum_{l} z_{l} c_{l}$. The interaction of the electrostatic potential and the unbalanced charge density ($\rho_{\rm e}\neq 0$) generates a force density 
\begin{align}
  \boldsymbol{f} &= -\rho_{\rm e}\nabla\phi , \label{force}
\end{align}
and this drives fluid flow via the Stokes equations for low-inertia, incompressible flow
\begin{align}
  \eta\nabla^2\boldsymbol{u} &= \nabla p - \boldsymbol{f} , \label{eq:stokes} \\
  \nabla\cdot\boldsymbol{u} &= 0 .
\end{align}
Finally, the swimmer is not held in place, so fluid flow around it will cause it to move with some propulsion velocity $\boldsymbol{U}$. This propulsion velocity is determined by the condition that there is no net force acting on the total system of swimmer plus fluid out to infinity~\cite{stone96}. This force-free condition is simply a reflection of the fact that all the forces are internal to this total system --- there are no long-range, external forces like gravity. The force-free condition can be translated into an expression for $\boldsymbol{U}$ by using the Lorentz reciprocal theorem, which is a restatement of the Stokes equations in integral form. This gives a closed-form expression for the propulsion velocity~\cite{teubner82}
\begin{align}
  \boldsymbol{U} &= -\frac{\hat{\boldsymbol{z}}}{6\pi\eta a}\int_V \left[ \left(\frac{3a}{2r}-\frac{a^3}{2r^3}-1\right)\cos\theta\hat{\boldsymbol{r}} -\right. \nonumber \\
                 &\quad -\left. \left(\frac{3a}{4r}+\frac{a^3}{4r^3}-1\right)\sin\theta\hat{\boldsymbol{\uptheta}} \right]\cdot\boldsymbol{f} \mathrm{d}V , \label{lorentz}
\end{align}
where $r$ is the distance from the particle centre, $\hat{\boldsymbol{r}}$ and $\hat{\boldsymbol{\uptheta}}$ are unit vectors in the $r$ and $\theta$ directions, and the scalar speed $U$ is defined by $\boldsymbol{U}=U\hat{\boldsymbol{z}}$. The volume integral is over the region outside the particle.

%%%%%%%%%%%%
\subsection{\label{sub:solve}Numerical Solution}

We solve the full non-linear model numerically using the finite element method (FEM), implemented in COMSOL. To do this, we make several modifications to the above equation system. In particular, we define a new force density $\boldsymbol{f}_{\rm FEM}$ to replace $\boldsymbol{f}$ in Eq.~\eqref{force}. The two quantities are related by
\begin{align}
  \label{eq:forceFEM} \boldsymbol{f}_{\rm FEM} &= \boldsymbol{f}- k_BT\sum_{z_{l} \ne 0}\nabla c_{l}\,.
\end{align}
This redefinition does not influence the result of the simulation, but limits spurious flows related to numerical artifacts in the electrostatics~\cite{rempfer16}. Our calculations are performed on an axisymmetric spherical domain of radius $L = 10a + 25 \kappa^{-1}$, which we verified to be sufficient to eliminate most finite size effects in our speed calculations. This frame co-moves with the colloid. We impose no-slip at the colloid surface, and on the edge of the domain, we employ the same boundary conditions as the theory has at infinity. For the fluid velocity we impose a no-stress condition on the edge of the domain
\begin{align}
  \label{eq:nostressFEM} \left[ \eta \left( \nabla \boldsymbol{u} + \left( \nabla \boldsymbol{u} \right)^\star \right) - p \mathbb{I} \right] \cdot \hat{\boldsymbol{n}}\,=\,0
\end{align}
with $\hat{\boldsymbol{n}}$ the normal to the domain, $\star$ denoting transposition, and $\mathbb{I}$ the 3D identity matrix. This is equivalent to imposing a force-free condition on the swimmer-fluid system~\cite{degraaf15,kreissl16}.

Our technique is to first obtain approximate numerical solutions for the electrostatic and concentration fields in the absence of advection, so neglecting the first term in Eq.~\eqref{flux}. This approach is justified because experimental swimmers generally have low P{\'e}clet numbers, i.e., molecular diffusion dominates over advection. The P{\'e}clet number is defined as $\mathrm{Pe} = U a / D$, with $U \approx 10~\umpers$, $a \approx 1 \um$, and $D \approx 10^{-9}~\mathrm{m^2 s^{-1}}$ typical for experiments on microswimmers, leading to $\mathrm{Pe} \approx 0.01$. The flow field is computed self-consistenly on the domain by employing the force density, Eq.~\eqref{eq:forceFEM}, following from the concentration and potential fields. The speed of the swimmer is then determined by taking the average of the fluid velocity on the edge of the domain: $\boldsymbol{U} = - \langle \boldsymbol{u} \rangle_{r = L}$, where $\boldsymbol{U}$ is in the lab frame and $\langle \boldsymbol{u}\rangle$ in the co-moving frame. We subsequently verified the low-P{\'e}clet-number approximation by solving the fully coupled equations (with advection) directly in a limited number of cases, which gave agreement to within a few percent. See Appendix~\ref{app:finite} for full details of the numerical calculations.

\subsection{Analytical Solution \label{sub:solveAnal}}

We also linearize the model to provide an analytical solution. To do this we assume that the fields $\phi$ and $c_l$ have only small deviations from their values in the uncharged, unreactive state where $\phi=0$ and $c_l=c_l^\infty$ everywhere (for $\phi$, this assumption corresponds to the usual Debye-H{\"u}ckel approximation, $\phi \ll k_{B}T/e$). We then expand the model to linear order in the small dimensionless parameters $\psi=\phi e/(k_BT)$ and $x_l=(c_l-c_l^\infty)/c_l^\infty$. Applying this linearization to Eq.~\eqref{flux} gives
\begin{align}
  \boldsymbol{j}_{l} = -c_{l}^\infty D_l\left[\nabla x_{l} + z_{l} \nabla\psi\right] , \label{flux linear}
\end{align}
where the advection term has been dropped entirely because $\boldsymbol{u}$ scales quadratically with the small parameters (to see this, note that Eq.~\eqref{force} contains a product of $\rho_e$ and $\phi$, which are both small). We must also Taylor expand the production rates $R_l$ to linear order, i.e.,
\begin{align}
  R_l &= \sum_m k_{lm}x_m + \mathcal{O}(x_m^2) , \label{Taylor}
\end{align}
where $\mathcal{O}(\cdot)$ means `of order $\cdot$', and the elements
\begin{align}
  k_{lm}=\frac{\partial R_l}{\partial x_m}\bigg|_{x_1,\,x_2...\, x_N=0 } . \label{K}
\end{align}
are components of a matrix $\matrify{k}$ which we can call the linear reaction matrix. Its meaning will become clearer when we consider specific reactions. From Eq.~\eqref{flux linear} and Eq.~\eqref{production}, we have, to linear order
\begin{align}
  \sum_m k_{lm}x_m = -c_{l}^\infty D_l\left[\nabla^2 x_{l} + z_{l} \nabla^2\psi\right] . \label{flux linear div}
\end{align}
This set of $N$ equations, together with the Poisson equation, which we rewrite as
\begin{align}
  \frac{k_BTe^2}{\epsilon}\sum_l c_l^\infty z_l x_l &= -\nabla^2 \psi , \label{rho linear}
\end{align}
makes up a system of $N+1$ linear differential equations in $N+1$ fields ($x_l$, $l\in 1, \dots N$ and $\psi$). 

This system of equations is soluble in a spherical geometry by standard spectral methods, and the electrostatic potential field so obtained can then be used to calculate the propulsion speed by evaluating the integral in Eq.~\eqref{lorentz}. In doing this, we make the further usual assumption of a relatively small driving field~\cite{henry31}. That is, if we define $\phi= \phi^\mathrm{eq} + \phi^\mathrm{sr}$ where $\phi^\mathrm{eq}$ is the electrostatic potential in the absence of surface reactions and $\phi^\mathrm{sr}$ is the additional potential generated by these reactions then $\phi^\mathrm{sr}\ll \phi^\mathrm{eq}$. As a result, the surface reaction rates $j_l^\mathrm{s}$, which only come into $\phi^\mathrm{sr}$, contribute linearly to the final velocity. The algebra required to solve Eq.~\eqref{flux linear div}-\eqref{rho linear} is significant, so we go through this explicitly in Appendix~\ref{sec:linear}.

%%%%%%%%%%%%
\subsection{Specific \ce{H2O2} Reaction Model \label{sec:reactionModel}}

The chemical reaction system we consider is the simplified version of the \ce{H2O2} reaction system described in Section~\ref{Summary}. On the particle surface, \ce{H2O2} decomposes into \ce{O2} and \ce{H2O}. For simplicity, however, we ignore both products of this reaction: \ce{O2} because it is electrically neutral and does not dissociatiate, and \ce{H2O} because it dissociates much less than \ce{H2O2} --- the respective equilibrium constants are $K_{\mathrm{eq},\ce{H2O}} = 1.0\times 10^{-14}~\M$ ($\mathrm{pH} = 7$) and $K_{\mathrm{eq},\ce{H2O2}} = 2.5\times 10^{-12}~\M$~[\!\!\citenum{everett53}]. In the bulk, we ignore any slow decomposition of \ce{H2O2} via reaction (R1), and the only bulk reaction we consider is the ionic dissociation (R4) which we rewrite here 
\begin{fleqn}
\begin{align}
  \nonumber \mathrm{(R4)} &\quad \ce{H2O2 <=>[k_\mathrm{forward}][k_\mathrm{reverse}] H+ + HO2-} .
\end{align}
\end{fleqn}
We therefore have only three chemically active species, with associated subscripts in brackets: \ce{H2O2} ($\circ$), \ce{H+} (+), and {\ce{HO2-} (-)}. Protonation reactions like (R4) are normally extremely rapid, with kinetics controlled by the diffusion and collison of the ions~\cite{caldin64}, and with simple first order rate expressions
\begin{align}
  k_\mathrm{forward} &= k_\mathrm{dis}c_{\circ} \,,\nonumber \\
  k_\mathrm{reverse} &= k_\mathrm{as}c_+c_- \,, \label{reaction rates}
\end{align}
where we estimate the association rate constant to be $k_\mathrm{as}=4.9\times 10^{10}~\M^{-1}\pers$ using the Smoluchowski-Debye theory for diffusion-limited reactions, see Appendix~\ref{sec:rates}. The dissociation rate constant $k_\mathrm{dis}=0.12 \pers$ is then determined from the equilibrium constant $K_\mathrm{eq}=k_\mathrm{dis}/k_\mathrm{as}=2.5\times 10^{-12}~\mathrm{M}$~[\!\!\citenum{everett53}]. Far from the particle, the system is in equilibrium, so we have
\begin{align}
  c_{+}^ {\infty} c_{-}^{\infty}&=K_\mathrm{eq}c_{\circ}^{\infty} , \label{eq}
\end{align}
The production rates are $R_+=R_-=k_\mathrm{forward}-k_\mathrm{reverse}$ and $R_\circ=k_\mathrm{reverse}-k_\mathrm{forward}$, and linearizing using Eq.~\eqref{K} gives the linear reaction matrix
\begin{align}
  \matrify{k} &= k_\mathrm{dis}c_\circ^\infty \left(
  \begin{array}{rrr}
    -1& 1& 1\\
     1&-1&-1\\
     1&-1&-1\\
  \end{array}
  \right) , \label{kik}
\end{align}
where the order of rows and columns is $\circ$, $+$, $-$.

The three reactive species have diffusivities $D_{\circ}=1.7\times 10^{-9}~\mathrm{m^2 s^{-1}}$~[\!\!\citenum{kern54}], $D_{+}=9.3 \times 10^{-9}~\mathrm{m^2 s^{-1}}$~[\!\!\citenum{haynes13}], and $D_{-}=0.9\times 10^{-9}~\mathrm{m^2 s^{-1}}$~[\!\!\citenum{vandenbrink84}]. We also have two unreactive ions, which we take to be \ce{Na+} and \ce{Cl-} with diffusivities $D_\ce{Na+}=1.3\times 10^{-9}~\mathrm{m^2 s^{-1}}$ and $D_\ce{Cl-}=2.0\times 10^{-9}~\mathrm{m^2 s^{-1}}$~[\!\!\citenum{samson03}]. Because these ions are not involved in chemical reactions at the surface or in the bulk, their concentration fields are in equilibrium with the electrostatic potential. The implication is that the diffusivity of these ions does not contribute to the propulsion speed in the linear regime. We show this mathematically in Appendix~\ref{linear appendix}.

The chemical concentrations at infinity are determined by the chemical equilibrium, Eq.~\eqref{eq} and by charge balance
\begin{align}
  c_{-}^{\infty}+c_{\ce{Cl-}}^{\infty}&=c_{+}^{\infty}+c_{\ce{Na+}}^{\infty} .
\end{align}
These two equations connect five concentrations, so we can set three concentrations freely. In practice, we choose instead to set the \ce{H2O2} concentration, the total ionic strength, and the pH. The reaction scheme presented here is the simplest possible that gives the necessary freedom: bulk ionic dissociation reactions require at least three reactive species, and the two non-reactive ions are necessary to allow the ionic strength to be modified independent of other parameters. 

For the variable parameters, our base set, used unless specified otherwise, is 1 $\mM$ salt, i.e., $c^\infty_{\ce{Na+}}=c^\infty_{\ce{Cl-}}=1~\mM$, ${a=500~\mathrm{nm}}$, and $c^\infty_{\circ}=3~\M$. For these parameters, $\kappa^{-1}=10~\mathrm{nm}$, and $c^\infty_{+}=c^\infty_{-}=3\times10^{-6}~\M$. These parameters were chosen because micron-sized particles and \ce{H2O2} concentrations of order 3$~\M$ are experimentally typical~\cite{brown14}, while the ${1~\mM}$ baseline salt concentration allows us to scan a wide range of the important parameter $\kappa a$ for realistically sized particles.

Meanwhile, the surface reactions are specified by surface fluxes $j^\mathrm{s}_l$, $l\in\{{\circ},+,-\}$, of the three active species. We consider the three model swimmers shown in Fig.~\ref{fig:reactionEffect}, referred to as: $S_\circ$, the nominally neutral swimmer; $S_+$, powered by a proton current; and $S_=$, powered by ionic diffusiophoresis. As mentioned above, only the dipolar part of the fluxes, that is the 1$^\mathrm{st}$ Legendre component, contributes to the propulsion speed of uniformly charged swimmers~\cite{golestanian07}, so we include only this term by setting $j^\mathrm{s}_l\equiv j^\mathrm{s}_{l,1}\cos{\theta}$ for each surface flux where $j^{\rm s}_{l,1}$ is a constant coefficient. For $S_\circ$, only $j_{\circ,1}^\mathrm{s}$ is finite; for $S_+$, only $j_{+,1}^\mathrm{s}$ is finite; and for $S_=$, $j_{+,1}^\mathrm{s}$ and $j_{-,1}^\mathrm{s}$ are equal and finite, with $j_{\circ,1}^\mathrm{s}=0$.

Our model makes a number of simplifications. This includes those chemical simplifications already discussed, as well as the neglect of potential contaminants such as \ce{CO2}, which also undergo ionic dissociation. The main purpose of this paper is to illustrate the physical principles behind the effect of bulk reactions on self-propulsion, and these physical principles will also apply to a more complex and realistic \ce{H2O2} reaction system, as well as to other chemical systems~\cite{ibele09, gao14}. We also neglect any dependence of the surface parameters on environmental conditions, so, for example, we take $\sigma={\rm constant}$, independent of pH, salt concentration etc. This does not imply that surface parameters are independent of the environment; it is just that detailed knowledge of this dependence is currently lacking. The `pure' effect of bulk reactions which we capture will occur in addition to any such interdependence.

%%%%%%%%%
\section{\label{sec:ephorNBR}Electrophoresis without Bulk Reactions}

Before discussing the effect of bulk ionic reactions, it is important to set out the basic theory for propulsion by self-electrophoresis without such reactions. While this theory has been set out multiple times before~\cite{moran10, moran11, golestanian07, sabass12b,moran14} for the limit of vanishing electrostatic screening length, $\kappa a \gg 1$, we give here a version that includes the effect of a finite $\kappa a$.

For comparison, we first write down the standard results for electrophoresis in an external, linear field~\cite{kim13}, Fig.~\ref{henry}a (top). We align the external field along the $z$ axis and consider the velocity $U_\mathrm{ext} \hat{\boldsymbol{z}}$ of a uniformly charged spherical particle with radius $a$ and surface ($\zeta$) potential~\cite{kim13}
\begin{align}
  \zeta = \frac{\sigma a}{\epsilon(1+\kappa a)} , \label{zeta sigma}
\end{align}
in an externally imposed electric field $\boldsymbol{E}= - \nabla \phi$. Far from the particle, $\boldsymbol{E}$ is a constant linear field $\boldsymbol{E} = E^\infty \hat{\boldsymbol{z}}$. The particle is suspended in an aqueous solution of a monovalent salt, e.g., \ce{NaCl}. In a weak field, particle velocity is proportional to electric field strength, $U_\mathrm{ext}= \mu_\mathrm{E} E^\infty$, with $\mu_\mathrm{E}$ called the electrophoretic mobility. The standard expression for $\mu_\mathrm{E}$ for small $\zeta$ is~\cite{henry31, kim13}
\begin{align}
	\mu_\mathrm{E} &= \frac{\zeta \epsilon}{\eta}f(\kappa a) , \label{external phoresis}	
\end{align}
where $f(\kappa a)$ is Henry's function~\cite{henry31} which accounts for electrostatic screening and depends only on $\kappa a$, the ratio between particle radius $a$ and the electrostatic screening length $\kappa^{-1}$. The function $f$ is plotted in Fig.~\ref{henry}b: it has constant limits of $f(\infty)=1$, corresponding to high salt concentration or large particles, and $f(0)=2/3$, corresponding to small particles or non-polar solvents. Eq.~\eqref{external phoresis}, typically in either the high or low $\kappa a$ limit, is the expression commonly used to compute colloidal $\zeta$ potentials from mobility measurements in, e.g., commercial Zetasizers.

In self-electrophoresis, the independent parameters are the surface reaction rates, and therefore the ionic fluxes, rather than the electric field. To facilitate understanding, we translate the expression for external electrophoresis into these terms. We write down expressions for the inverse electrostatic screening length
\begin{align}
  \kappa^2 = \frac{e^2}{\epsilon k_B T} \sum_l c_l^\infty\,,
\end{align}
the ionic conductivity
\begin{align}
  K &= \frac{e^2}{k_BT}\sum_l D_l c_l^\infty ,
\end{align}
and the concentration averaged diffusivity
\begin{align}
  \bar{D}=\left(\sum_l D_l c_l^\infty\right) /\left(\sum_l c_l^\infty\right)\,.
\end{align}
together with Ohm's law 
\begin{align}
  E^\infty &= \frac{i^\infty}{K} ,\label{average conductivity}
\end{align}
which relates the electric field to the ionic current density at infinity $i^{\infty}\hat{\boldsymbol{z}}$, and which we can rewrite as
\begin{align}
  \nabla\phi &= -\frac{i^\infty\boldsymbol{\hat{z}}}{\epsilon\kappa^2\bar{D}} \,,\label{ohm 2}
\end{align}
Combining Eq.~\eqref{ohm 2} with Eq.~\eqref{external phoresis} we then have
\begin{align}
  U_\mathrm{ext} = i^\infty \frac{1}{\eta\kappa^2\bar{D}} \zeta f(\kappa a) \,. \label{external field}
\end{align}
Note in particular that for electrophoresis in an external field, the particle speed is inversely proportional to the concentration-averaged diffusivity $\bar{D}$.

We now compare Eq.~\eqref{external field} with the analogous expression for the most well-studied self-electrophoretic swimmer, a proton-powered bimetallic swimmer~\cite{moran11, sabass12b}. Consider a spherical swimmer of radius $a$, with surface charge density $\sigma$, a surface proton flux $j^\mathrm{s}_+(\theta)=j^\mathrm{s}_{+,1}\cos\theta$ and no bulk reactions: we call this model $S_+^{\rm NBR}$. In this case, Eq.~\eqref{flux linear div}-\eqref{rho linear} can be easily solved to yield, after some algebra
\begin{align}
  \phi &= \frac{\sigma a}{\epsilon(1+\kappa a)}\left(\frac{a}{r}\right)e^{-\kappa (r-a)}+\frac{e a j^\mathrm{s}_{+,1}}{2\kappa^2D_+\epsilon}\left(\frac{a}{r}\right)^2\cos{\theta}+ \dots , \label{phi NBR}
\end{align}
where the first and second terms are $\phi^\mathrm{eq}$ and $\phi^\mathrm{sr}$, the potentials generated respectively by the surface charge and the surface reactions, and $ \dots$ indicates additional, electrostatically screened terms that are necessary to match the electrostatic boundary conditions, but which make no contribution to the propulsion, see Appendix~\ref{sec:electro}. The propulsion speed is obtained by evaluating Eq.~\eqref{lorentz} with Eq.~\eqref{phi NBR} to give
\begin{align}
  U^\mathrm{NBR}_+ &= \frac{-j^\mathrm{s}_{+,1}e}{3}\frac{1}{\eta \kappa^{2}D_+}\frac{\sigma}{\epsilon\kappa}F(\kappa a) , \label{standard model}
\end{align}
where $F$ is, like $f$, a function of $\kappa a$ only. The full form of $F$ is given in Appendix~\ref{app: ff calc}. 

Eq.~\eqref{standard model} corresponds closely to Eq.~\eqref{external field}, the particle velocity with external electrophoresis, and we compare these expressions factor by factor:

I: The relevant current density $i^\infty$ becomes $-j^\mathrm{s}_{+,1}e/3$ because of the exclusive dependence of the propulsion speed on the first Legendre component of the flux~\cite{golestanian07} discussed in Section~\ref{sec:reactionModel}.

II: The relevant diffusivity $\bar{D}$ becomes $D_+$ because, for self-electrophoresis in steady state, the ionic current can only be carried by the active ion involved in reactions at the particle surface, in this case \ce{H+}. There can be no net flux of the other ions, or they would build up at the particle surface. In fact, the other ions are in local equilibrium with the electrostatic potential $\phi$, i.e., $c_{\ce{Na+}} = c_{\ce{Na+}}^{\infty} \exp(-\phi / k_B T)$ etc., from standard Debye-H{\"u}ckel theory~\cite{kim13}, and the swimmer behavior therefore cannot depend on their dynamic properties at all. The appropriate version of Ohm's law for the self-electrophoretic swimmer is thus not Eq.~\eqref{ohm 2}, but instead~\cite{esplandiu16}
\begin{align}
\nabla\phi=-\frac{\boldsymbol{i}}{\kappa^2\epsilon D_+}\,, \label{Ohm NBR}
\end{align}
which depends only on the mobility of the active ion, and the electrostatic potential, Eq.~\eqref{phi NBR} and the propulsion speed have the same dependencies. This difference between external- and self-electrophoresis is commonly overlooked~\cite{golestanian07, brown14}, but has been confirmed in numerical calculations~\cite{moran14, esplandiu16}. The inverse scaling of self-propulsion speed with the diffusivity of the active ion will be crucial for understanding the effects of bulk ionic reactions in Section~\ref{sec:elphoWB}.

III: We have chosen to parametrize our model in terms of $\sigma$ rather than $\zeta$ because this is the most natural choice from a microscopic point of view. Much of the charge on the surface, both of conducting and dielectric particles, is due to surface-absorbed groups, leaving $\sigma$ fixed as other parameters, such as $\kappa$, vary. This has been demonstrated experimentally for dielectric particles~\cite{midmore96}. Nevertheless, the experimental evidence indicates that self-electrophoretic propulsion speeds scale with $\kappa^{-2}$~[\!\!\citenum{paxton06, palacci13, brown14}] which is consistent with a fixed $\zeta$, not a fixed $\sigma$, though to our knowledge there is no microscopic justification for this. Since we are most interested in bulk effects, we do not insist upon a particular surface parametrization, and Eq.~\ref{zeta sigma} can be used to translate our results into a parametrization where $\zeta$ is fixed. We have checked that this makes little qualitative difference to our results. Note that this point is distinct from the choice between conducting and dielectric boundary conditions on the particle surface, which is discussed in Section~\ref{Gen Mod}.

IV: we have replaced $f(\kappa a)$ with an equivalent expression for self-electrophoresis, $F(\kappa a)$, shown in Fig.~\ref{henry}b. In the thin-screening limit, $F(\kappa a\rightarrow \infty)=1$, and Eq.~\ref{standard model} then agrees with previous self-electrophoresis results in the thin-screening limit~\cite{golestanian07}, except that Ref.~[\!\!\citenum{golestanian07}] incorrectly assumes that the propulsion is controlled by the total ionic diffusivity $K$, as mentioned above under point II. In the opposite limit, $F(\kappa a\rightarrow 0)=(\kappa a/2)^3$, so that for small $\kappa a$ the propulsion speed scales with $a^3$
\begin{align}
  U^\mathrm{NBR}_+ &= -\frac{j^\mathrm{s}_{+,1}ea^3\sigma}{24\epsilon \eta D_+} . \label{standard model low ka}
\end{align}
The reason that $F(0)\rightarrow 0$, while $f(0)$ is finite, is the different geometry of the driving currents, as illustrated in Fig.~\ref{henry}a. For self-electrophoresis, the driving potential is a local, dipolar field which decays over a length of order $a$, see Eq.~\eqref{phi NBR}, whereas in external electrophoresis the driving potential is infinite in extent. Therefore, in self-electrophoresis, additional factors of $a$ in the propulsion speed are to be expected\footnote{The precise $a^3$ factor can be obtained from a scaling argument. We do not include this, as it is involved, and barely more informative than this qualitative explanation.}, as in Eq.~\eqref{standard model low ka}. 

Several of the features of Eq.~\eqref{standard model} have been verified experimentally for bimetallic swimmers~\cite{paxton04, wang06} like the Au-Pt swimmers in Fig.~\ref{fig:example}b, which explains the wide-acceptance of the self-electrophoretic model for this system. As discussed above, this equation has also been found to be applicable to single-catalyst swimmers such as Pt-polystyrene Janus particles~\cite{palacci13, brown14, ebbens14}, suggesting that these swimmers are also powered by proton currents~\cite{brown14, ebbens14}. 

The additional screening parameter $F(\kappa a)$, is more problematic. It predicts that the speed of a swimmer will drop off sharply as $\kappa a$ decreases, Fig.~\ref{henry}. This drop-off is significant for surprisingly large $\kappa a$: $F(10)\approx 0.5$, while $F(1)<0.1$. This shows that the thin-screening limit, which is commonly employed~\cite{golestanian07, brown14, ebbens14, brown16, das15}, is \textit{not} justified even for the common situation of a 1$~\um$ radius swimmer in 3~$\M$ \ce{H2O2}, where $\kappa a\approx 10$~[\!\!\citenum{brown14}]. However, to the best of our knowledge, there is no experimental evidence for this drop off. In fact, a small number of experiments show a larger speed for nanoswimmers~\cite{xuan14, lee14} than is typical for microswimmers~\cite{wheat10, howse07}. We discuss this experimental comparison in more detail in Section~\ref{sec:compare}.

%%%%%%%%%
\section{\label{sec:elphoWB}Electrophoresis with Bulk Reactions}

We now examine the effect of bulk reactions on the propulsion of model swimmers. We will examine the effect of bulk reaction (R4), \ce{H2O2 <=> H+ + HO2-}, on the three model swimmers which are depicted again for convenience in Fig.~\ref{fig:mobil}a. In Section~\ref{generalForm} we write down the general form of expressions for the swimming speed when bulk reactions are included, before focussing on the effect of two experimentally relevant parameters --- swimmer radius and \ce{H+} concentration --- on the swimming speed, in Sections~\ref{sub:result}-\ref{sub:diss}. In Section~\ref{sec:compare} we will compare our predictions with experimental observations.

%%%%%%%%%%%%
\subsection{General Form of the Solutions \label{generalForm}}

Mathematically, the bulk reactions make it impractical to solve even the linearized problem by hand. Instead, we solve the system of equations, Eq.~\eqref{flux linear div}-\eqref{rho linear} symbolically in MATLAB, see Appendix~\ref{sec:linear}. The final solution is very similar to the reactionless solution, with an extra bulk reaction factor $B(qa, \dots)$, and for each of our model swimmers, we can write
\begin{align}
  U_\dag &= \left[\frac{e\sigma j^{\rm s}_\dag}{3 \eta \epsilon\kappa^{3}D_+}F(\kappa a)\right]B_\dag \,, \label{speedDecompositionStandard}
\end{align} 
in the form of Eq.~\eqref{general result}. Here, $\dag$ indicates a particular swimmer type, i.e., ${\dag \in \{ \circ,+, = \}}$ and $j^{\rm s}_{\rm \dag}$ is the appropriate surface flux density for that swimmer. We define $j^{\rm s}_{\rm \dag}=j^{\rm s}_{\circ,1}$ for the $S_\circ$ swimmers, and $j^{\rm s}_{\rm \dag}=j^{\rm s}_{+,1}$ for the $S_+$ and $S_=$ swimmers. The use of $D_+$ in the denominator of Eq.~\eqref{speedDecompositionStandard} is an arbitrary definition. Under this definition, the bulk reaction factors in the absence of bulk reactions have the constant values $B^\mathrm{NBR}_\circ = 0$, $B^\mathrm{NBR}_+ = -1$ and $B^\mathrm{NBR}_{=} = (d_+ - d_-)/(d_+ d_-) = 9.3$. Here, $d_l$ is a rescaled diffusivity, $d_l=D_l/D_+$. By definition, $d_+=1$, but we retain $d_+$ for symmetry of notation.

Note that the expression in square brackets in Eq.~\eqref{speedDecompositionStandard} is indentical to Eq.~\eqref{standard model}. This emphasizes that all the propulsion mechanisms, $S_\circ$, $S_+$, and $S_=$, are really forms of self-electrophoresis, and display all the responses to, e.g., salt-concentration, particle radius, and surface charge which standard self-electrophoresis models predict. The inclusion of bulk reactions just adds a new layer of phenomena on top of this behavior.

%%%%%%%%%%%%
\subsection{\label{sub:result}Influence of Bulk Reactions on Swimming}

\begin{figure}
  \centering
  \includegraphics[width=8.5 cm]{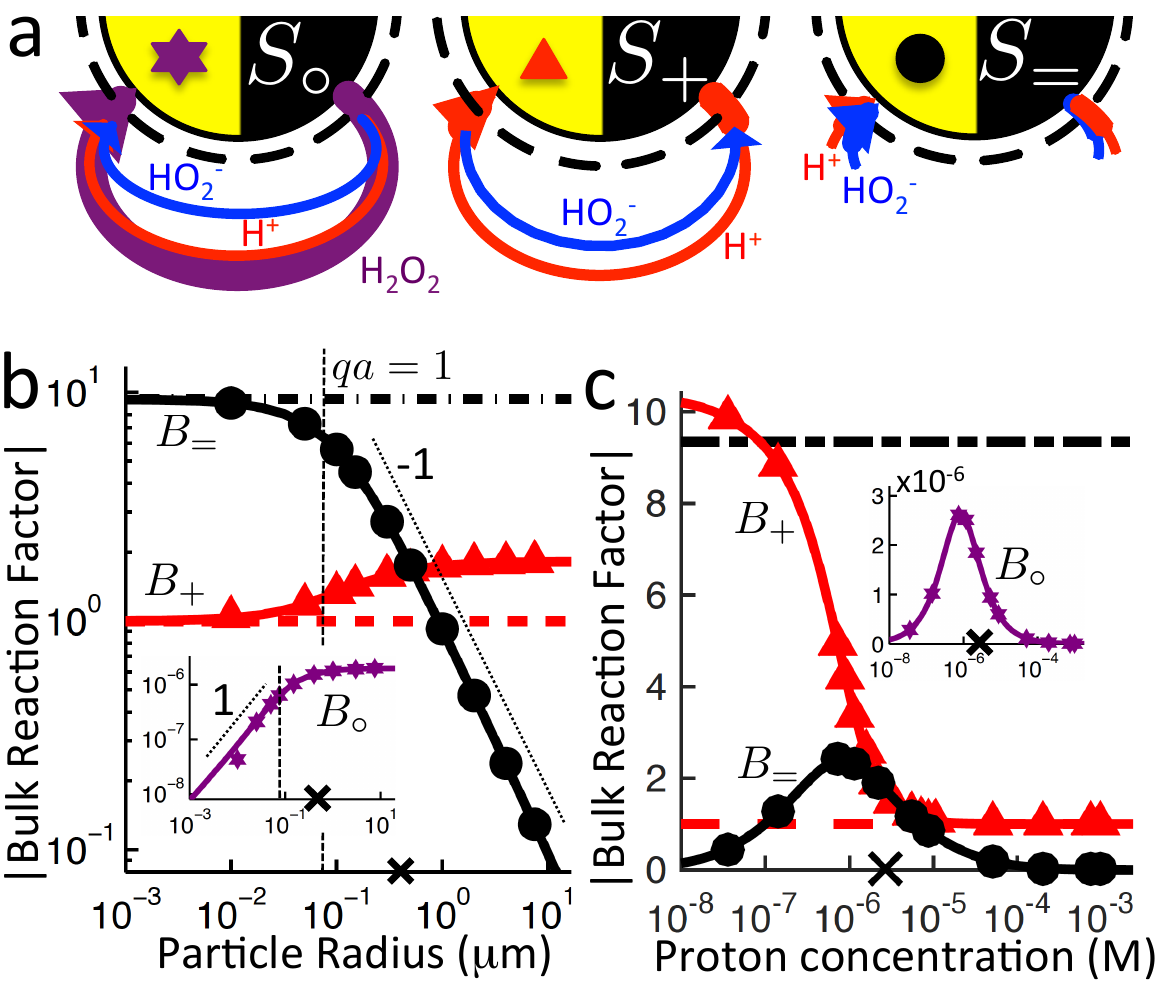}
  \caption{\label{fig:mobil} a) Recap of the model swimmers $S_\circ$, $S_+$, and $S_=$ and the effect of bulk reactions, from Fig.~\ref{fig:reactionEffect}. b-c) Magnitude of the dimensionless bulk reaction factors $|B|$ for type $S_\circ$ (\textcolor{ABpurple}{$\boldsymbol{\bigstar}$}, insets), $S_+$ (\textcolor{ABred}{$\boldsymbol{\blacktriangle}$}), and $S_{=}$ (\newmoon) propulsion, from analytical theory with (solid curves) and without (broken curves) bulk reactions; and FEM simulations (symbols). $\boldsymbol{\times}$ indicates the base parameter set defined in text. For (b) the particle radius and (c) the proton concentration $c_+^\infty$, at fixed $\kappa$.}
\end{figure}

We study the bulk reactions by varying two common experimental parameters: particle radius, and proton concentration, i.e., pH, Fig.~\ref{fig:mobil}b-c respectively. Including bulk reactions (solid curves $=$ analytic; solid symbols $=$ FEM numerics) introduces a range of effects compared to the case with no bulk reactions (broken, horizontal lines), with qualitatively different behavior for the three swimmer models.

\begin{table}[t]
  \caption{\label{table2} The bulk mobility factors predicted in the thin-screening $\kappa a\gg 1$, $\kappa a\gg qa$, and low dissociation $c_+^\infty,c_-^\infty\ll c_\circ^\infty$ limits, for low, $qa\ll 1$ and high $qa\gg 1$ reaction rates. In both limits the prefactor should be multiplied by the relevant expression in the right-hand columns. The full expressions are given in Appendix~\ref{app: ff calc}.}
  \begin{tabular}{c|c|c|c}
                      & Prefactor $\times$                                                                   & $qa\ll 1$           & $qa\gg 1$         \\
  \hline
                      &                                                                                      &                     &                   \\
  $B_\circ$ & $\alpha\dfrac{d_+-d_-}{d_+d_-}$ & $\dfrac{(qa)^2}{2}$ & $1$               \\
                      &                                                                                      &                     &                   \\
  $B_+$     & $1$                                                                                  & $-\dfrac{1}{d_+}$   & $-\dfrac{1}{d^*}$ \\
                      &                                                                                      &                     &                   \\
  $B_=$     & $\dfrac{d_+-d_-}{d_+d_-}$                                                            & $1$                 & $\dfrac{2}{qa}$   \\
                      &                                                                                      &                     &                   \\
  \end{tabular}
\end{table}

Examining Fig.~\ref{fig:mobil}b first, the bulk reactions permit propulsion of the neutral swimmer $S_\circ$ (inset), and $B_\circ$ increases with radius, saturating for large radii. However, the magnitude of $B_\circ$ always remains smaller than that of the other swimmers by a factor of order $10^{-6}$. In practice, this is typically partially compensated for by the much larger flux of the neutral species. For the proton-current-driven swimmer $S_+$, $B_+$ shows plateaux at both large and small radius, with the large-radius plateau approximately twice as high. For the ionic-diffusiophoretic swimmer $S_=$, $B_=$ scales inversely with radius for large radius but has a plateau at small radius. 

Meanwhile, varying the proton concentration $c^\infty_+$, as in Fig.~\ref{fig:mobil}c, produces a peak in $B_\circ$ and $B_=$, and decreases the overall value of $B_=$ by at least a factor of 5 compared to without reactions. For $S_+$ there are again two plateaux, at high and low $c_+^\infty$, with the low $c^\infty_+$ plateau now a factor of approximately 10 higher than the other.
 
The main control parameter for all these effects is $qa$, and there is a qualitative change of behaviour for all three swimmers at $qa\approx 1$: the vertical lines on Fig.~\ref{fig:mobil}b are for $qa=1$. In Table~\ref{table2}, we write down the bulk parameters for each of the model swimmers in the limits $qa\ll 1$ and $qa\gg1$. The full analytical expressions, which are lengthy, are provided in Appendix~\ref{speedModel}, but the basic physics can be understood from the limiting behaviour. For the Table, we have also assumed weak ionic dissociation, i.e., $K_\mathrm{eq}\ll c_\circ^\infty$ which is valid here, and thin electrostatic screening, $\kappa a\gg 1$. These assumptions also apply to the analytical expressions given in the rest of this section. Table~\ref{table2} matches Fig.~\ref{fig:mobil} in all but one respect, which is the scaling of $B_\circ$ at $qa\ll 1$, and this difference occurs because the assumption $\kappa a\gg 1$ does not hold for small $a$ in Fig.~\ref{fig:mobil}b. The parameters $\alpha$ and $d^*$ will be defined below. For Fig.~\ref{fig:mobil}c, $qa\approx 7$ or larger, so we will assume that this figure is always in the $qa\gg 1$ limit.

To understand the results shown in Fig.~\ref{fig:mobil} and Table~\ref{table2}, we will examine the bulk reactions in terms of three physical principles: reactive screening, the composition of the electrical current, and the dissociation of the neutral flux. Though we focus on these underlying principles, which are crucial for understanding the effect of bulk chemical reactions on \textit{any} swimmer, this structure also allows us to discuss the three model swimmers in a logical order: $S_=$, $S_+$ and $S_\circ$.

%%%%%%%%%%%%
\subsection{\label{sub:principles}Reactive Screening (Model $\mathbf{S_=}$)}

If an ion is released from the particle surface, it will react and come into local equilibrium with the surrounding solution. The characteristic distance over which this approach to equilibrium occurs can be called a `reactive screening length' $q^{-1}$. As for the simple model discussed in Section~\ref{bulk chemistry}, the reactive screening length is a balance between molecular diffusion and the reaction rate. However, the expression for $q$ is more complex than in the simple model. We find
\begin{align}
  q  &=  \sqrt{k_\mathrm{as}\left(\frac{c^\infty_{+}D_{+}+c^\infty_{-}D_{-}}{D_{+}D_{-}}\right)} .\label{qf}
\end{align} 
Mathematically, $q$ corresponds to one of the eigenvalues of the linear system of equations, Eq.~\eqref{flux linear div}-\eqref{rho linear}, see Appendix~\ref{ConcPot}. For our base parameter set, we obtain a screening length $q^{-1}=74~\nm$.

\begin{figure}
  \centering
  \includegraphics[width=8.5 cm]{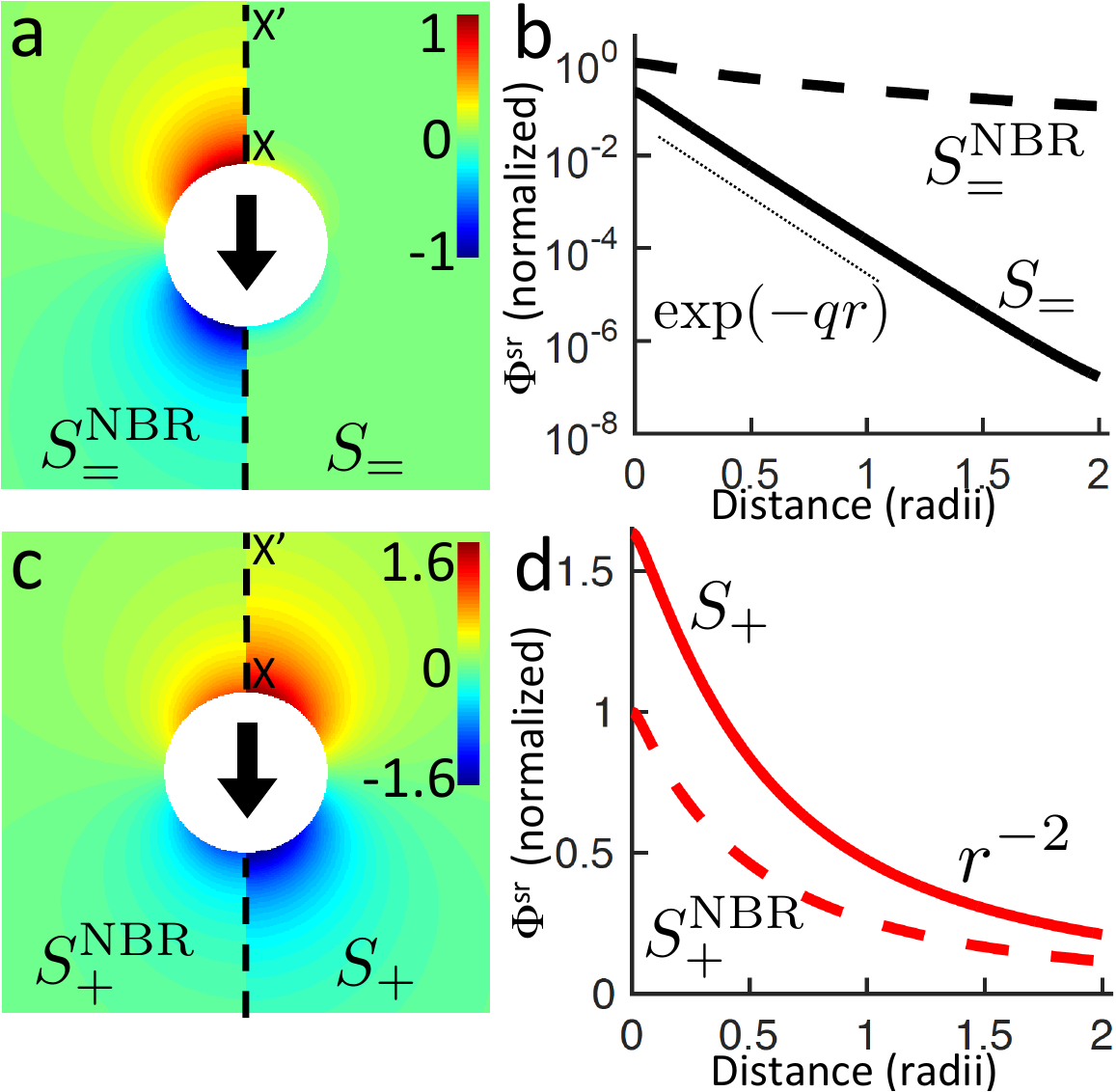}
  \caption{\label{fig:potflow} Normalized surface-reaction-generated potential $\phi^\mathrm{sr}$ for (a) $S_=$ and (c) $S_+$ swimmers, with (right) and without (left) bulk reactions. In each case, the potentials are normalized by the largest value of $|\phi^\mathrm{sr}|$ without bulk reactions. (b) Normalized radial decay of $\phi^\mathrm{sr}$ along X-X$'$ for (b) $S_=$ and (d) $S_+$. Solid curves are with bulk reactions, dashed curves without.}
\end{figure}

Just as for the simple model, reactive screening gives an exponential decay of chemical concentrations with distance from the particle surface. The inclusion of charged species means that the electrostatic potential can now also become screened. However, we observe such electrostatic screening only for $S_=$ swimmers, where the two ions released from the surface both react with oppositely charged ions in the bulk solution, causing an exponential decay in the resulting diffusion potential, see Fig.~\ref{fig:potflow}a.

For $S_+$ swimmers, no such screening is observed, see Fig.~\ref{fig:potflow}b. This is because the electrical current is conserved, so cannot be screened, and hence the associated electrical field also retains its unscreened dipolar form. Similarly, the \ce{H2O2} concentration field $c_\circ$ remains unscreened because this field is approximately conserved in the weak-dissociation limit. This results in an unscreened electrostatic potential field for $S_\circ$ swimmers (not shown).

For $S_=$ swimmers, seactive screening also explains the $1/(qa)$ scaling of $B_=$ at high $qa$, as we show with a simple scaling argument: From Eqs.~\eqref{j boundary}-\eqref{flux}, we expect a fixed ratio between the surface reaction rates and the concentration gradients normal to the surface. For example, at $r=a$
\begin{align}
  \frac{\partial c_+}{\partial r}\approx - \frac{j^\mathrm{s}_+}{D_+}\,, \label{prop}
\end{align}
independent of other parameters. For $qa\gg 1$, the concentration decays exponentially away from the surface
\begin{align}
  c_+\propto\exp(-qr) .
\end{align}
Differentiating this equation with respect to $r$ gives ${\partial c_+/\partial r\approx - qc_+}$ and comparing this with Eq.~\eqref{prop} yields 
\begin{align}
  c_+ (r=a)\approx \frac{j^\mathrm{s}_+}{D_+ q} . \label{c_+}
\end{align}
Since the diffusion potential is proportional to the ionic concentrations ($c_+$, $c_-$), and the propulsion speed $U_=$ is proportional to the diffusion potential, we have $U_=\propto q^{-1}$. On the other hand, without bulk reactions the only relevant length scale is $a$, so a similar argument gives $U_=^\mathrm{NBR} \propto a$. Therefore, one obtains $B_=\propto U_= / U^\mathrm{NBR}_= = 1/(qa)$. Physically, for $qa\gg 1$, the concentration flux only has the small screening length $q^{-1}$ over which to set up a diffusion potential, whereas without bulk reactions a length of order $a$ is available. 

This $1/(qa)$ scaling immediately explains the $B_=\propto 1/a$ scaling in Fig.~\ref{fig:mobil}b at high $qa$. We can also understand the peak in $B_=$ in Fig.~\ref{fig:mobil}c by noting that the screening length $q^{-1}$ vanishes both for high $c_+^\infty$, and for high $c_-^\infty$, Eq.~\eqref{qf}. Since $c_-^\infty$ scales inversely with $c_+^\infty$ due to the ionic equilibrium of Eq.~\eqref{eq}, this means that $q^{-1}$ vanishes at either end of the $c_+^\infty$ scale. As $B_=\propto q^{-1}$, it too vanishes at either extreme and is peaked for intermediate $c_+^\infty$. Physically, at either end of the $c_+^\infty$ scale, the high concentration of ions screens electric fields, preventing the formation of a diffusion potential.

%%%%%%%%%%%%
\subsection{Composition of the Electrical Current (Model $\mathbf{S_+}$)}

The total electrical current in the bulk is a conserved quantity, and is therefore not screened. However, the individual ionic fluxes making up that current are not conserved, and the bulk reactions modify the identity of the current-carrying ions. As discussed in Section~\ref{sec:ephorNBR}, this is important because the swimming speed scales inversely with the diffusivity of the current-carrying ion or ions. In our system, an initially pure proton current will be partially replaced by \ce{HO2-} ions travelling in the opposite direction, as illustrated in Fig.~\ref{fig:kappaq}. In the reactive, $qa\gg 1$ limit we can calculate the composition of this electrical current relatively simply. From this, we will obtain the propulsion speed of the $S_+$ swimmer.

In the $qa\gg 1$ limit, at any point outside the thin reactive screening layer, the ions released from the surface will have had time to come into equilibrium with each other. This is equivalent to requiring that the chemical production rates vanish, i.e., ${R_l=0}$. In the linear approximation, see Eq.~\eqref{Taylor}, this means
\begin{align}
  \sum_mk_{lm}x_m &= 0 , \label{eq coupling}
\end{align}
In other words, the deviations in concentration $x_l$ of each of the reactive chemical species are coupled by the reaction matrix $\matrify{k}$ given in Eq.~\eqref{kik}. This concentration coupling also implies a coupling of the chemical fluxes, in the same way that charge conservation implies the conservation of electrical current. Consider the linearized flux equation, Eq.~\eqref{flux linear}. Multiplying both sides by $k_{ml}/(D_lc_l^\infty)$ and summing over $l$ yields
\begin{align}
	-\sum_l\frac{k_{ml}\boldsymbol{j}_l}{D_lc_l^\infty} &= \nabla\sum_lk_{ml}x_l + \nabla\psi \sum_lk_{ml}z_l \,.
\end{align}
Then, from Eq.~\eqref{eq coupling}, with $l$ and $m$ exchanged, the first term on the right vanishes, while, because total charge is conserved for all chemical reactions, $\sum_l k_{ml}z_l=0$, see Eq.~\eqref{charge conservation} and the second term vanishes too. Hence, the general flux coupling equation is
\begin{align}
	\sum_l\frac{k_{ml}\boldsymbol{j}_l}{D_lc_l^\infty} &= 0 . \label{A current}
\end{align} 
For our specific system, substituting the expression for $\matrify{k}$ from Eq.~\eqref{kik} into Eq.~\eqref{A current} then gives
\begin{align}
	\frac{\boldsymbol{j}_+}{D_+c_+^\infty}+\frac{\boldsymbol{j}_-}{D_-c_-^\infty} &= \frac{\boldsymbol{j}_\circ}{D_\circ c_\circ^\infty} . \label{equilibrium current}
\end{align}
The physical meaning of Eq.~\eqref{equilibrium current} is that each of the molecular fluxes has a characteristic scale set by $D_lc_l^\infty$, and that, with this scaling, the relationship between the currents is set by the stoichiometry of the bulk reactions. Eq.~\ref{equilibrium current} can be rearranged to give each of the ionic fluxes $\boldsymbol{j}_\pm$ in terms of the conserved quantities $\boldsymbol{i}$ and $\boldsymbol{j}_\circ$
\begin{align}
	\boldsymbol{j}_{\pm} &= \pm\frac{\boldsymbol{i}}{e}\frac{D_\pm c_\pm^\infty}{D^*(c_+^\infty+c_-^\infty)}+\boldsymbol{j}_\circ\alpha ,\label{j+-}
\end{align}
where $D^*$ is the concentration-averaged diffusivity of the active ions
\begin{align}
  D^* &= \frac{D_+c_+^\infty+D_-c_-^\infty}{c_+^\infty+c_-^\infty} , \label{D*}
\end{align}
and the dimensionless factor $\alpha$, which specifies the equilibrium decomposition of a neutral current into ionic currents, is
\begin{align}
  \alpha &= \frac{(D_\circ c_\circ^\infty)^{-1}}{(D_+ c_+^\infty)^{-1}+(D_- c_-^\infty)^{-1}} . \label{alpha2}
\end{align}
The meaning of the first term in Eq.~\eqref{equilibrium current}, which is relevant for $S_+$ swimmers, is that the electric current is carried by a fixed proportion of \ce{H+} ions travelling in one (positive) direction, and a countercurrent of \ce{HO2-} ions in the opposite (negative) direction. In the second term, which is relevant for $S_\circ$ swimmers, the neutral flux $\boldsymbol{j}_\circ$ continuously dissociates into \ce{H+} and \ce{HO2-} ions, producing small, equal fluxes of these ions, which travel with the neutral flux.

If we are also outside the electrostatic screening length, which is the case in the $\kappa a\gg 1$ limit, we also have a zero-charge-density condition, which reads 
\begin{align}
  \sum_lc_l^\infty x_lz_l=0 \,.
\end{align}
Just as above, but now multiplying Eq.~\eqref{flux linear} by $z_l/D_l$, we can derive a direct relationship between the electric field and the chemical fluxes~\cite{esplandiu16}
\begin{align}
  \nabla\phi &= -\frac{e}{\kappa^2\epsilon}\sum_l\frac{\boldsymbol{j}_lz_l}{D_l}\,,\label{generalized} 
\end{align}
which, combined with Eq.~\eqref{j+-}, yields a version of Ohm's law for the reactive limit
\begin{align}
  \nabla\phi &= \frac{1}{\kappa^2\epsilon}\left[-\left(\frac{1}{D^*}\right)\boldsymbol{i} +\left(\alpha\frac{D_+-D_-}{D_+ D_-}\right)e\boldsymbol{j}_\circ\right] \,. \label{phi}
\end{align}
Comparing the first term of this equation with Eq.~\eqref{Ohm NBR} for self-electrophoresis without bulk reactions, we see that they are identical apart from the switch from $D_+$ to $D^*$.

For an $S_+$ swimmer, $\boldsymbol{j}_\circ=0$, and for high $qa$ all of the electric field outside a thin screening layer will be determined by the first term of Eq.~\eqref{phi}. From this we can understand the $1/D^*$ factor which appears in $B_+$, Table~\ref{table2}. Just as without bulk reactions, see Eq.~\eqref{standard model}, the propulsion speed is inversely proportional to the diffusivity of the current-carrying ion. However, the current is now made up of two ions, with a total effective diffusivity $D^*$. This explains the 2$\times$ speed increase in Fig.~\ref{fig:mobil}b: $D_+$ at low $a$ is replaced by $D^*$ at high $a$, and for $c_+^\infty=c_-^\infty$, which is the case in Fig.~\ref{fig:mobil}b, $D^*=(D_++D_-)/2\approx D_+/2$. 

To understand the effect of varying $c_+^\infty$, Fig.~\ref{fig:mobil}c, we examine the form of $D^*$ in Eq.~\eqref{D*}. At high $c_+^\infty$, $D^*\approx D_+$, while at low $c_+^\infty$ ($=$ high $c_-^\infty$), $D^*\approx D_-$. Physically, this is again simply a result of the relative number of each ionic species: if there is an overwhelming number of protons in solution, then the ionic current must be carried predominately by protons. This explains why there is a factor of $D_+/D_-\approx10$ speed difference between the two plateaux for $B_+$ in Fig.~\ref{fig:mobil}c.

%%%%%%%%%%%%
\subsection{\label{sub:diss}Dissociation of the Neutral Flux (Model $\mathbf{S_\circ}$)}

To understand the dissociation of a purely neutral flux, we must look in detail at the parameter $\alpha$ in Eq.~\eqref{alpha2}. The form of $\alpha$ can be explained by the fact that in the absence of a net electrical current, e.g., for $S_\circ$ swimmers, the ionic currents are constrained by $\boldsymbol{j}_+=\boldsymbol{j}_-$. This means that the total ionic flux will be limited by whichever ion has the lower value of $D_lc_l^\infty$, as this ion will contribute most to the flux balance in Eq.~\eqref{equilibrium current}. Hence the parameter $\alpha$, like $q^{-1}$, vanishes at the extreme ends of the $c_+^\infty$ scale: at low $c_+^\infty$ it is limited by the low proton concentration, and at high $c_+^\infty$, by the low \ce{HO2-} concentration.

The dissociation of the neutral flux generates a diffusion potential. Hence, the prefactor for $B_\circ$ in Table~\ref{table2} is made up of two factors: $\alpha$ and $(d_+-d_-)/(d_+d_-)$, the latter of which controls the diffusion potential just as for $B_=$. The peak in $B_\circ$ as a function of $c_+^\infty$ then follows directly from the behaviour of $\alpha$. 

Interestingly, both $S_\circ$ and $S_=$ show peaks in speed at intermediate $c_+^\infty$, but for two different reasons. For $S_=$, the reason is that the reassociation of ions is slowest at intermediate concentrations. For $S_\circ$, the reason is that the least conductive fraction of the solution limits the total carrying capacity, and this effect is strongest at either extreme in pH.

%%%%%%%%%
\section{\label{sec:compare}Comparison with Experiments}

\begin{figure}
  \centering
  \includegraphics[width=8.5 cm]{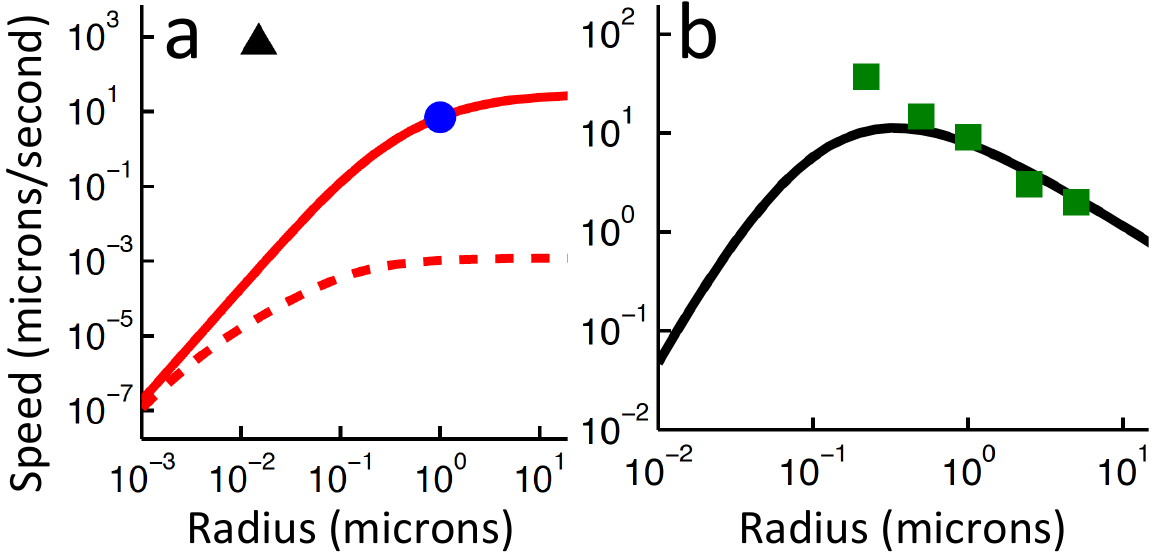}
  \caption{Comparison between theory and experiments. (a) The predicted speed of a swimmer powered by a proton current, in the presence of chemical reactions, with parameters chosen to match typical measurements on microparticles (red solid, theory; blue circle, experiment~\cite{wheat10}); and, with the same surface parameters, but other parameters chosen to match experiments on a nanoswimmer (red dashed, theory; black triangle, experiment~\cite{lee14}) (b) The speed of an $S_=$ swimmer plotted against experimental data on Janus-Pt microswimmers~\cite{ebbens12}. The experimental error bars are smaller than the data points.}
\label{fig:exp}
\end{figure}

We now compare our theoretical predictions with experimental results, in so far as this is possible at present. We stress here again the lack of understanding of the surface chemistry, and in particular of the effect of experimental parameters on the rate of surface reaction rates. Since we do not attempt to predict these reaction rates here, we cannot yet really test our theoretical predictions. In this section, however, we will \textit{assume} that the surface properties, i.e., surface reaction rates and surface charge densities, are constant. Under this assumption, $U\propto B$ for all swimmers, so long as the fuel and salt concentrations remain fixed, and this will allow us to make some suggestive comparisons with experiments.

Independent of bulk reactions and swimmer type, we predict a speed scaling with $a^3$ for small particles. In particular, for a proton powered swimmer $U_+^\mathrm{NBR}$, the predicted speed is as shown in Fig.~\ref{fig:exp}a (solid curve), due mostly to the new screening parameter $F(\kappa a)$ --- the bulk reactions do not significantly modify the form of this curve. Here, we have matched the solution parameters to those of Ref.~[\!\!\citenum{wheat10}] and have chosen the (constant) surface parameters so that the speed for a 1$~\um$ radius bimetallic sphere is 7$~\umpers$ (blue circle), as found in Ref.~[\!\!\citenum{wheat10}] (see Appendix~\ref{sec:rates} for the surface parameters used). We also plot data from Ref.~[\!\!\citenum{lee14}], which found that a swimmer of radius 10 nm has $U=650~\umpers$ (black triangle). These two experiments used similar concentrations of \ce{H2O2}, but differed in other features of the measurements, such as the salt concentration and the swimmer manufacturing technique. We thus also plot a theoretical curve (dashed), with the environmental parameters modified to match those of the nanoswimmer experiment~\cite{lee14}, keeping the surface parameters the same as found for the match to Ref.~[\!\!\citenum{wheat10}]. 

The resulting theoretical prediction is the dashed red curve in Fig.~\ref{fig:exp}a. If the assumption of constant surface properties holds true, then this curve should agree with the experimental value for nanoswimmers. Instead, there is a clear disparity amounting to several orders of magnitude (black arrow). We verified that this discrepancy is not the result of our linear approximation: we find a good match between analytics and numerics for values of $\sigma$ and $j^\mathrm{s}_+$ even higher than those used in plotting Fig.~\ref{fig:exp}a, see Appendix~\ref{app:finite}. The discrepancy could be explained in at least two ways. It may be that self-electrophoresis is not the correct propulsion mechanism for bimetallic nano-swimmers. It has recently been found that nanometer scale biological enzymes also exhibit self-propulsion~\cite{muddana10, sengupta13}, and a range of mechanisms has been proposed for this propulsion~\cite{golestanian15}, some of which might also apply to bimetallic nano-swimmers. Alternatively, it may be that the assumption of constant surface properties is inappropriate. That is, the proton current density could be much higher for these nano-swimmers than for micro-swimmers. Whatever the explanation, this discrepancy highlights the need for more systematic studies of identical or comparable swimmers over wide parameter ranges, as in Refs.~[\!\!\citenum{ebbens12, brown14, ebbens14}], and for independent measurements of the relevant ionic reaction rates. 

For Pt-polystyrene Janus particles, such systematic studies do exist~\cite{ebbens12}. These show a $U\propto a^{-1}$ scaling for $0.2~\um<a<5~\um$ (this scaling has also been observed over a narrower range for some bimetallic swimmers~\cite{wheat10}). Self-electrophoresis $S_+$, reaction (R2), is currently the preferred mechanism for Pt-polystyrene Janus swimmers~\cite{brown14, ebbens14}, but comparison of this $1/a$ scaling with Fig.~\ref{fig:mobil}b suggests self-ionic diffusiophoresis $S_=$ as an alternative mechanism, corresponding to reaction (R3). This is plausible: ion release without net electrical currents, which would correspond to reaction (R3), has previously been observed for \ce{H2O2} decomposition on \ce{Pt}~[\!\!\citenum{ono77}]. This mechanism would also avoid the conceptual difficulty of producing a net ionic current in single-catalyst systems~\cite{brown14, ebbens14}. However, when we plot the experimental data from Ref.~[\!\!\citenum{ebbens12}] against our theoretical predictions for $S_=$ propulsion, Fig.~\ref{fig:exp}b --- which is again scaled to match the experimental data for $1~\um$ radius swimmers, see Appendix~\ref{sec:rates} --- we see that the fit fails at small $a$, again due to the $F(\kappa a)$ parameter. It is possible that evaluation of the complete \ce{H2O2}-\ce{H2O} reaction system would provide a better fit, but this goes beyond the scope of this work. 

Note that the $1/a$ scaling has previously been explained by postulating that the overall surface reaction rate $j^\mathrm{s}_\circ$ is limited by diffusion~\cite{ebbens12}, and therefore scales as $1/a$ just from geometrical arguments. However, the diffusion-limit implies a large flux density $j^\mathrm{s}_\circ\approx D_\circ c_\circ^\infty/a$, which for a 1 $\um$ radius swimmer in ${3~\mathrm{M}~\ce{H2O2}}$, as in Ref.~[\!\!\citenum{ebbens12}] requires $j^\mathrm{s}_\circ\approx 3\times 10^{24}~\mathrm{m^{-2}}\pers$. So far, only much smaller rates, $j_\circ^\mathrm{s}\approx 10^{22}~\mathrm{m^{-2}}\pers$, have been measured, both by us~\cite{brown14} and by the authors~\cite{ebbens14} of Ref.~[\!\!\citenum{ebbens12}]. Therefore, these swimmers do not appear to be in the diffusion limited regime, so this explanation for the $1/a$ scaling cannot hold.

Next, we have previously calculated the values of the uncharged flux $j^\mathrm{s}_\circ$ and the charge density $\sigma$ for Pt-coated Janus swimmers~\cite{brown14}. We estimated that the propulsion speed of such swimmers was too high to be explained by a purely uncharged reaction like (R1)~\cite{brown14}. This estimate did not allow for bulk ionic reactions. However, including these reactions, we calculate in Appendix~\ref{app:finite} that such a mechanism could still only account for $\approx~5\%$ of the observed speed of these swimmers. Hence, a model with just surface reaction (R1), even with bulk dissociation, cannot explain the propulsion of \ce{H2O2}-powered swimmers, so that such swimmers probably still require more complex ionic surface reaction schemes like (R2) and (R3). Nevertheless, purely neutral-surface-reaction mechanisms could still be relevant for swimmers powered by more dissociative fuels, such as hydrazine~\cite{gao14}.

Finally, turning to the effect of pH, there have been two suggestive studies~\cite{brown14,gao14}, but no systematic investigation. First, we found that \ce{NaOH} reduced the swimming speed of Pt-polystyrene Janus swimmers, but that this effect was much weaker than the speed reduction due to \ce{NaCl}~\cite{brown14}. This is consistent with our prediction that increasing pH at fixed Debye length should raise the swimming speed for any of the 3 swimming mechanisms discussed. Raising the pH corresponds to moving left from the $\boldsymbol{\times}$ symbol in Fig.~\ref{fig:mobil}c; for all swimming mechanisms the value of $B$ increases in this direction. 

Second, the silica-iridium swimmers of Ref.~[\!\!\citenum{gao14}] show a clear spike in speed as a function of fuel (hydrazine) concentration similar in form to the peaks in Fig.~\ref{fig:mobil}c. This spike could be due to modulation of pH by the reaction product ammonia, which would imply that either neutral self-diffusiophoresis or self-ionic diffusiophoresis dominates this swimmer's self-propulsion. In both these experiments, however, variation of the reaction rates with pH could also explain the results~\cite{mckee69}, so further systematic study is necessary.

%%%%%%%%%
\section{\label{sec:concl}Conclusion}

In this article, we have theoretically explored the behaviour of chemically-propelled, synthetic microswimmers in their most usual chemical environment --- water. We have focussed on two unavoidable properties of aqueous solutions --- electrostatic screening and ionic dissociation --- and calculated their effect on the swimming speed of a wide range of microswimmers propelled by chemical reactions on their surfaces. These effects have not been studied systematically before; nevertheless, they are highly significant, and including these effects can modify predicted swimming speeds by several orders of magnitude. 

By ionic dissociation, we mean the breaking up of neutral molecules, including water, into charged species, for example, \ce{H2O2 <=> H+ + HO2-}. One of our main prediction is that this kind of ionic dissociation reaction allows even microswimmers whose surface chemistry does not involve any ions, e.g., swimmers propelled by the simple decomposition of hydrogen peroxide, \ce{2H2O2 -> 2H2O + O2}, to generate ionic gradients and thereby electric fields. The implication of this is that all microswimmers in water should experience some degree of self-electrophoresis, i.e., propulsion via self-generated electric fields. This is significant because self-electrophoresis is much more efficient than other putative propulsion mechanisms and is likely to dominate over them. Put simply: our results imply that all swimmers in aqueous solution are likely to be self-electrophoretic to a major degree.

The second major prediction of our work is that for some types of chemically-propelled swimmers, ionic dissociation reactions will result in a kind of exponential `reactive-screening'~\cite{banigan16}. Electrical and chemical concentration fields generated by surface reactions on microswimmers are usually taken to decay slowly into the bulk solution, that is, as a power-law with distance. Ionic dissociation can instead produce a short-ranged exponential decay of these fields, just as in electrostatic screening. This is significant because these chemical and electrical fields are implicated in inter-swimmer interactions and collective behaviour, and the interaction range will play a crucial role in this behaviour. 

Our third prediction relates to electrostatic screening itself. Most theoretical treatment of microswimmers has focussed on the thin-screening limit, where the electrostatic screening length is much smaller than the swimmer size. For very small swimmers, this limit does not apply, and we find that this massively reduces the predicted swimming speed. This is important because experiments on nanoscale swimmers~\cite{lee14} show that these in fact swim faster than microswimmers, in apparent contradiction to our predictions. This opens up the exciting possibility that nanoscale swimmers move by entirely novel mechanisms compared to their microscopic counterparts.

Finally, the general conclusion that we draw from our results is that much more experimental work is required to understand self-propulsion mechanisms. The effect of ionic dissociation in particular depends crucially on the type of surface reactions which are responsible for propulsion --- and the details of these reactions remain almost universally unknown. What is most urgently required in this regard is the independent measurement of surface reaction rates, which is challenging, and has so far only been achieved in the simplest of cases. However, recent results with electroosmotic pumps~\cite{afsharfarniya13} suggest that such measurements will not long remain beyond our reach. We particularly hope that our theoretical results will lead to renewed efforts in this direction.

More generally, our results suggest that a deeper understanding of self-propulsion will lead to greater insights into swimmer-swimmer interactions and collective effects. This is particularly relevant to synthetic swimmers, as their propulsion is closely coupled to their interactions through self-generated electrostatic, chemical, and hydrodynamic flow fields. We have shown here that reactive screening can qualitatively change the electrostatic interactions between swimmers. A detailed follow-up study will look explicitly at such interactions. Further theoretical work will focus on applying our calculations to fully realized experimental systems,~e.g., mixed metal-dielectric swimmers.

{\it Acknowledgements:} This work was funded by UK EPSRC grant EP/J007404/1 (WP); ERC Advanced Grant ERC-2013-AdG 340877-PHYSAP (AB and WP); NWO Rubicon Grant \#680501210 and Marie Sk{\l}odowska-Curie Intra European Fellowship (G.A. No. 654916) within Horizon 2020 (JdG); and DFG SPP 1726 `Microswimmers --- from single particle motion to collective behavior' (JdG and CH). We thank Patrick Kreissl for checking many of the calculations; and Mike Cates, Michael Kuron, Davide Marenduzzo, Alexander Morozov, Mihail Popescu, Georg Rempfer, Joakim Stenhammar, and Teun Vissers for useful comments on the manuscript.

\appendix

%%%%%%%%%
\section{\label{sec:linear}Calculation of the Analytical Solution}

In this appendix, we explicitly calculate the propulsion speed for a general swimmer. This calculation is based on the linearized model described in Section~\ref{sec:math}, but now with a more concise notation,~\ref{linear appendix}. From this linearized model, we determine first the electrostatic potential fields,~\ref{ConcPot}, then the propulsion speed,~\ref{app: ff calc}, as set out in Section~\ref{sec:math}. In ~\ref{speedModel}, we apply this general calculation to determine the speed of the model swimmers presented in the main text. Finally, in~\ref{sec:electro}, we demonstrate the equivalence of uniform charge and uniform potential boundary conditions for the calculation of the propulsion speed.

\subsection{The Linearized Model ~\label{linear appendix}}

We begin with the linearized model described in Section~\ref{sub:solveAnal}. For notational convenience, we define a composite dimensionless parameter $y_l$ by combining the linearized potential $\psi$ and concentration $x_l$ fields
\begin{align}
y_l  &=  \left\{
  \begin{array}{ll}\psi        \,,            & l=0 , \\
                                              x_l \,,& l=1,2,\dots,N .
  \end{array}
\right. \label{I}
\end{align}
With this notation, the linear system of equations, Eq.~\eqref{flux linear div} and Eq.~\eqref{rho linear} is given by
\begin{align}
  \nabla^2 y_l  &=  \left\{
  \begin{array}{ll}
    -\dfrac{e^2}{\epsilon k_BT}\displaystyle\sum_{m=1}^Nz_mc^{\infty}_m y_m\,, & l=0 , \\
    -z_l\nabla^2 y_0 - \dfrac{1}{D_lc^{\infty}_l }\displaystyle\sum_{m=1}^Nk_{lm}y_m\,, & l=1,2,\dots,N\,.
  \end{array}
  \right. \label{L}
\end{align}
Equation~\eqref{L} represents a system of $N+1$ linear equations. However, several of the species, typically inactive ions such as \ce{Na+} or \ce{Cl-}, may not be involved in any bulk or surface reactions, and we will now show that these inactive species can be eliminated. With $N'$ reactive species, where $N'< N$, we specify that the first $N'$ indices correspond to the reactive species. For the remaining, unreactive species, all the bulk reaction coefficients, $k_{lm}$ are zero, and there is no surface flux, so Eq.~\eqref{L} can only be satisfied if
\begin{align}
	y_l &= -z_l y_0\,, \quad l > N' . \label{M}
\end{align}
This is the linear approximation to the Boltzmann distribution, which one expects, since these unreactive species should be in equilibrium with the electric field. Using Eq.~\eqref{M}, these ions can be eliminated from the remaining $N'+1$ parts of Eq.~\eqref{L} to yield
\begin{widetext}
  \begin{align}
    \kappa^{-2}\nabla^2 y_l &= \left\{
    \begin{array}{ll}
      -\displaystyle\sum_{m=1}^{N'}\dfrac{\chi_my_m}{z_m}+\left(1-\displaystyle\sum_{m=1}^{N'}\chi_m\right) y_0, & l=0 , \\
	~\\
      \displaystyle\sum_{m=1}^{N'}\left(\dfrac{z_l\chi_m}{z_m}-\dfrac{k_{lm}\kappa^{-2}}{D_lc^{\infty}_l }\right)y_m -z_l\left(1-\displaystyle\sum_{m=1}^{N'}\chi_m\right) y_0, & l=1,2,\dots,N' ,
    \end{array}
    \right. \label{N}
  \end{align}
\end{widetext}
where $\kappa$ is the inverse Debye screening length
\begin{align}
  \kappa &= \left(4\pi l_B\sum_{l=1}^N z_l^2c_l^{\infty}\right)^{\frac{1}{2}} . \label{kappa}
\end{align}
with the Bjerrum length, $l_B=e^2/(4\pi\epsilon k_BT)$, and where $\chi_l$ is a dimensionless ionic concentration
\begin{align}
  \chi_l &= 4\pi l_B\kappa^{-2}z_l^2c^{\infty}_l .\label{alphasup}
\end{align}
Eliminating the inactive ions makes it clear that the motion of the swimmer cannot depend on the diffusivity of these ions, and is only affected by them through the value of $\kappa$ and through charge balance.

Finally, linearizing the boundary conditions in Eqs.~\eqref{G} and~\eqref{j boundary} gives
\begin{align}
	\hat{\boldsymbol{n}}\cdot\left(\nabla y_l(\boldsymbol{s}) + z_l\nabla y_0(\boldsymbol{s})\right) &= -\frac{j^\mathrm{s}_l}{D_lc^{\infty}_l } ,\nonumber\\
	\hat{\boldsymbol{n}}\cdot\nabla y_0(\boldsymbol{s}) &= -\frac{\sigma e}{k_BT\epsilon} \label{O} .
\end{align}

%%%%%%%%%%%%
\subsection{\label{ConcPot}Calculation of the Electrostatic Potential}

Equation~\eqref{N} has the form of a matrix equation with components corresponding to the chemical concentrations and the electrostatic potential, so it is convenient to introduce some additional matrix notation. The bold font is reserved for real-space vectors, such as the fluid velocity $\boldsymbol{u}$, while vectors in this concentration-potential space will be underlined. A general vector $\vectify{t}$ will have $N'+1$ components labelled $t_l$, while a matrix $\matrify{T}$ will have $(N'+1)\times(N'+1)$ components labelled $T_{lp}$. A point in the concentration-potential space is specified by the vector $\vectify{y}$, with components $y_l$, as defined in Eq.~\eqref{I}. Using this notation, we can rewrite Eq.~\eqref{N} as
\begin{align}
	\nabla^2 \vectify{y} &= \matrify{M}\vectify{y} , \label{P}
\end{align}
which can be solved by finding the $N'+1$ eigenvectors of the matrix $\matrify{M}$, with eigenvalues $\mu_p$. These eigenvectors define a new basis, in which $\matrify{M}$ is diagonal. Defining $\vectify{w}$ as the representation of $\vectify{y}$ in this basis, we have
\begin{align}
	\nabla^2 \vectify{w} &= \matrify{G}^2\vectify{w} , \label{QQ}
\end{align}
where the matrix $\matrify{G}$ is diagonal, with components $G_{lp}=\delta_{lp}g_p$, where $\delta_{lp}$ is the Kronecker delta and $g_p=\sqrt{\mu_p}$ is an inverse screening length. For the model system described in the main text, the unique values of $g_p$ are $\kappa$, 0 and $q$, as given in Eq.~\eqref{qf}. Here, for clarity, we use the index $p$ to refer to the screening lengths, and the indices $l$ or $m$ to refer to the concentrations and potentials, even where these are dummy indices. 

Equation~\eqref{QQ} is a series of $N'+1$ independent Helmholtz equations, and the full solution to this equation is just a vector of individual solutions to the Helmholtz equation. In spherical polar coordinates, these solutions have the form~\cite{kim03, banigan16}
\begin{align}
  w_p &= \sum_{n} w_{p,n}P_n(\cos{\theta})\left(\frac{a}{r}\right)^{n+1}\frac{T_n(g_pr)}{T_n(g_pa)}e^{-g_p(r-a)} ,\label{R}
\end{align}
with $w_{p,n}$ an as yet undetermined surface coefficient, $P_n$ the Legendre polynomial of order $n$~[\!\!\citenum{riley06}], $\theta$ the polar angle, and
\begin{align}
	T_n(x) &= \sum_{s=0}^n\frac{2^sn!(2n-s)!}{s!(2n)!(n-s)!}x^s . \label{S}
\end{align}
We refer to the Legendre components by the subscript $n$ throughout, and where we have multiple subscripts, the Legendre subscript shall be preceded by a comma. Transforming back into the original coordinate frame linearly combines the solutions in Eq.~\eqref{R}, so that the final form for the electrostatic potential is
\begin{align}
  \phi &= \sum_{p,n} \phi_{p,n}P_n(\cos{\theta})\left(\frac{a}{r}\right)^{n+1}\frac{T_n(g_pr)}{T_n(g_pa)}e^{-g_p(r-a)} , \label{AB}
\end{align}
with analogous expressions for each concentration field. Here, $\phi_{p,n}$ are surface coefficients which we will now determine.

First, transformation back into the original coordinate system is achieved with a transformation matrix $\matrify{K}$
\begin{align}
	\vectify{y} &= \matrify{K}\vectify{w}, \label{T}
\end{align}
where each element $K_{lp}$ of $\matrify{K}$ is equal to the $l^\mathrm{th}$ component (in the original coordinate system) of the $p^\mathrm{th}$ eigenvector. Applying this transformation to Eq.~\eqref{R} gives
\begin{align}
	y_l &= \sum_{p,n}K_{lp} w_{p,n}P_n(\cos{\theta})\left(\frac{a}{r}\right)^{n+1}\frac{T_n(g_pr)}{T_n(g_pa)}e^{-g_p(r-a)} . \label{U}
\end{align}
The boundary conditions specified in Eq.~\eqref{O} can also be re\-arranged into a matrix equation
\begin{align}
	\matrify{B}\hat{\boldsymbol{n}}\cdot\nabla \vectify{y}\bigg|_{r=a} &= \vectify{b} , \label{V}
\end{align}
where $\vectify{b}$ is a vector specifying each of the boundary fluxes or charge density. We define the harmonic components $\vectify{b}_{n}$ of $\vectify{b}$ by ${\vectify{b}=\sum_nP_{n}(\cos\theta)\vectify{b}_{n}}$, with analogous expressions defining $\matrify{B}_{n}$. The solution to the boundary conditions is found by inverting Eq.~\eqref{V} to yield
\begin{align}
	\vectify{w}_{n} &= \matrify{L}_{n}\vectify{b}_{n} , \label{W}
\end{align}
where
\begin{align}
	\matrify{L}_{n} &= \left[\matrify{B}_{n}\matrify{K}\matrify{D}_{n}\right]^{-1} , \label{X}
\end{align}
in which the diagonal matrix $\matrify{D}_{n}$ has elements
\begin{align}
	D_{lp,n} &= \delta_{lp}\left[ g_p\frac{\partial \log{T_n(x)}}{\partial x}\bigg|_{x=g_pa}-\left(\frac{n+1}{a}+g_p\right)\right] . \label{Y}
\end{align} 
Inserting the boundary conditions into Eq.~\eqref{U} then gives
\begin{align}
	y_l &= \sum_{p,n} y_{lp,n}P_n(\cos{\theta})\left(\frac{a}{r}\right)^{n+1}\frac{T_n(g_pr)}{T_n(g_pa)}e^{-g_p(r-a)} , \label{Z}
\end{align}
where the surface coefficients are
\begin{align}
	y_{lp,n} &= K_{lp}\sum_{m}L_{pm,n}b_{m,n} . \label{AA}
\end{align}
In particular, this yields for the surface coefficients of the electrostatic potential (for which the index $l=0$)
\begin{align}
\phi_{p,n} &= \frac{k_BT}{e}K_{0p}\sum_m L_{pm,n} b_{m,n} \,. \label{AC}
\end{align}
Eq.~\eqref{AC}, together with Eq.~\eqref{AB} completely determines the electrostatic potential field.

We can also determine the equilibrium and non-equilibrium components of the potential. Writing $\phi=\phi^\mathrm{eq}+\phi^\mathrm{sr}$, where $\phi^\mathrm{eq}=\phi\left(\{j^\mathrm{s}_l\}\rightarrow 0\right)$ is the equilibrium potential distribution without any surface chemical reactions (here, $\{j^\mathrm{s}_l\}$ is the complete set of surface fluxes), and $\phi^\mathrm{sr}=\phi\left(\sigma\rightarrow 0\right)$ is the additional potential generated by the surface reactions, we have
\begin{align}
  \phi^\mathrm{eq} &= \sum_{p,n} \phi^\mathrm{eq}_{p,n}P_n(\cos{\theta})\left(\frac{a}{r}\right)^{n+1}\frac{T_n(g_pr)}{T_n(g_pa)}e^{-g_p(r-a)} \,,\\ 
  \phi^\mathrm{sr} &= \sum_{p,n} \phi^\mathrm{sr}_{p,n}P_n(\cos{\theta})\left(\frac{a}{r}\right)^{n+1}\frac{T_n(g_pr)}{T_n(g_pa)}e^{-g_p(r-a)} \,, \label{ABSplit}
\end{align}
with surface coefficients
\begin{align}
\phi^\mathrm{eq}_{p,n} &= \frac{k_BT}{e}K_{0p}\sum_m L_{pm,n} b^\mathrm{eq}_{m,n} \,. \label{ACeq}
\end{align}
\begin{align}
\phi_{p,n}^\mathrm{sr} &= \frac{k_BT}{e}K_{0p}\sum_m L_{pm,n} b^\mathrm{sr}_{m,n} \,. \label{ACsr}
\end{align}
Here $\vectify{b}^\mathrm{eq}=\vectify{b}(\{j^\mathrm{s}_l\}\rightarrow 0)$ is the vector specifying the boundary conditions for a charged but unreactive particle, and $\vectify{b}^\mathrm{sr}=\vectify{b}(\sigma\rightarrow 0)$ specifies the boundary conditions for an uncharged but reactive particle.

%%%%%%%%%%%%
\subsection{\label{app: ff calc}Calculation of the Propulsion Speed}

Having determined the electrostatic potential, we calculate the fluid flow by making use of the Lorentz reciprocal theorem~\cite{sabass12}. This allows one to transform the Stokes equation, Eq.~\eqref{eq:stokes}, from a 3D partial differential equation into an integral equation on the 2D domain boundary (the swimmer surface). Using this approach, a general formula for the propulsion velocity $\boldsymbol{U}$ of a non-slip sphere generated by an axisymmetric distribution of force density $\boldsymbol{f}$ has been derived~\cite{teubner82}
\begin{align}
  \boldsymbol{U} &= -\frac{\hat{\boldsymbol{z}}}{6\pi\eta a} \int_V \left[ \left(\frac{3a}{2r}-\frac{a^3}{2r^3}-1\right) \cos\theta \hat{\boldsymbol{r}} -\right. \nonumber \\
                 &\quad -\left. \left(\frac{3a}{4r}+\frac{a^3}{4r^3}-1\right) \sin\theta \hat{\boldsymbol{\uptheta}} \right] \cdot \boldsymbol{f} \mathrm{d}V ,
\end{align}
where the volume integral is over the region outside the sphere, and the scalar speed $U$ used in the main text is defined by $\boldsymbol{U} = U\hat{\boldsymbol{z}}$. Here, $\boldsymbol{f}=-\rho_e\nabla\phi$ from Eq.~\eqref{force}.

For a uniformly charged sphere, the equilibrium potential distribution is
\begin{align}
  \phi^\mathrm{eq} &= \frac{\sigma a^2 e^{-\kappa(r-a)}}{r\epsilon(1+\kappa a)} . \label{eq field}
\end{align}
Making the usual assumption of a small driving field, i.e., $\phi^\mathrm{sr}\ll \phi^\mathrm{eq}$ then gives
\begin{align}
	U &= \frac{2\sigma}{3\eta a}\sum_p\frac{\kappa-g_p}{(\kappa+g_p)^2} \phi^\mathrm{sr}_{p,1}F(\kappa a, g_pa) ,\label{final solution}
\end{align}
where the $\phi^\mathrm{sr}_{p,1}$ are to be read out from Eq.~\eqref{ACsr} and
\begin{align}
	F(x,y) &= \frac{(x+y)^3}{6(1+x)(1+y)}e^{x+y} \label{f}\\
	       &\times\int_1^\infty\frac{(t-1)^2(2t+1)}{t^5}(1+xt)(1+yt)e^{-t(x+y)}\mathrm{d}t , \nonumber
\end{align}
which is the self-electrophoretic equivalent of the Henry function for electrophoresis in an external field~\cite{henry31}. We have verified that Eq.~\eqref{final solution} is also obtained by solving the 3D Stokes equations directly, following Henry's methods~\cite{henry31, kim03, kim13}. We also write down a single-argument form of the self-electrophoretic function $F(x)=F(x, 0)$, which is useful when considering swimmers without bulk reactions, for which $y=qa=0$
\begin{align}
	F(x) &= \frac{x^3e^x}{6(1+x)}\int_1^\infty\frac{(t-1)^2(2t+1)}{t^5}(1+xt)e^{-tx}\mathrm{d}t .\label{f0}
\end{align}
This is the function discussed in Section~\ref{sec:elphoWB}.

\subsection{Propulsion Speed for the Model Swimmers \label{speedModel}}

We now write down the propulsion speed for the 3 model swimmers discussed in the main text. These expressions were determined by solving Eq.~\eqref{final solution} symbolically in MATLAB, and making the further assumption of weak ionic dissociation. We find that the bulk reaction factors, as defined in Eq.~\eqref{speedDecompositionStandard} are
\begin{align}
	B_{\circ} &= \dfrac{(d_+-d_-) K_\mathrm{eq}}{(c_+^\infty+c_-^\infty) d^*d_\circ}\left[1-\Theta\left(\kappa a, qa\right)\right] ,\nonumber\\
	B_{+} &= -\dfrac{1}{d^*}\left[1-\dfrac{c^\infty_- (d_+-d_-)}{ d_+(c_+^\infty+c_-^\infty) }\Theta\left(\kappa a, qa\right)\right] ,\label{mobPM}\\
	B_{=} &= \dfrac{d_+-d_-}{d_+d_-}\Theta\left(\kappa a, qa\right)\nonumber ,
\end{align}
where $\Theta(\kappa a, qa)$ depends on the relationship between the 3 length scales $a$, $\kappa^{-1}$ and $q^{-1}$ and is
\begin{align}
\Theta(\kappa a, qa) &= \left(\frac{\kappa a}{\kappa a+qa}\right)^3\frac{2(aq+1)}{(aq)^2+2aq+2}\frac{F(\kappa a, qa)}{F(\kappa a)} .\label{dimensionless}
\end{align}
With the limits $\kappa\gg q$, and either $qa\gg 1$ or $qa \ll 1$, we obtain the expressions given in Table~\ref{table2}. 

%%%%%%%%%%%%
\subsection{\label{sec:electro}A Note on Electrostatic Boundary Conditions}

In this section, we show that a particle with fixed, uniform surface charge $\sigma$ has the same propulsion velocity as an equivalent particle with fixed, uniform surface potential $\zeta$, as long as
\begin{align}
\zeta &= \frac{\sigma a}{\epsilon(1+\kappa a)} . \label{AD}
\end{align}
To do this, we first need to show that modifying the electrostatic boundary conditions of the particle has only a limited effect on the fields of concentration and potential; namely, that such modifications can only introduce electrostatic fields corresponding to the equilibrium Debye-H{\"u}ckel solutions around passive colloids. 

We take a swimmer, in a given chemical environment, and apply to it three sets of boundary conditions. Boundary conditions (1) and (2), with corresponding solutions $\vectify{y}^{(1)}$ and $\vectify{y}^{(2)}$, have equal chemical flux boundary conditions (equal surface reaction rates), but have arbitrary, different electrostatic boundary conditions. Boundary condition (3) consists of a no flux condition on all species (no surface reactions), and the electrostatic boundary condition
\begin{align}
\label{electro diff} y_0^{(3)}(\boldsymbol{s}) &= y_0^{(2)}(\boldsymbol{s})-y_0^{(1)}(\boldsymbol{s}) .
\end{align}
Since there are no fluxes through this particle's surface, each chemical species is in equilibrium, and the solution to this boundary condition is just the equilibrium, Debye-H{\"u}ckel solution
\begin{align}
y_l^{(3)} &= \left\{
  \begin{array}{ll}
     \dfrac{e\psi}{k_BT}    & l=0 , \\
                            & \\
    -\dfrac{ez_l\psi}{k_BT} & l=1,2,\dots,N',
  \end{array}
  \right. \label{y1}
\end{align}
where the equilibrium potential field $\psi$ must satisfy both the electrostatic boundary condition, Eq.~\eqref{electro diff}, and the Debye-H{\"u}ckel equation
\begin{align}
 \nabla^2\psi=\kappa^2\psi\,. \label{debye huckel}
\end{align}
One can then show by direct substitution of Eq.~\eqref{y1} into Eq.~\eqref{N}, that the solutions to the three boundary problems are related by $\vectify{y}^{(2)}-\vectify{y}^{(1)} = \vectify{y}^{(3)}$. In particular, $\phi^{(2)}-\phi^{(1)} = \psi$, which implies, from Eq.~\eqref{debye huckel} 
\begin{align}
  \nabla^2\left[\phi^{(2)}-\phi^{(1)}\right] &= \kappa^2\left[\phi^{(2)}-\phi^{(1)}\right] .\label{AF}
\end{align}
Hence the difference $\phi^{(2)}-\phi^{(1)}$ between the electric potential fields of particles (1) and (2) corresponds to an equilibrium Debye-H{\"u}ckel solution around a passive colloid. 

As before, we make the assumption of a small driving field, $\phi^\mathrm{sr}\ll\phi^\mathrm{eq}$, where $\phi=\phi^\mathrm{eq}+\phi^\mathrm{sr}$. Now, consider two particles (1') and (2'), with equal surface reactions, but where (1') has uniform surface charge density $\sigma$, and (2') has uniform surface potential $\zeta$, with $\sigma$ and $\zeta$ satisfying Eq.~\eqref{AD}. In this case, the two equilibrium fields are equal, i.e., $\phi^{(1') \mathrm{eq}}=\phi^{(2') \mathrm{eq}}$, and are given by Eq.~\eqref{eq field}. Subtracting this equality from Eq.~\eqref{AF} yields, for the remaining, non-equilibrium part of the potential
\begin{align}
  \nabla^2\left[\phi^{(2') \mathrm{sr}}-\phi^{(1') \mathrm{sr}}\right] &= \kappa^2\left[\phi^{(2') \mathrm{sr}}-\phi^{(1') \mathrm{sr}}\right] .\label{AG}
\end{align}
In other words, the difference in the reaction-generated electrostatic potential field between (1') and (2') is an equilibrium, Debye-H{\"u}ckel type field, which has an inverse screening length $g_p=\kappa$. Since there is a $(\kappa-g_p)$ factor in Eq.~\eqref{final solution}, we see that such a field can have no effect on the propulsion speed. This proves the assertion that, to linear order, a particle with fixed, uniform surface charge $\sigma$ will have the same propulsion velocity as an equivalent particle with fixed, uniform surface potential $\zeta$, as long as Eq.~\eqref{AD} is satisfied.

In fact, one can make a more general statement, which we will not prove. For any two particles (1') and (2'), with equal arbitrary shape, surface reactions, and equilibrium (possibly non-uniform) fields $\phi^\mathrm{eq}$, not only the propulsion speed but the entire flow field will be the same. A physical justification for this conclusion is that if the interaction between one equilibrium field ($\phi^\mathrm{eq}$) and another (the difference field between (1') and (2')) could generate fluid flow, then this would constitute a perpetual-motion machine. Analogous conclusions have also been drawn for electrophoresis in an external field~\cite{obrien78}.

%%%%%%%%%
\section{\label{sec:rates}Experimental Parameters}

\subsection{\label{ionic association} The Ionic Association Constant}

An important parameter in our calculations is the ionic reaction association constant $k_\mathrm{as}$ in reaction (R4), ${\ce{H+ + HO2- <=> H2O2}}$ in water, see Section~\ref{sec:reactionModel}. We were unable to find a value for this constant in the literature. However, reactions involving the transfer of a proton or a hydroxyl ion are normally sufficiently fast to be diffusion limited~\cite{caldin64}. It has been shown~\cite{debye42}, that the diffusion-limited rate constant between two species, $A$ and $B$, with diffusivities $D_A,~D_B$, and valences $z_A,~z_B$, which react at a short distance $r_{AB}$ is~\cite{laidler87}
\begin{align}
	k_\mathrm{as} &= \left[4\pi\left(D_A+D_B\right)r_{AB}\right]f(z_Az_B, r_{AB}) . \label{charged diffusion rate}
\end{align}
Here, $f(z_Az_B, r_{AB})$ is a modifier for charged species
\begin{align}
	f=\frac{z_Az_Be^2}{4\pi\epsilon r_{AB}k_BT}\left[\exp{\left(\frac{z_Az_Be^2}{4\pi\epsilon r_{AB}k_BT}\right)-1}\right]^{-1}\!\!\!\!\!\!.    \label{f_mod}
\end{align} 
For reactions between oppositely charged species, over a typical reaction distance in water of ${r_{AB}=0.2~\mathrm{nm}}$~[\!\!\citenum{laidler87}], $f(-1,r_{AB})=3.59$. For the reaction between \ce{H+} (species A) and \ce{HO2-} (species B), this yields, using the diffusivities quoted in the main text, $k_\mathrm{as}=4.9\times 10^{10}~\M^{-1}\pers$. This is consistent with measured rates for similar reactions~\cite{caldin64}, e.g., ${\ce{H+ + HCO3- <=> H2CO3}}$ in water has ${k_\mathrm{as}=5\times 10^{10}~\M^{-1}\pers}$~[\!\!\citenum{pocker77}].

\subsection{Comparison with Experiments}

For the comparison of self-electrophoretic micro- and nano- swimmers in Fig.~\ref{fig:exp}, we take experimental parameters from Ref.~[\!\!\citenum{wheat10}] (microswimmers) and Ref.~[\!\!\citenum{lee14}] (nanoswimmers). For microswimmers, we use $a=1~\um$, $c_\circ^\infty=1.5~\M$ and no added salt. For nanoswimmers, we use $a=15~\mathrm{nm}$, $c_\circ^\infty=1.5~\M$, pH 7 and $5~\mM$ \ce{NaCl} (this has the same Debye length as 1 $\mM$ trisodium citrate, which was used in practice~\cite{lee14}). For both swimmers, we take $\sigma=10^{-2}~e/\mathrm{nm^2}$~[\!\!\citenum{brown16}] and $j^\mathrm{s}_{+,1}=1.66\times10^{-7}~\mathrm{mol/(m^2 s)}$, which is chosen to match the microswimmer speed in Ref.~[\!\!\citenum{wheat10}].

For comparison between $S_=$ and polystyrene-Pt Janus particles, we take experimental parameters from Ref.~[\!\!\citenum{ebbens12}], which are $c_\circ^\infty=3~\M$ with no added salt. We used $\sigma=10^{-2}~e/\mathrm{nm^2}$~[\!\!\citenum{brown16}] and $j^\mathrm{s}_{+,1}=6.42\times10^{-6}~\mathrm{mol/(m^2 s)}$, which is chosen to match the experimental propulsion speed at $a=1~\umpers$.

%%%%%%%%%
\section{\label{app:finite}Finite Element Method Calculations}

\begin{figure}[!htb]
\centering
\includegraphics[width=8.5 cm]{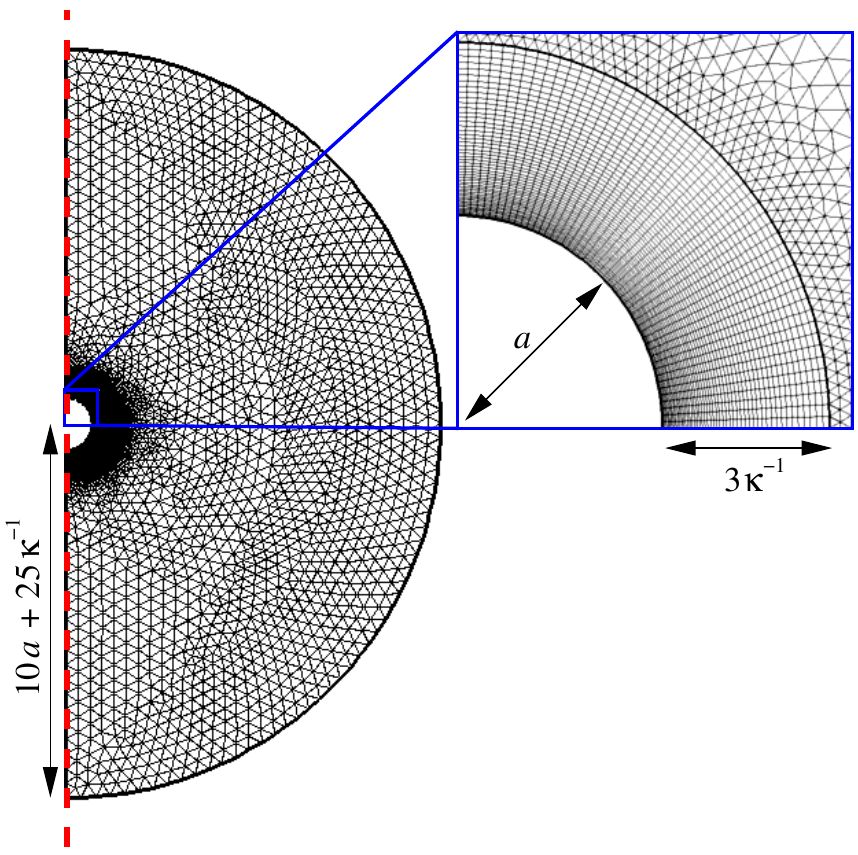}
\caption{The mesh on which the FEM calculations are performed. This particular mesh was generated for radius $a = 0.5~\um$ and a salt concentration of $10^{-5}$ mol/L, but illustrates the generic features of all the meshes. The rotational symmetry of the simulation domain is exploited to calculate on a quasi-2D domain: the symmetry axis is indicated by the dashed red line. The domain typically has a radius $L=10a + 25 \kappa^{-1}$ in size. This domain is subdivided into two pieces on which triangular and quadrilateral elements are used. In a range of $3\kappa^{-1}$ around the colloid the domain consists of quadrilaterals, which grow in size geometrically, see the zoom-in (blue box). Beyond this range the elements are triangular and are allowed to grow out linearly to best fit the domain boundary and reduce the overall number of elements.}
\label{plotsFEM}
\end{figure}

In this section we give additional details for the numerical finite element method (FEM) calculations discussed in Section~\ref{sub:solve}. FEM calculations are performed using the COMSOL 5.1 Multiphysics Modelling package. 

We employed the following strategies to accelerate the calculations and obtain high quality results. (i) The solutions were obtained in a 2D cylindrically symmetric geometry. (ii) We ignored the advective coupling term in Eq.~\eqref{flux}. This allowed us to split the problem into electrostatic plus hydrodynamic parts, as for the linear theory, and thus solve the uncoupled equations more efficiently. This approach is justified, since the P{\'e}clet number (Pe) $\lesssim 10^{-2}$ for typical experimental systems. We also verified this directly, by including the advective coupling term in a subset of the data points, finding good agreement. (iii) We created a physics-specific mesh, see Fig.~\ref{plotsFEM}, on which we solved the system. Quadrilateral elements were used out to a distance of 3$\kappa^{-1}$ from the colloid surface. These elements grow exponentially in size with increasing distance, whilst maintaining a constant number along the tangential direction. The remainder of the domain was meshed with triangular elements which grow larger with distance from the colloid. This approach is necessary to ensure convergence of the model. (iv) The following polynomial orders were employed for the test functions: electrostatics (3), diffusion (5) and hydrodynamics (2+3). These higher orders proved necessary to reduce spurious flow (see also Ref.~[\!\!\citenum{degraaf15}]). (v) Finite-size scaling was employed to check for artifacts arising from the finite extent of the simulation domain, we found that for $L = 10a + 25\kappa^{-1}$ the effects on the speed of the particle were negligible. (vi) Mesh refinement was used for several simulations to determine the dependence of our result on the element size. (vii) We also varied the tolerance on the residual for a few cases to verify that our solutions had sufficiently converged.

\begin{table*}
\small
\caption{The charge densities and the first Legendre components of the surface flux densities used in Fig.~\ref{fig:mobil} in the main text and Fig.~\ref{plotsNonLinSpeed} here. The flux densities have units $\mathrm{mol/(m^2 s)}$, and the charge densities have units $e/\mathrm{nm^2}$. The final column gives the product of $\sigma$ and the relevant non-zero flux density, with units $e\mathrm{mol/(m^2 nm^2 s)}$.\label{table1}}
\begin{tabular}{c|c|ccc|c|c}
 Fig.&Type&$j^\mathrm{s}_{\circ,1}$&$j^\mathrm{s}_{+,1}$&$j^\mathrm{s}_{-,1}$&$\sigma$&$\sigma j^\mathrm{s}$\\
\hline
~&$S_{\circ}$& $3\times10^{-1}$& $0$&$0$&$10^{-4}$&$3\times10^{-5}$\\
\ref{fig:mobil}&$S_{+}$& $0$& $3\times10^{-7}$& $0$&$10^{-4}$&$3\times10^{-11}$\\
~&$S_{=}$& $0$& $3\times10^{-5}$& $3\times10^{-5}$&$10^{-4}$&$3\times10^{-9}$\\ \hline
~&$S_{\circ}$& $1.5\times10^{-2}$& $0$&$0$&$10^{-2}$&$1.5\times10^{-4}$\\
\ref{plotsNonLinSpeed}&$S_{+}$& $0$& $1.5\times10^{-5}$& $0$&$10^{-2}$&$1.5\times10^{-7}$\\
~&$S_{=}$& $0$& $1.5\times10^{-5}$& $1.5\times10^{-5}$&$10^{-2}$&$1.5\times10^{-7}$\\\hline
\end{tabular}
\end{table*}

To verify the analytic results, we first performed calculations with sufficiently low values of the surface charge density and flux to remain in the linear regime. These $j^{s}$ and $\sigma$ are given in Table~\ref{table1}. Figure~\ref{fig:mobil} in the main text shows that there is excellent correspondence between the theory and FEM calculations in this regime. Different fluxes were used for the different propulsion models because the low efficiency of type $S_{\circ}$ and $S_{=}$ propulsion mean that numerical errors become significant more quickly as the flux density is reduced for these models. 

\begin{figure}
\centering
\includegraphics[width=8.5 cm]{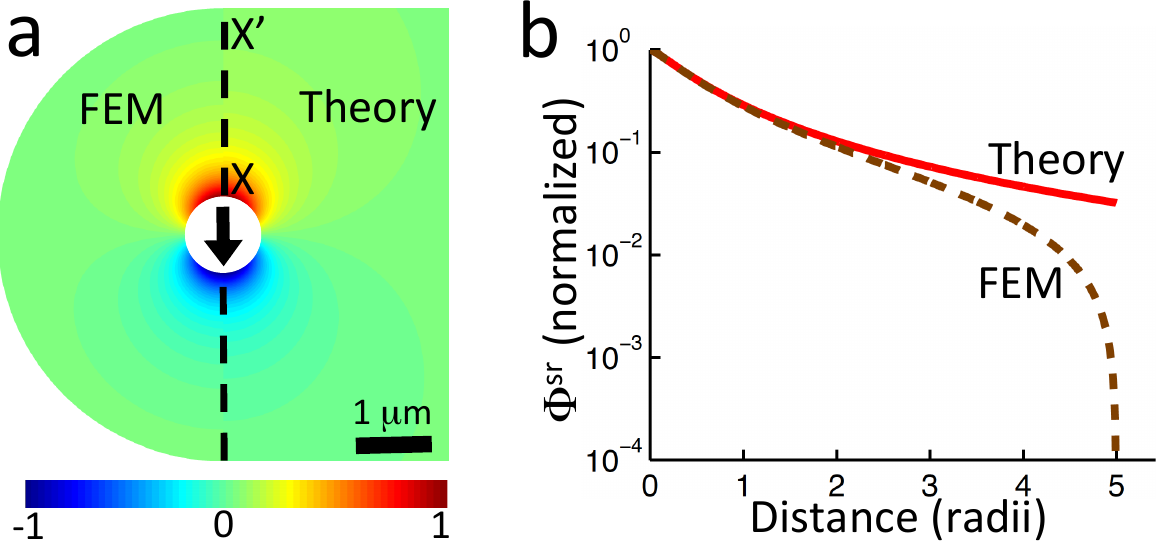}
\caption{a) Comparison of normalized surface-reaction-generated potential fields $\phi^\mathrm{sr}$ for (left) FEM and (right) linearized theory, for $S_{+}$ self-electrophoresis, with $j^\mathrm{s}_{+,1}=3\times10^{-7}~\mathrm{mol/(m^2s)}$, and with other conditions as in the base parameter set (see main text). The radius of the simulation domain $L=3~\um=6a$ here. b) Normalized radial decay of $\phi^\mathrm{sr}$ for linearized theory (\textcolor{ABred}{\textbf{---}}) and FEM calculations (\textcolor{ABbrown}{\textbf{\--\--\--}}) along X-X$'$ in a.}
\label{plotsFEMTheory}
\end{figure}

In addition, the FEM calculations and the linearized theory produce essentially identical electrostatic potential fields. Fig.~\ref{plotsFEMTheory}a illustrates this for type $S_{+}$ electrophoresis. Note that we had to use a much smaller computational domain than we typically use ($L = 6a$ here, rather than $L = 10a + 25\kappa^{-1}$), in order to show details in Fig.~\ref{plotsFEMTheory}a. This means that the deviation from the theory, which stems from the $\phi = 0$ boundary condition on the edge of the domain, occurs closer to the particle than in our regular calculations, see Fig.~\ref{plotsFEMTheory}b. However, the potential and flow-fields decay sufficiently rapidly that this does not affect the potential near the particle, or the propulsion speed beyond a few percent.

\begin{figure}
\centering
\includegraphics[width=8.5 cm]{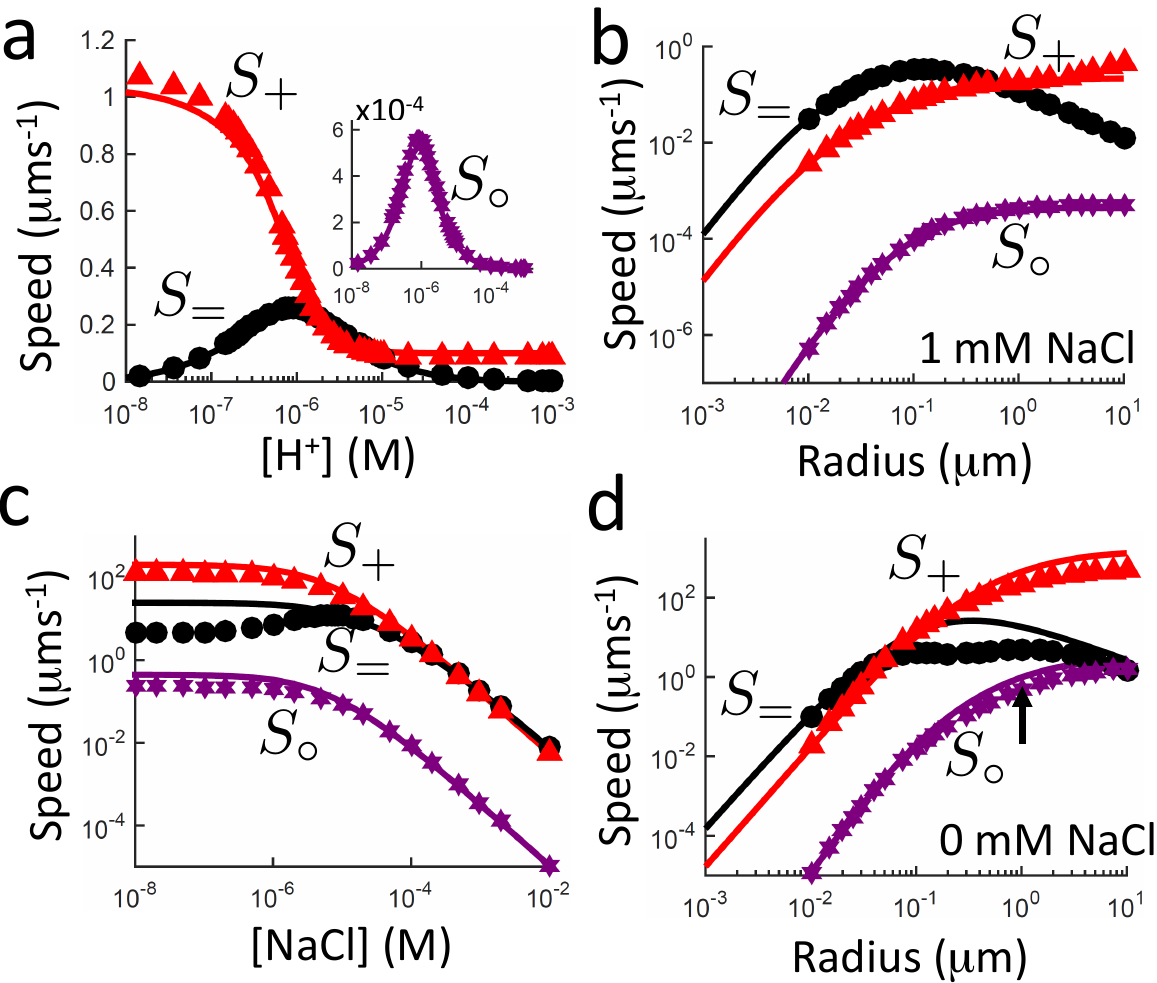}
\caption{Propulsion speed for realistic parameters (given in this text) for type $S_{\circ}$ (\textcolor{ABpurple}{$\boldsymbol{\bigstar}$}, inset), $S_{+}$ (\textcolor{ABred}{$\boldsymbol{\blacktriangle}$}), and $S_{=}$ (\newmoon) propulsion, from analytical theory (solid curves) and FEM simulations (symbols). For a) [\ce{H+}] at fixed $\kappa$ equivalent to 1 $\mM$ \ce{NaCl}, b) particle radius with 1 $\mM$ \ce{NaCl} c) \ce{NaCl} concentration, d) particle radius with 0 $\mM$ \ce{NaCl}. In d, the black arrow indicates the experimental point from Ref.~[\!\!\citenum{brown14}] referred to in the text.}
\label{plotsNonLinSpeed}
\end{figure}

We can also use the FEM to go beyond the linear approximation. We defer to future work a systematic investigation of the non-linear behavior, and here focus on the propulsion speed for selected experimentally relevant values of the surface charge density and chemical fluxes. These values are taken from measurements on the Pt-polystyrene Janus swimmers in Ref.~[\!\!\citenum{brown16}], and are listed in Table~\ref{table1}. The neutral flux density $j^\mathrm{s}_{\circ,1}$ is that which would be produced by a Janus particle which uniformly consumes \ce{H2O2} on one hemisphere at a rate $\Gamma=8\times 10^{10}$ molecules per second per particle. This rate was measured for $a=1~\um$ radius particles in 3 M \ce{H2O2}~[\!\!\citenum{brown14}]. The surface charge density is taken from the electrophoretic mobility measurements made on the same particles in Ref.~[\!\!\citenum{brown14}]. The ionic fluxes are unknown, but we arbitrarily set $j^\mathrm{s}_{\pm,1}=10^{-3}j^\mathrm{s}_{\circ,1}$, so that $S_{+}$ electrophoresis gives a speed of order $100~\umpers$, which is somewhat larger than typical experimental values, $\approx 10~\umpers$ for Au-Pt spherical microswimmers~\cite{wheat10, theurkauff12}. Hence, our results should overestimate the non-linear behavior of the propulsion speed. Note that though the ionic flux densities for the experimentally realistically case are sometimes lower than those for the linear case, the product of charge density and surface flux is always greater in the experimentally realistic case, Table~\ref{table1}.

Figure~\ref{plotsNonLinSpeed}a-b, both with 1 $\mM$ \ce{NaCl}, correspond to Fig.~\ref{fig:mobil} in the main text. We see that the analytical theory continues to match the FEM calculations well even for these realistic values of the flux and charge densities. However, many experiments are performed with no added salt, and as shown in Fig.~\ref{plotsNonLinSpeed}c, the agreement worsens as the salt concentration falls. This is to be expected, since it is low salt that generates a high-$\zeta$, large-screening-length regime where linear approximations break down~\cite{hunter01}. In fact, with 0 $\mM$ NaCl, the dimensionless zeta-potential $\zeta e/(k_BT)=5.6$ for these particles, well outside the Debye-H{\"u}ckel regime of $\zeta e/(k_BT)\ll 1$. Nevertheless, for all propulsion types, the agreement remains semi-quantitative between simulations and theory over the whole radius range for 0 $\mM$ NaCl, Fig.~\ref{plotsNonLinSpeed}d.

From Fig.~\ref{plotsNonLinSpeed}d, we obtain a speed of $0.5~\umpers$ for type $S_{\circ}$ electrophoresis with particles of radius ${a=1~\um}$, no salt, and 3 M \ce{H2O2} (the black arrow indicates the relevant data point). As stated in the main text, this predicted speed can account for at most $5\%$ of the experimentally measured propulsion speed of ${15-20~\umpers}$ obtained for Pt-Polystyrene Janus particles under the same conditions~\cite{brown14}.

%{\footnotesize
%\bibliography{ChemicalPropulsion} %your .bib file

\begin{mcitethebibliography}{74}
\providecommand*{\natexlab}[1]{#1}
\providecommand*{\mciteSetBstSublistMode}[1]{}
\providecommand*{\mciteSetBstMaxWidthForm}[2]{}
\providecommand*{\mciteBstWouldAddEndPuncttrue}
  {\def\EndOfBibitem{\unskip.}}
\providecommand*{\mciteBstWouldAddEndPunctfalse}
  {\let\EndOfBibitem\relax}
\providecommand*{\mciteSetBstMidEndSepPunct}[3]{}
\providecommand*{\mciteSetBstSublistLabelBeginEnd}[3]{}
\providecommand*{\EndOfBibitem}{}
\mciteSetBstSublistMode{f}
\mciteSetBstMaxWidthForm{subitem}
{(\emph{\alph{mcitesubitemcount}})}
\mciteSetBstSublistLabelBeginEnd{\mcitemaxwidthsubitemform\space}
{\relax}{\relax}

\bibitem[Einstein(1905)]{einstein05}
A.~Einstein, \emph{Ann. der Physik}, 1905, \textbf{17}, 549\relax
\mciteBstWouldAddEndPuncttrue
\mciteSetBstMidEndSepPunct{\mcitedefaultmidpunct}
{\mcitedefaultendpunct}{\mcitedefaultseppunct}\relax
\EndOfBibitem
\bibitem[Paxton \emph{et~al.}(2004)Paxton, Kistler, Olmeda, Sen, St.~Angelo,
  Cao, Mallouk, Lammert, and Crespi]{paxton04}
W.~F. Paxton, K.~C. Kistler, C.~C. Olmeda, A.~Sen, S.~K. St.~Angelo, Y.~Cao,
  T.~E. Mallouk, P.~E. Lammert and V.~H. Crespi, \emph{J. Am. Chem. Soc.},
  2004, \textbf{126}, 13424\relax
\mciteBstWouldAddEndPuncttrue
\mciteSetBstMidEndSepPunct{\mcitedefaultmidpunct}
{\mcitedefaultendpunct}{\mcitedefaultseppunct}\relax
\EndOfBibitem
\bibitem[Howse \emph{et~al.}(2007)Howse, Jones, Ryan, Gough, Vafabakhsh, and
  Golestanian]{howse07}
J.~R. Howse, R.~A.~L. Jones, A.~J. Ryan, T.~Gough, R.~Vafabakhsh and
  R.~Golestanian, \emph{Phys. Rev. Lett.}, 2007, \textbf{99}, 048102\relax
\mciteBstWouldAddEndPuncttrue
\mciteSetBstMidEndSepPunct{\mcitedefaultmidpunct}
{\mcitedefaultendpunct}{\mcitedefaultseppunct}\relax
\EndOfBibitem
\bibitem[Wang \emph{et~al.}(2006)Wang, Hernandez, Bartlett, Bingham, Kline,
  Sen, and Mallouk]{wang06}
Y.~Wang, R.~M. Hernandez, D.~J. Bartlett, J.~M. Bingham, T.~R. Kline, A.~Sen
  and T.~E. Mallouk, \emph{Langmuir}, 2006, \textbf{22}, 10451\relax
\mciteBstWouldAddEndPuncttrue
\mciteSetBstMidEndSepPunct{\mcitedefaultmidpunct}
{\mcitedefaultendpunct}{\mcitedefaultseppunct}\relax
\EndOfBibitem
\bibitem[Solovev \emph{et~al.}(2009)Solovev, Mei, E., Huang, and
  Schmidt]{solovev09}
A.~A. Solovev, Y.~Mei, B.~U. E., G.~Huang and O.~G. Schmidt, \emph{Small},
  2009, \textbf{5}, 1688\relax
\mciteBstWouldAddEndPuncttrue
\mciteSetBstMidEndSepPunct{\mcitedefaultmidpunct}
{\mcitedefaultendpunct}{\mcitedefaultseppunct}\relax
\EndOfBibitem
\bibitem[Jiang \emph{et~al.}(2010)Jiang, Yoshinaga, and Sano]{yoshinaga10}
H.-R. Jiang, N.~Yoshinaga and M.~Sano, \emph{Phys. Rev. Lett.}, 2010,
  \textbf{105}, 268302\relax
\mciteBstWouldAddEndPuncttrue
\mciteSetBstMidEndSepPunct{\mcitedefaultmidpunct}
{\mcitedefaultendpunct}{\mcitedefaultseppunct}\relax
\EndOfBibitem
\bibitem[Buttinoni \emph{et~al.}(2012)Buttinoni, Volpe, K{\"u}mmel, Volpe, and
  Bechinger]{buttinoni12}
I.~Buttinoni, G.~Volpe, F.~K{\"u}mmel, G.~Volpe and C.~Bechinger, \emph{J.
  Phys.: Cond. Mat.}, 2012, \textbf{24}, 284129\relax
\mciteBstWouldAddEndPuncttrue
\mciteSetBstMidEndSepPunct{\mcitedefaultmidpunct}
{\mcitedefaultendpunct}{\mcitedefaultseppunct}\relax
\EndOfBibitem
\bibitem[Ahmed \emph{et~al.}(2013)Ahmed, Wang, Mair, Fraleigh, Li, Castro,
  Hoyos, Huang, and Mallouk]{ahmed13}
S.~Ahmed, W.~Wang, L.~Mair, R.~Fraleigh, S.~Li, L.~Castro, M.~Hoyos, T.~Huang
  and T.~Mallouk, \emph{Langmuir}, 2013, \textbf{29}, 16113--16118\relax
\mciteBstWouldAddEndPuncttrue
\mciteSetBstMidEndSepPunct{\mcitedefaultmidpunct}
{\mcitedefaultendpunct}{\mcitedefaultseppunct}\relax
\EndOfBibitem
\bibitem[Ebbens and Howse(2010)]{ebbens10}
S.~J. Ebbens and J.~R. Howse, \emph{Soft Matter}, 2010, \textbf{6},
  726--738\relax
\mciteBstWouldAddEndPuncttrue
\mciteSetBstMidEndSepPunct{\mcitedefaultmidpunct}
{\mcitedefaultendpunct}{\mcitedefaultseppunct}\relax
\EndOfBibitem
\bibitem[Theurkauff \emph{et~al.}(2012)Theurkauff, Cottin-Bizonne, Palacci,
  Ybert, and Bocquet]{theurkauff12}
I.~Theurkauff, C.~Cottin-Bizonne, J.~Palacci, C.~Ybert and L.~Bocquet,
  \emph{Phys. Rev. Lett.}, 2012, \textbf{108}, 268303\relax
\mciteBstWouldAddEndPuncttrue
\mciteSetBstMidEndSepPunct{\mcitedefaultmidpunct}
{\mcitedefaultendpunct}{\mcitedefaultseppunct}\relax
\EndOfBibitem
\bibitem[Palacci \emph{et~al.}(2013)Palacci, Sacanna, Steinberg, Pine, and
  Chaikin]{palacci13}
J.~Palacci, S.~Sacanna, A.~P. Steinberg, D.~J. Pine and P.~M. Chaikin,
  \emph{Science}, 2013, \textbf{339}, 936--940\relax
\mciteBstWouldAddEndPuncttrue
\mciteSetBstMidEndSepPunct{\mcitedefaultmidpunct}
{\mcitedefaultendpunct}{\mcitedefaultseppunct}\relax
\EndOfBibitem
\bibitem[Buttinoni \emph{et~al.}(2013)Buttinoni, Bialk\'e, K\"ummel, L\"owen,
  Bechinger, and Speck]{buttinoni13}
I.~Buttinoni, J.~Bialk\'e, F.~K\"ummel, H.~L\"owen, C.~Bechinger and T.~Speck,
  \emph{Phys. Rev. Lett.}, 2013, \textbf{110}, 238301\relax
\mciteBstWouldAddEndPuncttrue
\mciteSetBstMidEndSepPunct{\mcitedefaultmidpunct}
{\mcitedefaultendpunct}{\mcitedefaultseppunct}\relax
\EndOfBibitem
\bibitem[Ginot \emph{et~al.}(2015)Ginot, Theurkauff, Levis, Ybert, Bocquet,
  Berthier, and Cottin-Bizonne]{ginot15}
F.~Ginot, I.~Theurkauff, D.~Levis, C.~Ybert, L.~Bocquet, L.~Berthier and
  C.~Cottin-Bizonne, \emph{Phys. Rev. X}, 2015, \textbf{5}, 011004\relax
\mciteBstWouldAddEndPuncttrue
\mciteSetBstMidEndSepPunct{\mcitedefaultmidpunct}
{\mcitedefaultendpunct}{\mcitedefaultseppunct}\relax
\EndOfBibitem
\bibitem[Soto and Golestanian(2014)]{soto14}
R.~Soto and R.~Golestanian, \emph{Phys. Rev. Lett.}, 2014, \textbf{112},
  068301\relax
\mciteBstWouldAddEndPuncttrue
\mciteSetBstMidEndSepPunct{\mcitedefaultmidpunct}
{\mcitedefaultendpunct}{\mcitedefaultseppunct}\relax
\EndOfBibitem
\bibitem[Uspal \emph{et~al.}(2015)Uspal, Popescu, Dietrich, and
  Tasinkevych]{uspal15}
W.~E. Uspal, M.~N. Popescu, S.~Dietrich and M.~Tasinkevych, \emph{Soft Matter},
  2015, \textbf{11}, 434--438\relax
\mciteBstWouldAddEndPuncttrue
\mciteSetBstMidEndSepPunct{\mcitedefaultmidpunct}
{\mcitedefaultendpunct}{\mcitedefaultseppunct}\relax
\EndOfBibitem
\bibitem[Banigan and Marko(2016)]{banigan16}
E.~J. Banigan and J.~F. Marko, \emph{Phys. Rev. E}, 2016, \textbf{93},
  012611\relax
\mciteBstWouldAddEndPuncttrue
\mciteSetBstMidEndSepPunct{\mcitedefaultmidpunct}
{\mcitedefaultendpunct}{\mcitedefaultseppunct}\relax
\EndOfBibitem
\bibitem[Brown \emph{et~al.}(2016)Brown, Vladescu, Dawson, Vissers,
  Schwarz-Linek, Lintuvuori, and Poon]{brown16}
A.~Brown, I.~Vladescu, A.~Dawson, T.~Vissers, J.~Schwarz-Linek, J.~Lintuvuori
  and W.~Poon, \emph{Soft matter}, 2016, \textbf{12}, 131\relax
\mciteBstWouldAddEndPuncttrue
\mciteSetBstMidEndSepPunct{\mcitedefaultmidpunct}
{\mcitedefaultendpunct}{\mcitedefaultseppunct}\relax
\EndOfBibitem
\bibitem[Vicsek \emph{et~al.}(1995)Vicsek, Czir\'ok, Ben-Jacob, Cohen, and
  Shochet]{vicsek95}
T.~Vicsek, A.~Czir\'ok, E.~Ben-Jacob, I.~Cohen and O.~Shochet, \emph{Phys. Rev.
  Lett.}, 1995, \textbf{75}, 1226--1229\relax
\mciteBstWouldAddEndPuncttrue
\mciteSetBstMidEndSepPunct{\mcitedefaultmidpunct}
{\mcitedefaultendpunct}{\mcitedefaultseppunct}\relax
\EndOfBibitem
\bibitem[Stenhammar \emph{et~al.}(2013)Stenhammar, Tiribocchi, Allen,
  Marenduzzo, and Cates]{Stenhammar13}
J.~Stenhammar, A.~Tiribocchi, R.~J. Allen, D.~Marenduzzo and M.~E. Cates,
  \emph{Phys. Rev. Lett.}, 2013, \textbf{111}, 145702\relax
\mciteBstWouldAddEndPuncttrue
\mciteSetBstMidEndSepPunct{\mcitedefaultmidpunct}
{\mcitedefaultendpunct}{\mcitedefaultseppunct}\relax
\EndOfBibitem
\bibitem[Zheng \emph{et~al.}(2013)Zheng, ten Hagen, Kaiser, Wu, Cui, Silber-Li,
  and L\"owen]{zheng13}
X.~Zheng, B.~ten Hagen, A.~Kaiser, M.~Wu, H.~Cui, Z.~Silber-Li and H.~L\"owen,
  \emph{Phys. Rev. E}, 2013, \textbf{88}, 032304\relax
\mciteBstWouldAddEndPuncttrue
\mciteSetBstMidEndSepPunct{\mcitedefaultmidpunct}
{\mcitedefaultendpunct}{\mcitedefaultseppunct}\relax
\EndOfBibitem
\bibitem[Matas-Navarro \emph{et~al.}(2014)Matas-Navarro, Golestanian,
  Liverpool, and Fielding]{matas14}
R.~Matas-Navarro, R.~Golestanian, T.~Liverpool and S.~Fielding, \emph{Phys.
  Rev. E}, 2014, \textbf{90}, 032304\relax
\mciteBstWouldAddEndPuncttrue
\mciteSetBstMidEndSepPunct{\mcitedefaultmidpunct}
{\mcitedefaultendpunct}{\mcitedefaultseppunct}\relax
\EndOfBibitem
\bibitem[Z\"ottl and Stark(2014)]{zottl14}
A.~Z\"ottl and H.~Stark, \emph{Phys. Rev. Lett.}, 2014, \textbf{112},
  118101\relax
\mciteBstWouldAddEndPuncttrue
\mciteSetBstMidEndSepPunct{\mcitedefaultmidpunct}
{\mcitedefaultendpunct}{\mcitedefaultseppunct}\relax
\EndOfBibitem
\bibitem[Yang and Marchetti(2015)]{yang15}
X.~Yang and M.~Marchetti, \emph{Phys. Rev. Lett.}, 2015, \textbf{115},
  258101\relax
\mciteBstWouldAddEndPuncttrue
\mciteSetBstMidEndSepPunct{\mcitedefaultmidpunct}
{\mcitedefaultendpunct}{\mcitedefaultseppunct}\relax
\EndOfBibitem
\bibitem[Thakur and Kapral(2012)]{thakur12}
S.~Thakur and R.~Kapral, \emph{Phys. Rev. E}, 2012, \textbf{85}, 026121\relax
\mciteBstWouldAddEndPuncttrue
\mciteSetBstMidEndSepPunct{\mcitedefaultmidpunct}
{\mcitedefaultendpunct}{\mcitedefaultseppunct}\relax
\EndOfBibitem
\bibitem[Pohl and Stark(2014)]{pohl14}
O.~Pohl and H.~Stark, \emph{Phys. Rev. Lett.}, 2014, \textbf{112}, 238303\relax
\mciteBstWouldAddEndPuncttrue
\mciteSetBstMidEndSepPunct{\mcitedefaultmidpunct}
{\mcitedefaultendpunct}{\mcitedefaultseppunct}\relax
\EndOfBibitem
\bibitem[Bickel \emph{et~al.}(2014)Bickel, Zecua, and W{\"u}rger]{bickel14}
T.~Bickel, G.~Zecua and A.~W{\"u}rger, \emph{Phys. Rev. E}, 2014, \textbf{89},
  050303\relax
\mciteBstWouldAddEndPuncttrue
\mciteSetBstMidEndSepPunct{\mcitedefaultmidpunct}
{\mcitedefaultendpunct}{\mcitedefaultseppunct}\relax
\EndOfBibitem
\bibitem[Paxton \emph{et~al.}(2006)Paxton, Baker, Kline, Wang, Mallouk, and
  Sen]{paxton06}
W.~F. Paxton, P.~T. Baker, T.~R. Kline, Y.~Wang, T.~E. Mallouk and A.~Sen,
  \emph{J. Am. Chem. Soc.}, 2006, \textbf{128}, 14881--14888\relax
\mciteBstWouldAddEndPuncttrue
\mciteSetBstMidEndSepPunct{\mcitedefaultmidpunct}
{\mcitedefaultendpunct}{\mcitedefaultseppunct}\relax
\EndOfBibitem
\bibitem[Golestanian \emph{et~al.}(2007)Golestanian, Liverpool, and
  Ajdari]{golestanian07}
R.~Golestanian, T.~B. Liverpool and A.~Ajdari, \emph{New J. Phys.}, 2007,
  \textbf{9}, 126\relax
\mciteBstWouldAddEndPuncttrue
\mciteSetBstMidEndSepPunct{\mcitedefaultmidpunct}
{\mcitedefaultendpunct}{\mcitedefaultseppunct}\relax
\EndOfBibitem
\bibitem[Moran \emph{et~al.}(2010)Moran, Wheat, and Posner]{moran10}
J.~Moran, P.~Wheat and J.~Posner, \emph{Phys. Rev. E}, 2010, \textbf{81},
  065302\relax
\mciteBstWouldAddEndPuncttrue
\mciteSetBstMidEndSepPunct{\mcitedefaultmidpunct}
{\mcitedefaultendpunct}{\mcitedefaultseppunct}\relax
\EndOfBibitem
\bibitem[Moran and Posner(2011)]{moran11}
J.~L. Moran and J.~D. Posner, \emph{J. Fluid Mech.}, 2011, \textbf{680},
  31--66\relax
\mciteBstWouldAddEndPuncttrue
\mciteSetBstMidEndSepPunct{\mcitedefaultmidpunct}
{\mcitedefaultendpunct}{\mcitedefaultseppunct}\relax
\EndOfBibitem
\bibitem[Ebbens \emph{et~al.}(2012)Ebbens, Tu, Howse, and
  Golestanian]{ebbens12}
S.~Ebbens, M.-H. Tu, J.~R. Howse and R.~Golestanian, \emph{Phys. Rev. E}, 2012,
  \textbf{85}, 020401\relax
\mciteBstWouldAddEndPuncttrue
\mciteSetBstMidEndSepPunct{\mcitedefaultmidpunct}
{\mcitedefaultendpunct}{\mcitedefaultseppunct}\relax
\EndOfBibitem
\bibitem[Brown and Poon(2014)]{brown14}
A.~T. Brown and W.~C.~K. Poon, \emph{Soft Matter}, 2014, \textbf{10},
  4016--4027\relax
\mciteBstWouldAddEndPuncttrue
\mciteSetBstMidEndSepPunct{\mcitedefaultmidpunct}
{\mcitedefaultendpunct}{\mcitedefaultseppunct}\relax
\EndOfBibitem
\bibitem[Ebbens \emph{et~al.}(2014)Ebbens, Gregory, Dunderdale, Howse, Ibrahim,
  Liverpool, and Golestanian]{ebbens14}
S.~Ebbens, D.~A. Gregory, G.~Dunderdale, J.~R. Howse, Y.~Ibrahim, T.~B.
  Liverpool and R.~Golestanian, \emph{Euro. Phys. Lett.}, 2014, \textbf{106},
  58003\relax
\mciteBstWouldAddEndPuncttrue
\mciteSetBstMidEndSepPunct{\mcitedefaultmidpunct}
{\mcitedefaultendpunct}{\mcitedefaultseppunct}\relax
\EndOfBibitem
\bibitem[Gileadi(2011)]{gileadi11}
E.~Gileadi, \emph{Physical Electrochemistry: Fundamentals, Techniques and
  Applications}, Wiley-VCH Weinheim, Germany, 2011\relax
\mciteBstWouldAddEndPuncttrue
\mciteSetBstMidEndSepPunct{\mcitedefaultmidpunct}
{\mcitedefaultendpunct}{\mcitedefaultseppunct}\relax
\EndOfBibitem
\bibitem[Sabass and Seifert(2012)]{sabass12b}
B.~Sabass and U.~Seifert, \emph{J. Chem. Phys.}, 2012, \textbf{136},
  214507\relax
\mciteBstWouldAddEndPuncttrue
\mciteSetBstMidEndSepPunct{\mcitedefaultmidpunct}
{\mcitedefaultendpunct}{\mcitedefaultseppunct}\relax
\EndOfBibitem
\bibitem[Lee \emph{et~al.}(2014)Lee, Alarcón-Correa, Miksch, Hahn, Gibbs, and
  Fischer]{lee14}
T.-C. Lee, M.~Alarcón-Correa, C.~Miksch, K.~Hahn, J.~G. Gibbs and P.~Fischer,
  \emph{Nano Lett.}, 2014, \textbf{14}, 2407--2412\relax
\mciteBstWouldAddEndPuncttrue
\mciteSetBstMidEndSepPunct{\mcitedefaultmidpunct}
{\mcitedefaultendpunct}{\mcitedefaultseppunct}\relax
\EndOfBibitem
\bibitem[Henry(1931)]{henry31}
D.~Henry, \emph{Proc. Roy. Soc. Lond. A Mat.}, 1931, \textbf{133},
  106--129\relax
\mciteBstWouldAddEndPuncttrue
\mciteSetBstMidEndSepPunct{\mcitedefaultmidpunct}
{\mcitedefaultendpunct}{\mcitedefaultseppunct}\relax
\EndOfBibitem
\bibitem[Kim and Karrila(2013)]{kim13}
S.~Kim and S.~J. Karrila, \emph{Microhydrodynamics: principles and selected
  applications}, Courier Dover Publications, 2013\relax
\mciteBstWouldAddEndPuncttrue
\mciteSetBstMidEndSepPunct{\mcitedefaultmidpunct}
{\mcitedefaultendpunct}{\mcitedefaultseppunct}\relax
\EndOfBibitem
\bibitem[Muddana \emph{et~al.}(2010)Muddana, Sengupta, Mallouk, Sen, and
  Butler]{muddana10}
H.~S. Muddana, S.~Sengupta, T.~E. Mallouk, A.~Sen and P.~J. Butler, \emph{J.
  Am. Chem. Soc.}, 2010, \textbf{132}, 2110\relax
\mciteBstWouldAddEndPuncttrue
\mciteSetBstMidEndSepPunct{\mcitedefaultmidpunct}
{\mcitedefaultendpunct}{\mcitedefaultseppunct}\relax
\EndOfBibitem
\bibitem[Sengupta \emph{et~al.}(2013)Sengupta, Dey, Muddana, Tabouillot, Ibele,
  Butler, and Sen]{sengupta13}
S.~Sengupta, K.~K. Dey, H.~S. Muddana, T.~Tabouillot, M.~E. Ibele, P.~J. Butler
  and A.~Sen, \emph{J. Am. Chem. Soc.}, 2013, \textbf{135}, 1406\relax
\mciteBstWouldAddEndPuncttrue
\mciteSetBstMidEndSepPunct{\mcitedefaultmidpunct}
{\mcitedefaultendpunct}{\mcitedefaultseppunct}\relax
\EndOfBibitem
\bibitem[Golestanian \emph{et~al.}(2005)Golestanian, Liverpool, and
  Ajdari]{golestanian05}
R.~Golestanian, T.~B. Liverpool and A.~Ajdari, \emph{Phys. Rev. Lett.}, 2005,
  \textbf{94}, 220801\relax
\mciteBstWouldAddEndPuncttrue
\mciteSetBstMidEndSepPunct{\mcitedefaultmidpunct}
{\mcitedefaultendpunct}{\mcitedefaultseppunct}\relax
\EndOfBibitem
\bibitem[de~Graaf \emph{et~al.}(2015)de~Graaf, Rempfer, and Holm]{degraaf15}
J.~de~Graaf, G.~Rempfer and C.~Holm, \emph{IEEE Trans. Nanobiosci.}, 2015,
  \textbf{14}, 272\relax
\mciteBstWouldAddEndPuncttrue
\mciteSetBstMidEndSepPunct{\mcitedefaultmidpunct}
{\mcitedefaultendpunct}{\mcitedefaultseppunct}\relax
\EndOfBibitem
\bibitem[Anderson(1989)]{anderson89}
J.~L. Anderson, \emph{Ann. Rev. Fluid Mech.}, 1989, \textbf{21}, 61--99\relax
\mciteBstWouldAddEndPuncttrue
\mciteSetBstMidEndSepPunct{\mcitedefaultmidpunct}
{\mcitedefaultendpunct}{\mcitedefaultseppunct}\relax
\EndOfBibitem
\bibitem[Duan \emph{et~al.}(2012)Duan, Ibele, Liu, and Sen]{duan12}
W.~Duan, M.~Ibele, R.~Liu and A.~Sen, \emph{Euro. Phys. J. E}, 2012,
  \textbf{35}, 1--8\relax
\mciteBstWouldAddEndPuncttrue
\mciteSetBstMidEndSepPunct{\mcitedefaultmidpunct}
{\mcitedefaultendpunct}{\mcitedefaultseppunct}\relax
\EndOfBibitem
\bibitem[Stone and Samuel(1996)]{stone96}
H.~A. Stone and A.~D.~T. Samuel, \emph{Phys. Rev. Lett.}, 1996, \textbf{77},
  4102--4104\relax
\mciteBstWouldAddEndPuncttrue
\mciteSetBstMidEndSepPunct{\mcitedefaultmidpunct}
{\mcitedefaultendpunct}{\mcitedefaultseppunct}\relax
\EndOfBibitem
\bibitem[Teubner(1982)]{teubner82}
M.~Teubner, \emph{J. Chem. Phys.}, 1982, \textbf{76}, 5564--5573\relax
\mciteBstWouldAddEndPuncttrue
\mciteSetBstMidEndSepPunct{\mcitedefaultmidpunct}
{\mcitedefaultendpunct}{\mcitedefaultseppunct}\relax
\EndOfBibitem
\bibitem[Rempfer \emph{et~al.}(2016)Rempfer, Davies, Holm, and
  de~Graaf]{rempfer16}
G.~Rempfer, G.~Davies, C.~Holm and J.~de~Graaf, \emph{J. Chem. Phys.}, 2016,
  \textbf{145}, 044901\relax
\mciteBstWouldAddEndPuncttrue
\mciteSetBstMidEndSepPunct{\mcitedefaultmidpunct}
{\mcitedefaultendpunct}{\mcitedefaultseppunct}\relax
\EndOfBibitem
\bibitem[Kreissl \emph{et~al.}(2016)Kreissl, Holm, and de~Graaf]{kreissl16}
P.~Kreissl, C.~Holm and J.~de~Graaf, \emph{J. Chem. Phys.}, 2016, \textbf{144},
  204902\relax
\mciteBstWouldAddEndPuncttrue
\mciteSetBstMidEndSepPunct{\mcitedefaultmidpunct}
{\mcitedefaultendpunct}{\mcitedefaultseppunct}\relax
\EndOfBibitem
\bibitem[Everett and Minkoff(1953)]{everett53}
A.~J. Everett and G.~J. Minkoff, \emph{Trans. Faraday Soc.}, 1953, \textbf{49},
  410\relax
\mciteBstWouldAddEndPuncttrue
\mciteSetBstMidEndSepPunct{\mcitedefaultmidpunct}
{\mcitedefaultendpunct}{\mcitedefaultseppunct}\relax
\EndOfBibitem
\bibitem[Caldin(1964)]{caldin64}
E.~Caldin, \emph{Fast reactions in solution}, Blackwell Scientific
  Publications, Oxford, 1964\relax
\mciteBstWouldAddEndPuncttrue
\mciteSetBstMidEndSepPunct{\mcitedefaultmidpunct}
{\mcitedefaultendpunct}{\mcitedefaultseppunct}\relax
\EndOfBibitem
\bibitem[Kern(1954)]{kern54}
D.~M.~H. Kern, \emph{J. Am. Chem. Soc.}, 1954, \textbf{76}, 4208--4214\relax
\mciteBstWouldAddEndPuncttrue
\mciteSetBstMidEndSepPunct{\mcitedefaultmidpunct}
{\mcitedefaultendpunct}{\mcitedefaultseppunct}\relax
\EndOfBibitem
\bibitem[Haynes(2013)]{haynes13}
\emph{CRC handbook of chemistry and physics, 93rd ed.}, ed. W.~M. Haynes, CRC
  press, Boca Raton, U.S.A., 2013\relax
\mciteBstWouldAddEndPuncttrue
\mciteSetBstMidEndSepPunct{\mcitedefaultmidpunct}
{\mcitedefaultendpunct}{\mcitedefaultseppunct}\relax
\EndOfBibitem
\bibitem[{van den}~Brink \emph{et~al.}(1984){van den}~Brink, Visscher, and
  Barendrecht]{vandenbrink84}
F.~{van den}~Brink, W.~Visscher and E.~Barendrecht, \emph{J. Electroanal.
  Chem.}, 1984, \textbf{172}, 301--325\relax
\mciteBstWouldAddEndPuncttrue
\mciteSetBstMidEndSepPunct{\mcitedefaultmidpunct}
{\mcitedefaultendpunct}{\mcitedefaultseppunct}\relax
\EndOfBibitem
\bibitem[Samson \emph{et~al.}(2003)Samson, Marchand, and Snyder]{samson03}
E.~Samson, J.~Marchand and K.~Snyder, \emph{Materials and Structures}, 2003,
  \textbf{36}, 156\relax
\mciteBstWouldAddEndPuncttrue
\mciteSetBstMidEndSepPunct{\mcitedefaultmidpunct}
{\mcitedefaultendpunct}{\mcitedefaultseppunct}\relax
\EndOfBibitem
\bibitem[Ibele \emph{et~al.}(2009)Ibele, Mallouk, and Sen]{ibele09}
M.~Ibele, T.~E. Mallouk and A.~Sen, \emph{Angew. Chem. Int. Ed.}, 2009,
  \textbf{48}, 3308--3312\relax
\mciteBstWouldAddEndPuncttrue
\mciteSetBstMidEndSepPunct{\mcitedefaultmidpunct}
{\mcitedefaultendpunct}{\mcitedefaultseppunct}\relax
\EndOfBibitem
\bibitem[Gao \emph{et~al.}(2014)Gao, Pei, Dong, and Wang]{gao14}
W.~Gao, A.~Pei, R.~Dong and J.~Wang, \emph{J. Am. Chem. Soc.}, 2014,
  \textbf{136}, 2276\relax
\mciteBstWouldAddEndPuncttrue
\mciteSetBstMidEndSepPunct{\mcitedefaultmidpunct}
{\mcitedefaultendpunct}{\mcitedefaultseppunct}\relax
\EndOfBibitem
\bibitem[Moran and Posner(2014)]{moran14}
J.~L. Moran and J.~D. Posner, \emph{Phys. Fluids}, 2014, \textbf{26},
  042001\relax
\mciteBstWouldAddEndPuncttrue
\mciteSetBstMidEndSepPunct{\mcitedefaultmidpunct}
{\mcitedefaultendpunct}{\mcitedefaultseppunct}\relax
\EndOfBibitem
\bibitem[Esplandiu \emph{et~al.}(2016)Esplandiu, Farniya, and
  Reguera]{esplandiu16}
M.~Esplandiu, A.~Farniya and D.~Reguera, \emph{J. Chem. Phys.}, 2016,
  \textbf{144}, 124702\relax
\mciteBstWouldAddEndPuncttrue
\mciteSetBstMidEndSepPunct{\mcitedefaultmidpunct}
{\mcitedefaultendpunct}{\mcitedefaultseppunct}\relax
\EndOfBibitem
\bibitem[Midmore \emph{et~al.}(1996)Midmore, Pratt, and Herrington]{midmore96}
B.~R. Midmore, G.~V. Pratt and T.~M. Herrington, \emph{J. Colloid Interface
  Sci.}, 1996, \textbf{184}, 170--174\relax
\mciteBstWouldAddEndPuncttrue
\mciteSetBstMidEndSepPunct{\mcitedefaultmidpunct}
{\mcitedefaultendpunct}{\mcitedefaultseppunct}\relax
\EndOfBibitem
\bibitem[Das \emph{et~al.}(2015)Das, Garg, Campbell, Howse, Sen, Velegol,
  Golestanian, and Ebbens]{das15}
S.~Das, A.~Garg, A.~I. Campbell, J.~Howse, A.~Sen, D.~Velegol, R.~Golestanian
  and S.~J. Ebbens, \emph{Nature communications}, 2015, \textbf{6}, 8999\relax
\mciteBstWouldAddEndPuncttrue
\mciteSetBstMidEndSepPunct{\mcitedefaultmidpunct}
{\mcitedefaultendpunct}{\mcitedefaultseppunct}\relax
\EndOfBibitem
\bibitem[Xuan \emph{et~al.}(2014)Xuan, Shao, Lin, Dai, and He]{xuan14}
M.~Xuan, J.~Shao, X.~Lin, L.~Dai and Q.~He, \emph{ChemPhysChem}, 2014,
  \textbf{15}, 2255--2260\relax
\mciteBstWouldAddEndPuncttrue
\mciteSetBstMidEndSepPunct{\mcitedefaultmidpunct}
{\mcitedefaultendpunct}{\mcitedefaultseppunct}\relax
\EndOfBibitem
\bibitem[Wheat \emph{et~al.}(2010)Wheat, Marine, Moran, and Posner]{wheat10}
P.~M. Wheat, N.~A. Marine, J.~L. Moran and J.~D. Posner, \emph{Langmuir}, 2010,
  \textbf{26}, 13052--13055\relax
\mciteBstWouldAddEndPuncttrue
\mciteSetBstMidEndSepPunct{\mcitedefaultmidpunct}
{\mcitedefaultendpunct}{\mcitedefaultseppunct}\relax
\EndOfBibitem
\bibitem[Golestanian(2015)]{golestanian15}
R.~Golestanian, \emph{Phys. Rev. Lett.}, 2015, \textbf{115}, 108102\relax
\mciteBstWouldAddEndPuncttrue
\mciteSetBstMidEndSepPunct{\mcitedefaultmidpunct}
{\mcitedefaultendpunct}{\mcitedefaultseppunct}\relax
\EndOfBibitem
\bibitem[Ono \emph{et~al.}(1977)Ono, Matsumura, Kitajima, and Fukuzumi]{ono77}
Y.~Ono, T.~Matsumura, N.~Kitajima and S.~Fukuzumi, \emph{J. Phys. Chem.}, 1977,
  \textbf{81}, 1307--1311\relax
\mciteBstWouldAddEndPuncttrue
\mciteSetBstMidEndSepPunct{\mcitedefaultmidpunct}
{\mcitedefaultendpunct}{\mcitedefaultseppunct}\relax
\EndOfBibitem
\bibitem[McKee(1969)]{mckee69}
D.~W. McKee, \emph{J. Catalysis}, 1969, \textbf{14}, 355\relax
\mciteBstWouldAddEndPuncttrue
\mciteSetBstMidEndSepPunct{\mcitedefaultmidpunct}
{\mcitedefaultendpunct}{\mcitedefaultseppunct}\relax
\EndOfBibitem
\bibitem[Afshar~Farniya \emph{et~al.}(2013)Afshar~Farniya, Esplandiu, Reguera,
  and Bachtold]{afsharfarniya13}
A.~Afshar~Farniya, M.~J. Esplandiu, D.~Reguera and A.~Bachtold, \emph{Phys.
  Rev. Lett.}, 2013, \textbf{111}, 168301\relax
\mciteBstWouldAddEndPuncttrue
\mciteSetBstMidEndSepPunct{\mcitedefaultmidpunct}
{\mcitedefaultendpunct}{\mcitedefaultseppunct}\relax
\EndOfBibitem
\bibitem[Kim and Yoon(2003)]{kim03}
J.~Y. Kim and B.~J. Yoon, \emph{J. Colloid Interf. Sci.}, 2003, \textbf{262},
  101--106\relax
\mciteBstWouldAddEndPuncttrue
\mciteSetBstMidEndSepPunct{\mcitedefaultmidpunct}
{\mcitedefaultendpunct}{\mcitedefaultseppunct}\relax
\EndOfBibitem
\bibitem[Riley \emph{et~al.}(2006)Riley, Hobson, and Bence]{riley06}
K.~F. Riley, M.~P. Hobson and S.~J. Bence, \emph{Mathematical methods for
  physics and engineering: a comprehensive guide}, Cambridge University Press,
  2006\relax
\mciteBstWouldAddEndPuncttrue
\mciteSetBstMidEndSepPunct{\mcitedefaultmidpunct}
{\mcitedefaultendpunct}{\mcitedefaultseppunct}\relax
\EndOfBibitem
\bibitem[Sabass and Seifert(2012)]{sabass12}
B.~Sabass and U.~Seifert, \emph{J. Chem. Phys.}, 2012, \textbf{136},
  064508\relax
\mciteBstWouldAddEndPuncttrue
\mciteSetBstMidEndSepPunct{\mcitedefaultmidpunct}
{\mcitedefaultendpunct}{\mcitedefaultseppunct}\relax
\EndOfBibitem
\bibitem[O'Brien and White(1978)]{obrien78}
R.~W. O'Brien and L.~R. White, \emph{J. Chem. Soc.- Faraday T. 2}, 1978,
  \textbf{74}, 1607--1626\relax
\mciteBstWouldAddEndPuncttrue
\mciteSetBstMidEndSepPunct{\mcitedefaultmidpunct}
{\mcitedefaultendpunct}{\mcitedefaultseppunct}\relax
\EndOfBibitem
\bibitem[Debye(1942)]{debye42}
P.~Debye, \emph{T. Electrochem. Soc.}, 1942, \textbf{82}, 265--272\relax
\mciteBstWouldAddEndPuncttrue
\mciteSetBstMidEndSepPunct{\mcitedefaultmidpunct}
{\mcitedefaultendpunct}{\mcitedefaultseppunct}\relax
\EndOfBibitem
\bibitem[Laidler(1987)]{laidler87}
\emph{Chemical Kinetics, 3rd ed.}, ed. K.~J. Laidler, Harper Collins, New York,
  U.S.A., 1987\relax
\mciteBstWouldAddEndPuncttrue
\mciteSetBstMidEndSepPunct{\mcitedefaultmidpunct}
{\mcitedefaultendpunct}{\mcitedefaultseppunct}\relax
\EndOfBibitem
\bibitem[Pocker and Bjorkquist(1977)]{pocker77}
Y.~Pocker and D.~W. Bjorkquist, \emph{J. Am. Chem. Soc.}, 1977, \textbf{99},
  6537--6543\relax
\mciteBstWouldAddEndPuncttrue
\mciteSetBstMidEndSepPunct{\mcitedefaultmidpunct}
{\mcitedefaultendpunct}{\mcitedefaultseppunct}\relax
\EndOfBibitem
\bibitem[Hunter(2001)]{hunter01}
R.~J. Hunter, \emph{Foundations of colloid science}, Oxford University Press,
  2001\relax
\mciteBstWouldAddEndPuncttrue
\mciteSetBstMidEndSepPunct{\mcitedefaultmidpunct}
{\mcitedefaultendpunct}{\mcitedefaultseppunct}\relax
\EndOfBibitem
\end{mcitethebibliography}
%\bibliographystyle{rsc} %the RSC's .bst file
%}

\providecommand*{\mcitethebibliography}{\thebibliography}
\csname @ifundefined\endcsname{endmcitethebibliography}
{\let\endmcitethebibliography\endthebibliography}{}

\end{document}